\documentclass[floatfix, superscriptaddress, longbibliography, twocolumn, a4paper]{revtex4-1}

\usepackage{times}
\usepackage[table]{xcolor}
\usepackage{graphicx} 

\usepackage{bm, amsmath, amssymb, mathtools}
\usepackage{siunitx}

\usepackage[table]{xcolor}
\definecolor{green}{HTML}{008900}

\newcommand{\ket}[1]{\left|#1\right>}

\newcommand{\qop}{{\bf q}}
\newcommand{\pop}{{\bf p}}
\newcommand{\SX}{{\bf S_a}}
\newcommand{\SZ}{{\bf S_b}}

\newcommand{\SXH}{{\bf S_a}}
\newcommand{\SYH}{{\bf S_b}}
\newcommand{\SZH}{{\bf S_c}}
\newcommand{\SXZ}{{\bf S_{a, b}}}
\newcommand{\X}{{\bf X}}
\newcommand{\Z}{{\bf Z}}
\newcommand{\Y}{{\bf Y}}
\newcommand{\bsigma}{\boldsymbol{\sigma}}
\newcommand{\rhoo}{\boldsymbol{\rho}}
\newcommand{\PP}{\boldsymbol{\mathcal{P}}}

\newcommand{\D}{{\bf D}}
\newcommand{\I}{{\bf I}}
\newcommand{\Ha}{{\bf H_1}}
\newcommand{\Hb}{{\bf H_2}}
\newcommand{\Hc}{{\bf H_3}}
\newcommand{\CD}{{\bf CD}}
\newcommand{\U}{{\bf U}}
\newcommand{\G}{{\bf G}}
\newcommand{\V}{{\bf V}}
\newcommand{\betap}{{\bf \boldsymbol{\beta}_{\perp}}}
\newcommand{\M}{{\bf M}}
\newcommand{\N}{{N}}

\newcommand{\R}{{\bf R}}
\newcommand{\br}{{\bf r}}

\newcommand{\LL}{{\bf L}}

\usepackage{soul}

\newcommand{\sigmax}{{\bf \bsigma_x}}
\newcommand{\sigmay}{{\bf \bsigma_y}}
\newcommand{\sigmaz}{{\bf \bsigma_z}}
\newcommand{\sigmam}{{\bf \bsigma_-}}

\newcommand{\aop}{{\bf a}}
\newcommand{\adag}{{\bf a^{\dag}}}

\usepackage[a4paper, paperwidth=21.0cm, paperheight=29.7cm, inner=0.75in, outer=0.5in, top=0.5in, bottom=0.5in, includefoot]{geometry}

\usepackage{fancyhdr}
\begin{document} 


\title{Quantum error correction of a qubit encoded in grid states of an oscillator}

\author{P. Campagne-Ibarcq}
\thanks{These authors contributed equally.}
\affiliation{Department of Applied Physics and Physics, Yale University, New Haven, CT 06520, USA}
\author{A. Eickbusch}
\thanks{These authors contributed equally.}
\affiliation{Department of Applied Physics and Physics, Yale University, New Haven, CT 06520, USA}
\author{S. Touzard}
\thanks{These authors contributed equally.}
\affiliation{Department of Applied Physics and Physics, Yale University, New Haven, CT 06520, USA}
\author{E. Zalys-Geller}
\affiliation{Department of Applied Physics and Physics, Yale University, New Haven, CT 06520, USA}
\author{N.E. Frattini}
\affiliation{Department of Applied Physics and Physics, Yale University, New Haven, CT 06520, USA}
\author{V.V. Sivak}
\affiliation{Department of Applied Physics and Physics, Yale University, New Haven, CT 06520, USA}
\author{P. Reinhold}
\affiliation{Department of Applied Physics and Physics, Yale University, New Haven, CT 06520, USA}
\author{S. Puri}
\affiliation{Department of Applied Physics and Physics, Yale University, New Haven, CT 06520, USA}
\author{S. Shankar}
\affiliation{Department of Applied Physics and Physics, Yale University, New Haven, CT 06520, USA}
\author{R.J. Schoelkopf}
\affiliation{Department of Applied Physics and Physics, Yale University, New Haven, CT 06520, USA}
\author{L. Frunzio}
\affiliation{Department of Applied Physics and Physics, Yale University, New Haven, CT 06520, USA}
\author{M. Mirrahimi}
\affiliation{Quantic Team, Inria Paris,
2 rue Simone Iff, 75012 Paris, France}
\author{M.H. Devoret}
\affiliation{Department of Applied Physics and Physics, Yale University, New Haven, CT 06520, USA}

\date{\today}

\maketitle

\textbf{Quantum bits are more robust to noise when they are encoded non-locally \cite{Shor1996}. In such an encoding, errors affecting the underlying physical system can then be detected and corrected before they corrupt the encoded information. In 2001, Gottesman, Kitaev and Preskill (GKP) proposed a hardware-efficient instance of such a qubit, which is delocalised in the phase-space of a single oscillator \cite{gottesman2001encoding}. However, implementing measurements that reveal error syndromes of the oscillator while preserving the encoded information \cite{travaglione2002preparing, Pirandola2006, vasconcelos2010all, terhal2016encoding, motes2017encoding} has proved experimentally challenging: the only realisation so far relied on post-selection \cite{fluhmann2018sequential, fluhmann2018encoding}, which is incompatible with quantum error correction (QEC). The novelty of our experiment is precisely that it implements these non-destructive error-syndrome measurements for a superconducting microwave cavity. We design and implement an original feedback protocol that incorporates such measurements to prepare square and hexagonal GKP code states. We then demonstrate QEC of an encoded qubit with unprecedented suppression of all logical errors, in quantitative agreement with a theoretical estimate based on the measured imperfections of the experiment. Our protocol is applicable to other continuous variable systems and, in contrast with previous implementations of QEC \cite{Waldherr2014, Kelly2015, Cramer2016, ofek2016extending, hu2019quantum}, can mitigate all logical errors generated by a wide variety of noise processes, and open a way towards fault-tolerant quantum computation.}

\begin{figure}

\includegraphics{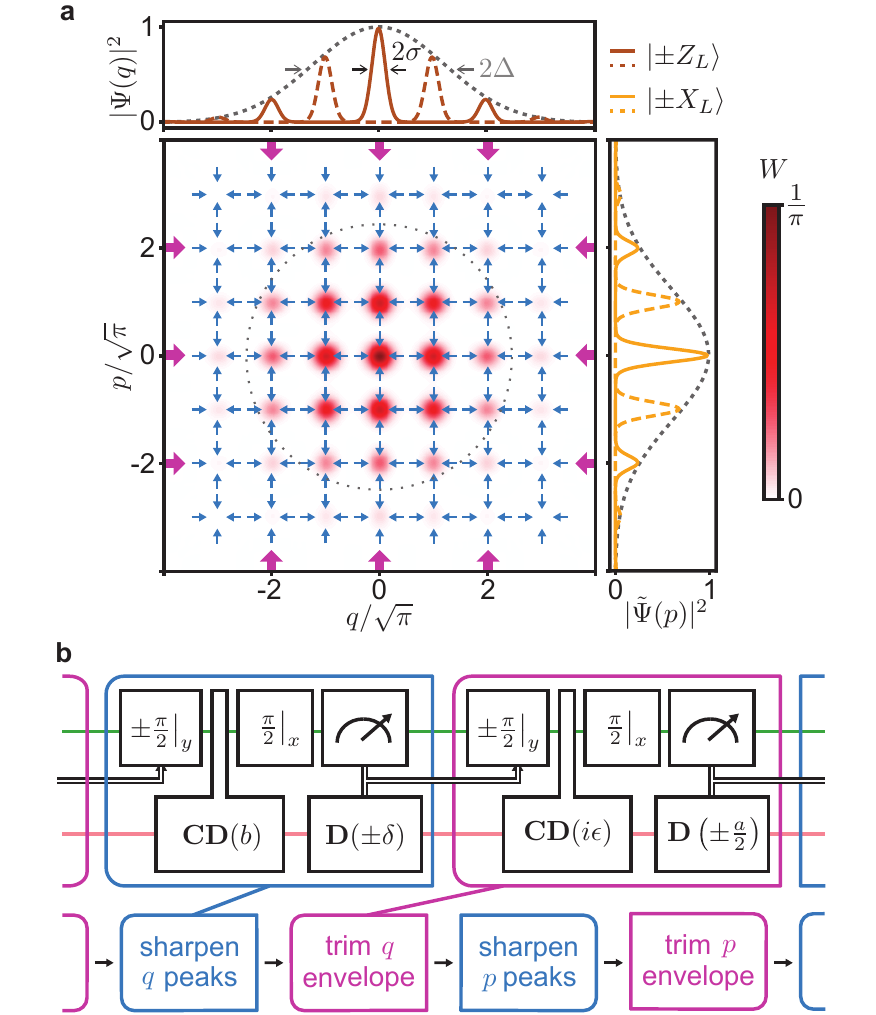}
    \caption{\label{fig1} {\bf Quantum error correction protocol.} {\bf a,} Simulated Wigner function of the fully mixed logical state in a code defined by a width $\sigma=0.25$ for the peaks and $\Delta=1/(2\sigma)=2$ for the normalising envelope. Our QEC protocol prevents the squeezed peaks from spreading (blue arrows) and the overall envelope from extending (purple arrows). Side panels represent the probability distributions of the $|\pm X_L \rangle$ and $|\pm Z_L \rangle$ states along each quadrature, which retain disjoint supports along $q$ or $p$  under stabilisation. {\bf b,} The full QEC protocol alternates indefinitely two peak-sharpening rounds and two envelope-trimming rounds to prevent spreading of the grid state peaks and envelope in phase-space (respectively blue and purple arrows in {\bf a}). Each round, a conditional displacement entangles the transmon (green line) and the storage oscillator (salmon line). A subsequent measurement of the transmon controls the sign of a feedback shift of the oscillator and of a $\pi/2$-rotation resetting the transmon (double-stroke black arrows). The peak-sharpening shift $\delta\simeq0.2$ maximises the stabiliser value in steady-state, and the envelope-trimming conditional displacement by $\epsilon\simeq0.2$ sets the width of the grid state envelope (see supplementary information), which is optimal given experimental constraints.}

\end{figure}

The qubit encoding proposed by GKP is based on grid patterns in phase-space, which only emerge by interfering periodically spaced position eigenstates with adequate phase relationships, as represented in Fig.~\ref{fig1}. The resulting ``grid state'' code belongs to the class of stabiliser codes. In the stabiliser formalism of QEC, the measurement of chosen operators - the stabilisers - reveals unambiguously the action of undesired noise without disturbing the state of the logical qubit. As a consequence of this latter condition, the stabilisers must commute with all observables of the logical qubit, which are combinations of the logical Pauli operators. For the grid state code, these operators are phase-space displacements, defined as $\D(\beta)=e^{-i \mathrm{Re}(\beta)\pop+i \mathrm{Im}(\beta)\qop}$, where $\qop$ and $\pop$ are the conjugated position and momentum operators, such that $[\qop, \pop]=i$. For example, the stabilisers of the canonical square grid state code are $\SX=\D(a=2\sqrt{\pi})$ and $\SZ=\D(b=2i\sqrt{\pi})$, and the Pauli operators are $\X=\D(a/2)$, $\Z=\D(b/2)$, and $\Y=\D((a+b)/2)$. The phase of the stabilisers encodes no information about the logical qubit but reveals momentum shifts modulo $2\pi/|a|$ and position shifts modulo $2\pi/|b|$. Thus, shifts that are smaller than a quarter of a grid period are unambiguously identified and can be corrected. Since common decoherence processes such as photon relaxation~\cite{albert2017performance, noh2019gkp}, pure dephasing or spurious non-linearities result in a continuous evolution of quasi-probability distribution in phase-space~\cite{Cahill1969, gottesman2001encoding}, shifts of order $a, b$ do not occur instantaneously. Therefore, if the stabilisers are measured frequently enough, noise-induced shifts can be detected and corrected, which inhibits all logical errors.

However, contrary to this description based on ideal position eigenstates, physically realisable code states do not extend infinitely in phase-space: they are superpositions of periodically spaced squeezed states of width $\sigma$, with a Gaussian overall envelope of width $\Delta=1/(2\sigma)$ (see Fig~\ref{fig1}a). These states are still approximate eigenstates of the stabilisers, such that $|\langle \SXZ \rangle|\simeq 1$, and are still approximate eigenstates of the Pauli operators. Any pair of orthogonal code states are shifted from one another in phase-space (e.g. by $a/2$ for $\ket{\pm Z_L}$, $b/2$ for $\ket{\pm X_L}$). For sufficient squeezing, their supports do not significantly overlap, the logical qubit is well-defined and a QEC protocol can be directly adapted from the ideal case.

\section{Measurement of displacement operators}

The expectation value of displacement operators $\D(\beta)$, such as the stabilisers and Pauli operators of the GKP code, are periodic functions of a generalised quadrature $\br = \mathrm{Re}(\beta)\pop+i \mathrm{Im}(\beta)\qop$. We measure these so-called ``modular variables'' \cite{Aharonov1969, Popescu2010, fluhmann2018sequential} by effectively coupling the quadrature of an oscillator to a generator of rotations of an ancillary physical qubit, here its Pauli Z ($\sigmaz)$. In our experiment, the oscillator is the fundamental mode of a reentrant coaxial microwave cavity made out of bulk aluminum \cite{reagor2015quantum}, which we call the storage mode, and the ancillary physical qubit is a transmon (see methods). The storage mode has a single-photon lifetime $T_s=245~\mu$s, while the transmon has energy and coherence lifetimes of $T_1=50~\mu$s and $T_{2E}=60~\mu$s and can be non-destructively measured in 700 ns via an ancillary low-Q resonator. Interestingly, the desired coupling $\br\sigmaz$ between the storage and transmon can be effectively activated with microwave drives in the presence of the naturally present dispersive interaction \cite{Wallraff2004}, even with arbitrarily weak interaction strength. Schematically, when the storage is displaced far from the origin of phase-space, the dispersive interaction results in two quickly separating trajectories, each corresponding to a different transmon eigenstate. We employ this evolution within a sequence of fast storage displacements intertwined with transmon rotations to engineer an arbitrary conditional displacement in 1.1$\mu$s, following the unitary evolution  $\CD(\beta)= e^{i \big( -\mathrm{Re}(\beta) \pop + \mathrm{Im}(\beta) \qop \big)\frac{\sigmaz}{2}}$ (see methods). This entangling gate can equivalently be viewed as a rotation of the transmon's Bloch vector around the $\sigmaz$ axis by an angle dependent on the phase-space distribution of the storage mode. When applied on the transmon initialised on the equator of its Bloch sphere, it leads to  $\langle \sigmax - i \sigmay \rangle = \langle \D(\beta)\rangle$~\cite{fluhmann2018sequential}. Intuitively, as the measurement of a displacement is a measurement of a quadrature modulo the grid pitch, the conditional displacement is such that two oscillator quadrature eigenstates separated by $2n\pi/\beta$ induce the same qubit rotation up to an integer number of turns $n$.

Conditional displacements embedded within a transmon Ramsey sequence enable the measurement of the code stabilisers and, therefore, lay at the heart of the quantum error correction of GKP codes \cite{kitaev1995quantum, svore2013faster, weigand2018generating}. Conveniently, this sequence is also employed to obtain the expectation value of any displacement operator $\langle \D(\beta) \rangle$ for an arbitrary state of the storage oscillator. This leads to the state characteristic function $C(\beta)$, which is the two-dimensional Fourier transform of the Wigner function \cite{Haroche2006, walls2007quantum}. This complex-valued representation fully characterises an arbitrary state. In our experiment, we measure $\mathrm{Re}(C(\beta))$, which contains the information about the symmetric component of the Wigner function, to characterise the generated grid states as represented in Fig.~\ref{fig2}--\ref{fig4}. The imaginary part $\mathrm{Im}(C(\beta))$ contains information about the anti-symmetric component of the Wigner function and is expected to take a uniform null value for the symmetric grid states we consider. We verify this property at critical points. 

\section{Convergence to the GKP code manifold}

\begin{figure}
\includegraphics{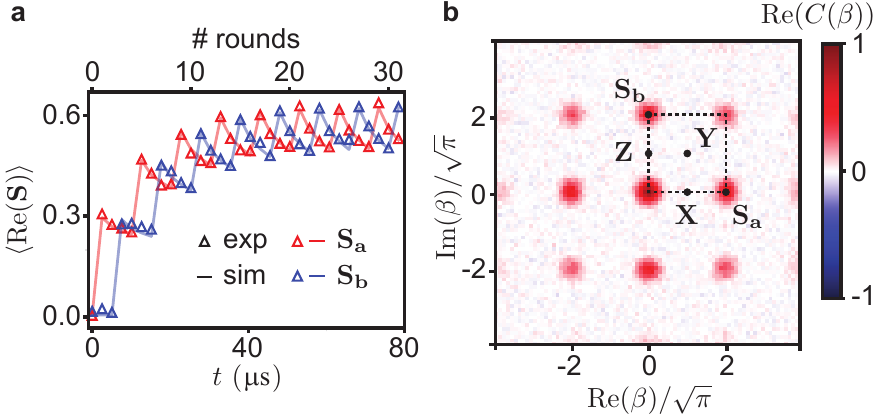}
    \caption{\label{fig2} {\bf Square code in the steady-state of the QEC protocol.} {\bf a,} Measured average value of the code stabilisers real part when turning on the QEC protocol from the vacuum state. Each stabiliser oscillates over a 4-round period as a result of the periodicity of the QEC protocol, and steady-state is reached in about 20~rounds. {\bf b,} Real part of the measured characteristic function of the storage mode in steady-state (after 200 rounds). Particular points corresponding to stabilisers and Pauli operators are indicated by black circles, and the dashed lines enclose an area of $4\pi$.}
\end{figure} 

We now derive a QEC protocol that employs the conditional displacement gate described above to protect finite-size grid states. Note that there exists an optimal width of the envelope $\Delta$, resulting from a trade off: more extended grid states have better resolved peaks and are thus more robust against shifts, but are more sensitive to dissipation (see supplementary information). Therefore, our protocol is designed, first, to keep the oscillator state probability distribution peaked in phase-space at $q,~p=0~\mathrm{mod}~2\pi/|a|,~2\pi/|b|$, and second, to prevent the overall envelope from drifting or expanding more than necessary. Given our experimental constraints, we work with a finite size GKP code with envelope width $\Delta\simeq 3.2$, chosen to maximise the coherence time of the logical qubit (see supplementary information). 

From the discussion above, maintaining the phase-space distribution peaked at the grid points involves mapping the stabilisers $\SX$ or $\SZ$ onto the ancilla transmon with conditional displacements and then performing actuating displacements based on transmon measurements. As the measurement of the transmon only yields a binary outcome, these steps are constructed to answer the simple questions ``has the grid moved up or down?'' when measuring $\SX$, and similarly ``left or right?'' when measuring $\SZ$. After each measurement, we apply a fixed-length displacement in the direction opposite to the one indicated by the answer. The combination of the back-action of the measurements and of our feedback sharpens the peaks of the grid states. Similar measurements of small displacement operators and feedback trim the envelope of the grid states to keep it from drifting and expanding (see supplementary information). The repeated action of this basic protocol forms a discrete-time Markovian sequence leading to an effective dissipative force that pushes the state of the storage oscillator toward the code manifold, as depicted in Fig.~\ref{fig1}a. This engineered dissipation counteracts the evolution due to noise, thereby inhibiting logical errors.

Starting from the ground state of the oscillator, we apply this protocol indefinitely as summarised in Fig.~\ref{fig1}b. In Fig.~\ref{fig2}a we plot the measured average values of $\mathrm{Re}(\SX)$ and $\mathrm{Re}(\SZ)$ after $n$ correction rounds. The stabiliser values increase rapidly to converge to a steady-state in about 20 rounds. On top of this trend, the mean value of each stabiliser oscillates over a period of 4 rounds by increasing to 0.62 when the peaks are sharpened in the corresponding phase-space quadrature and then decay down to 0.5 over the 3 next rounds. Beyond this periodic oscillation, the stabilisers do not evolve over hundreds of rounds (not shown), which indicates that our protocol has entered a steady-state. The characterisation of this steady-state can now reveal if it corresponds to the desired GKP manifold.

We plot the real part of the characteristic function of the steady state after 200 rounds in Fig.~\ref{fig2}b. This state is a maximally mixed state of the logical qubit, as can be seen from the null value of the points corresponding to the three logical Pauli operators. Note that this characteristic function representation is the Fourier conjugate of the theoretical Wigner representation given in Fig.~\ref{fig1}a. However, the two are similar for grid states since the Fourier transform of a grid is itself a grid. Our results are quantitatively reproduced by master-equation simulations (lines in Fig.~\ref{fig2}a) whose parameters are all independently calibrated. From these simulations, we estimate that the squeezing of the peaks of the generated grid states oscillates between $ 7.4$~dB and 9.5~dB in steady-state ---close to the level required for fault-tolerant quantum computation~\cite{glancy2006,fukui2018high,vuillot2019quantum}--- and the average photon number oscillates between 8.6 and 10.2 photons. 

\section{Logical qubit initialisation}

Once the oscillator has reached its steady state, it is in the code manifold and we initialise the logical qubit by replacing one of the QEC rounds with a measurement of $\X, \Y$ or $\Z$. To measure the logical Pauli operators, we first prepare the transmon in $|+x\rangle$ and then apply the conditional displacement $\CD(\beta)$ with $\beta= a/2, (a+b)/2, b/2$ respectively. After the sequence, $\langle \sigmax-i\sigmay\rangle= \langle \X\rangle, \langle \Y\rangle~\mathrm{or}~ \langle \Z\rangle$ and a subsequent $\sigmax$ readout of the transmon with outcome $\pm 1$  heralds the preparation of the approximately orthogonal states $|\pm X_L\rangle$, $|\pm Y_L\rangle$ or $|\pm Z_L\rangle$, up to a re-centering displacement (see supplementary information).

However, as $\X$ ,$\Y$ or $\Z$ differ from the Pauli operators of the finitely squeezed code we consider, the sequence described above results in a readout of the logical qubit with non-unit fidelity and in an imperfect initialisation. Fortunately, when this sequence is followed by a few QEC rounds projecting the generated state back onto the code manifold, this readout is non-demolition for the target logical state and can be repeated to increase its fidelity (see supplementary information). In Fig.~\ref{fig3}a (respectively b), we show the characteristic function of the storage state obtained when two $\X$ (respectively $\Y$) measurements, separated by four QEC rounds, yield the same outcome. The Pauli operators value in these two cases are respectively $\langle \mathrm{Re}(\X)\rangle =-0.8$ and $\langle \mathrm{Re}(\Y)\rangle =-0.63$. We insist here that these values do not reflect the preparation fidelity to the finitely squeezed logical states $|- X_L\rangle$ and $|-Y_L\rangle$, and the prepared state is as close, within experimental uncertainties, to the target state as allowed by the imperfect code correction (see supplementary information). The same methods are applied to prepare eigenstates of other Pauli operators (data not shown). In particular, the characteristic function of the $|- Z_L\rangle$ state is the same as the one of $| - X_L\rangle$ rotated by $90^{\circ}$ (not shown).

\section{Coherence of the error-corrected logical qubit}

\begin{figure}

\includegraphics{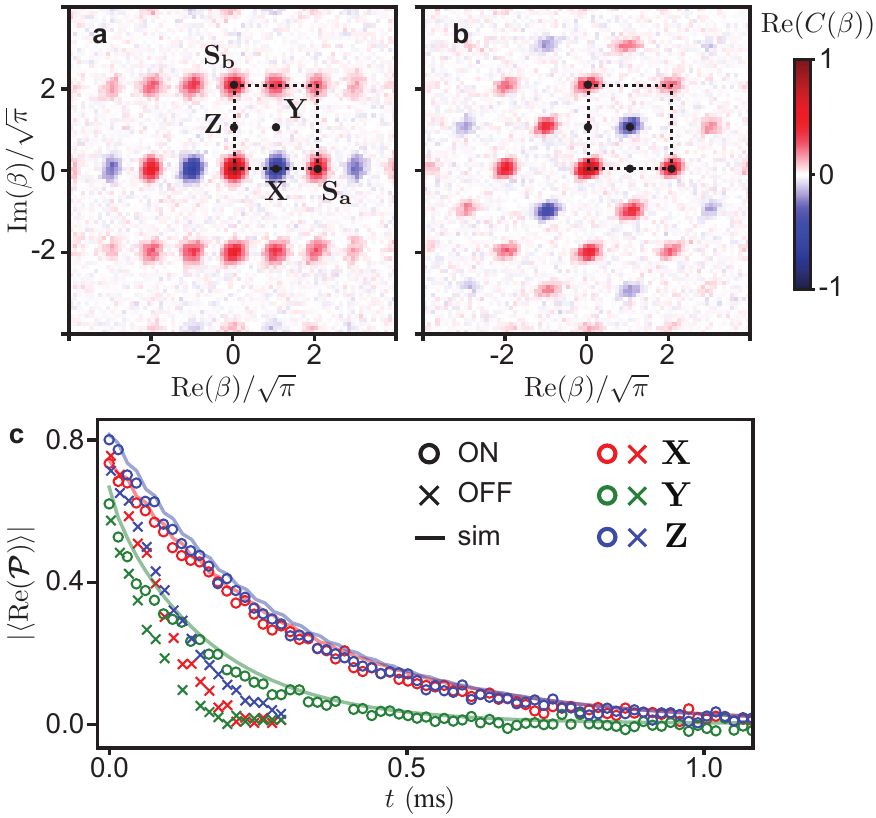}
    \caption{\label{fig3} {\bf Initialisation and coherence characterisation of the logical qubit in the square encoding.} {\bf a,} Characteristic function of $|-X_L\rangle$ prepared, in steady-state, by applying a feedback $\Z$-gate conditioned on the outcome $+1$ of a first single-round $\mathrm{Re}(\X)$ measurement, before heralding a higher fidelity state on the outcome $-1$ of a second identical measurement. {\bf b,} Same procedure to prepare $|-Y_L\rangle$. {\bf c,} After preparing $|-X_L\rangle$, $|-Y_L\rangle$ or $|-Z_L\rangle$, the time-decay of the real part of $\PP=\X,\Y$ or $\Z$, respectively, is measured when continuously applying the QEC protocol (ON) or not (OFF). The QEC protocol extends the lifetime of the 3 Bloch vector components to $T_X=T_Z=275~\mu\mathrm{s}$ and $T_Y=160~\mu\mathrm{s}$, and the results are quantitatively reproduced by master-equation simulations (lines). 
    }

\end{figure} 

In order to test the error-correction performance of our protocol, we prepare one of the logical states $| - X_L\rangle, |-Y_L\rangle$ or $|-Z_L\rangle$, and compare the decay of the mean value of the real part of the corresponding operator $\PP=\X,\Y$ or $\Z$ in time when performing QEC (open circles in Fig.~\ref{fig3}b) or not (crosses). In all three cases, our protocol extends the coherence of the logical qubit. We extract the coherence times of the error-corrected qubit $T_X=T_Z=275~\mu\mathrm{s}$ and $T_Y=160~\mu\mathrm{s}$. The shorter coherence time of the $\Y$ Pauli operator, also visible in the uncorrected case, is expected as the distance in phase-space from the probability peaks of the $|+ Y_L\rangle$ state to those of the $|- Y_L\rangle$ state is $\sqrt{2}$ shorter than in the case of $|\pm X_L\rangle$ and $|\pm Z_L\rangle$. Therefore, diffusive shifts in phase-space induced by photon dissipation cause more flips of the $\Y$ component of the logical qubit Bloch vector. Master-equation simulations reproduce these results quantitatively.

\section{Hexagonal code}

\begin{figure}

\includegraphics{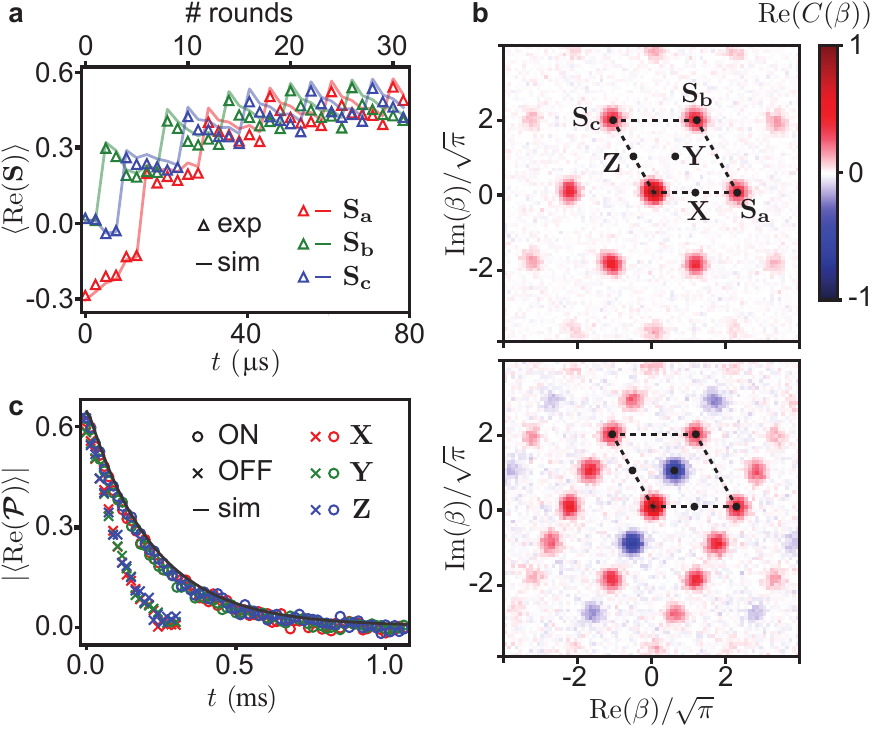}
    \caption{\label{fig4} {\bf Convergence to the code manifold, state-preparation and coherence in the hexagonal code.} {\bf a,} The grid state peaks and envelope are sequentially sharpened and trimmed along three directions. When turning on our protocol from the ground state of the oscillator, the real part of the stabilisers expectation values oscillate every six rounds and increase to rapidly reach a steady-state. {\bf b,} After 200 rounds, the oscillator state is a fully mixed logical state revealing the code structure (top). Eigenstate of a Pauli operator, such as $|-Y_L\rangle$  (bottom) can be prepared by single-round measurement of Re($\Y$) followed by a feedback displacement. {\bf c,} Due to the code symmetry, the decay of the logical Bloch vector is isotropic. An exponential fit (black line) indicates a lifetime of 205~$\mu$s, enhanced by QEC. }

\end{figure} 

We have executed a variant of the square code of Fig.~\ref{fig1} known as the hexagonal code, in which the decay times of all three Pauli operators are equal by symmetry. In general, a two-dimensional grid state code is defined as the common eigenspace of any two commuting stabilisers $\SXH=\D(a)$ and $\SYH=\D(b)$, as long as $\mathrm{Im}(a^*b)=4\pi$. Geometrically, this condition implies that the magnitude of the cross-product of the two vectors representing these stabilisers correspond to an area of $4\pi$ (see Fig.~\ref{fig2}b, \ref{fig4}b). In the hexagonal GKP code~\cite{gottesman2001encoding}, we have $b=a e^{i\frac{\pi}{3}}$, which respects the above area condition for $a=\sqrt{\frac{8\pi}{\sqrt{3}}}$. The Pauli operators correspond to displacements of equal length $\X=\D(a/2)$, $\Y=\D(b/2)$ and $\Z=\D(c/2)$ with $c=a e^{i\frac{2\pi}{3}}$. For symmetry reasons, we also define a third stabiliser $\SZH=\Z^2=\D(c)$ that commutes with the two others.

We perform QEC on this code by adapting the protocol described in Sec.~2. Here, measurement of the 3 hexagonal stabilisers followed by small corrective feedback displacements sharpen the peaks along three different directions. These are interleaved with measurement of 3 short displacement operators, which trim the envelope. When applying this protocol on the storage mode initialised in the ground state, the stabilisers mean values oscillate every 6 rounds as each of these displacement operators is measured in turn, and rapidly converge to a stationary regime for which their value oscillate between 0.4 and 0.55 (see Fig.~\ref{fig4}a). We measure the real part of the characteristic function of the fully mixed logical state reached after 200~rounds, which reveals the hexagonal structure of the code (Fig.~\ref{fig4}b). Here again, master equation simulations reproduce quantitatively these results and indicate that the generated grid states are characterised by the same squeezing for the peaks as in the square encoding (between 7.5~dB and 9.5~dB in steady-state). Note that the temporary negative value of $\mathrm{Re}(\SX)$ registered at short times originates from the particular programming of the feedback algorithm on the fast-electronics FPGA board: the oscillator state gets shifted at the beginning of the sequence, which is included in simulations.

We prepare the logical qubit in an eigenstate of each Pauli operator with a single-round measurement of $\mathrm{Re}(\X)$, $\mathrm{Re}(\Y)$ or $\mathrm{Re}(\Z)$. In Fig.~\ref{fig4}b, we show the measured characteristic function of the $| -Y_L \rangle$ state. Note that the characteristic functions of $| -X_L \rangle$ and $| -Z_L \rangle$ are equal to the one of $| -Y_L \rangle$ but rotated by $\pm 60^{\circ}$ (not shown). Finally we characterise the coherence of the error-corrected logical qubit by measuring the decay of the Pauli operator mean values in time. As expected, the decoherence of the logical qubit is now isotropic and significantly extended compared to the uncorrected case, with coherence times of $T_X=T_Y=T_Z=205~\mu$s.

\section{Logical errors and outlook}

The coherence of the logical qubit is limited by two factors. First, the duration of the error-correction rounds, despite being a factor of a hundred shorter than the storage mode single-photon lifetime, is not negligible. The transmon readout and its processing using fast electronics accounts for about half of this duration, and the conditional displacement gate for the other half. Although the gate speed is limited in our current implementation, alternative implementations could result in faster gates \cite{Touzard2019}. The second factor limiting the coherence of the logical qubit are transmon errors. Among these, $\sigmaz$ errors (phase-flips) commute with the storage-transmon interaction Hamiltonian and thus do not propagate to the logical qubit (see supplementary information). On the other hand, the $\sigmax$ and $\sigmay$ transmon errors (bit-flips), as well as excitations to the higher excited states of the transmon, propagate to the logical qubit as they lead to random displacements of the storage mode. Simulations indicate that bit-flips of the transmon and the finite correction rate each account for about half of the error rate of the logical qubit.

The coherence of the logical qubit could be further improved by replacing the transmon with a noise-biased ancillary qubit~\cite{puri2018stabilized, Grimm2019, shi2019fault} and by using a superconducting cavity with larger quality factor \cite{reagor2015quantum}. This multipronged effort at improving the GKP code based on superconducting circuits will be particularly rewarding since fault-tolerant single and multi-qubit Clifford gates can be implemented in a straightforward way \cite{gottesman2001encoding, Gao2018}, and such logical qubits can be embedded in further layers of protection~\cite{glancy2006, fukui2018high, vuillot2019quantum, baragiola2019all}.

\section*{Acknowledgments}
The authors thank Christa Fl\"uhmann, Jonathan Home, Steven Girvin, Liang Jiang and Kyungjoo Noh for helpful discussions, and M. Rooks for fabrication assistance. M. Mirrahimi thanks the Yale Quantum Institute for hosting him during the time he was collaborating on this project. Facilities use was supported by YINQE and the Yale SEAS cleanroom. This research was supported by ARO under Grant No. W911NF-18-1-0212 and ARO grant No. W911NF-16-1-0349. L.F., R.J.S. and M.H.D. are founders of QCI. L.F. and R.J.S. are shareholders of QCI. All authors but A.E. and E.Z.G. are inventors of patents related to the subject.

\section*{Data availability}

The experimental data and numerical simulations presented in this manuscript
are available from the corresponding authors upon request.

\section*{Author contributions}

P.C-I., A.E and S.T. designed and performed the experiment and analysed the data. E.Z-G. N.E.F., V.V.S., P.R., S.S., R.J.S. and L.F. contributed to the experimental apparatus, and S.P. and M.M. contributed theoretical support. M.H.D. supervised the project. P.C-I., A.E., S.T. and M.H.D. wrote the manuscript. All authors provided suggestions for the experiment, discussed the results and contributed to the manuscript.

\bibliography{GKP_bib}

\begin{thebibliography}{41}%
\makeatletter
\providecommand \@ifxundefined [1]{%
 \@ifx{#1\undefined}
}%
\providecommand \@ifnum [1]{%
 \ifnum #1\expandafter \@firstoftwo
 \else \expandafter \@secondoftwo
 \fi
}%
\providecommand \@ifx [1]{%
 \ifx #1\expandafter \@firstoftwo
 \else \expandafter \@secondoftwo
 \fi
}%
\providecommand \natexlab [1]{#1}%
\providecommand \enquote  [1]{``#1''}%
\providecommand \bibnamefont  [1]{#1}%
\providecommand \bibfnamefont [1]{#1}%
\providecommand \citenamefont [1]{#1}%
\providecommand \href@noop [0]{\@secondoftwo}%
\providecommand \href [0]{\begingroup \@sanitize@url \@href}%
\providecommand \@href[1]{\@@startlink{#1}\@@href}%
\providecommand \@@href[1]{\endgroup#1\@@endlink}%
\providecommand \@sanitize@url [0]{\catcode `\\12\catcode `\$12\catcode
  `\&12\catcode `\#12\catcode `\^12\catcode `\_12\catcode `\%12\relax}%
\providecommand \@@startlink[1]{}%
\providecommand \@@endlink[0]{}%
\providecommand \url  [0]{\begingroup\@sanitize@url \@url }%
\providecommand \@url [1]{\endgroup\@href {#1}{\urlprefix }}%
\providecommand \urlprefix  [0]{URL }%
\providecommand \Eprint [0]{\href }%
\providecommand \doibase [0]{http://dx.doi.org/}%
\providecommand \selectlanguage [0]{\@gobble}%
\providecommand \bibinfo  [0]{\@secondoftwo}%
\providecommand \bibfield  [0]{\@secondoftwo}%
\providecommand \translation [1]{[#1]}%
\providecommand \BibitemOpen [0]{}%
\providecommand \bibitemStop [0]{}%
\providecommand \bibitemNoStop [0]{.\EOS\space}%
\providecommand \EOS [0]{\spacefactor3000\relax}%
\providecommand \BibitemShut  [1]{\csname bibitem#1\endcsname}%
\let\auto@bib@innerbib\@empty
\bibitem [{\citenamefont {Shor}(1996)}]{Shor1996}%
  \BibitemOpen
  \bibfield  {author} {\bibinfo {author} {\bibfnamefont {P.}~\bibnamefont
  {Shor}},\ }in\ \href {\doibase 10.1109/SFCS.1996.548464} {\emph {\bibinfo
  {booktitle} {Proc. 37th Conf. Found. Comput. Sci.}}}\ (\bibinfo  {publisher}
  {IEEE Comput. Soc. Press},\ \bibinfo {year} {1996})\ pp.\ \bibinfo {pages}
  {56--65}\BibitemShut {NoStop}%
\bibitem [{\citenamefont {Gottesman}\ \emph {et~al.}(2001)\citenamefont
  {Gottesman}, \citenamefont {Kitaev},\ and\ \citenamefont
  {Preskill}}]{gottesman2001encoding}%
  \BibitemOpen
  \bibfield  {author} {\bibinfo {author} {\bibfnamefont {D.}~\bibnamefont
  {Gottesman}}, \bibinfo {author} {\bibfnamefont {A.}~\bibnamefont {Kitaev}}, \
  and\ \bibinfo {author} {\bibfnamefont {J.}~\bibnamefont {Preskill}},\
  }\href@noop {} {\bibfield  {journal} {\bibinfo  {journal} {Phys. Rev. A}\
  }\textbf {\bibinfo {volume} {64}},\ \bibinfo {pages} {012310} (\bibinfo
  {year} {2001})}\BibitemShut {NoStop}%
\bibitem [{\citenamefont {Travaglione}\ and\ \citenamefont
  {Milburn}(2002)}]{travaglione2002preparing}%
  \BibitemOpen
  \bibfield  {author} {\bibinfo {author} {\bibfnamefont {B.}~\bibnamefont
  {Travaglione}}\ and\ \bibinfo {author} {\bibfnamefont {G.~J.}\ \bibnamefont
  {Milburn}},\ }\href@noop {} {\bibfield  {journal} {\bibinfo  {journal} {Phys.
  Rev. A}\ }\textbf {\bibinfo {volume} {66}},\ \bibinfo {pages} {052322}
  (\bibinfo {year} {2002})}\BibitemShut {NoStop}%
\bibitem [{\citenamefont {Pirandola}\ \emph {et~al.}(2006)\citenamefont
  {Pirandola}, \citenamefont {Mancini}, \citenamefont {Vitali},\ and\
  \citenamefont {Tombesi}}]{Pirandola2006}%
  \BibitemOpen
  \bibfield  {author} {\bibinfo {author} {\bibfnamefont {S.}~\bibnamefont
  {Pirandola}}, \bibinfo {author} {\bibfnamefont {S.}~\bibnamefont {Mancini}},
  \bibinfo {author} {\bibfnamefont {D.}~\bibnamefont {Vitali}}, \ and\ \bibinfo
  {author} {\bibfnamefont {P.}~\bibnamefont {Tombesi}},\ }\href {\doibase
  10.1140/epjd/e2005-00306-3} {\bibfield  {journal} {\bibinfo  {journal} {Eur.
  Phys. J. D}\ }\textbf {\bibinfo {volume} {37}},\ \bibinfo {pages} {283}
  (\bibinfo {year} {2006})}\BibitemShut {NoStop}%
\bibitem [{\citenamefont {Vasconcelos}\ \emph {et~al.}(2010)\citenamefont
  {Vasconcelos}, \citenamefont {Sanz},\ and\ \citenamefont
  {Glancy}}]{vasconcelos2010all}%
  \BibitemOpen
  \bibfield  {author} {\bibinfo {author} {\bibfnamefont {H.~M.}\ \bibnamefont
  {Vasconcelos}}, \bibinfo {author} {\bibfnamefont {L.}~\bibnamefont {Sanz}}, \
  and\ \bibinfo {author} {\bibfnamefont {S.}~\bibnamefont {Glancy}},\
  }\href@noop {} {\bibfield  {journal} {\bibinfo  {journal} {Opt. Lett.}\
  }\textbf {\bibinfo {volume} {35}},\ \bibinfo {pages} {3261} (\bibinfo {year}
  {2010})}\BibitemShut {NoStop}%
\bibitem [{\citenamefont {Terhal}\ and\ \citenamefont
  {Weigand}(2016)}]{terhal2016encoding}%
  \BibitemOpen
  \bibfield  {author} {\bibinfo {author} {\bibfnamefont {B.}~\bibnamefont
  {Terhal}}\ and\ \bibinfo {author} {\bibfnamefont {D.}~\bibnamefont
  {Weigand}},\ }\href@noop {} {\bibfield  {journal} {\bibinfo  {journal} {Phys.
  Rev. A}\ }\textbf {\bibinfo {volume} {93}},\ \bibinfo {pages} {012315}
  (\bibinfo {year} {2016})}\BibitemShut {NoStop}%
\bibitem [{\citenamefont {Motes}\ \emph {et~al.}(2017)\citenamefont {Motes},
  \citenamefont {Baragiola}, \citenamefont {Gilchrist},\ and\ \citenamefont
  {Menicucci}}]{motes2017encoding}%
  \BibitemOpen
  \bibfield  {author} {\bibinfo {author} {\bibfnamefont {K.~R.}\ \bibnamefont
  {Motes}}, \bibinfo {author} {\bibfnamefont {B.~Q.}\ \bibnamefont
  {Baragiola}}, \bibinfo {author} {\bibfnamefont {A.}~\bibnamefont
  {Gilchrist}}, \ and\ \bibinfo {author} {\bibfnamefont {N.~C.}\ \bibnamefont
  {Menicucci}},\ }\href@noop {} {\bibfield  {journal} {\bibinfo  {journal}
  {Phys. Rev. A}\ }\textbf {\bibinfo {volume} {95}},\ \bibinfo {pages} {053819}
  (\bibinfo {year} {2017})}\BibitemShut {NoStop}%
\bibitem [{\citenamefont {Fl{\"u}hmann}\ \emph {et~al.}(2018)\citenamefont
  {Fl{\"u}hmann}, \citenamefont {Negnevitsky}, \citenamefont {Marinelli},\ and\
  \citenamefont {Home}}]{fluhmann2018sequential}%
  \BibitemOpen
  \bibfield  {author} {\bibinfo {author} {\bibfnamefont {C.}~\bibnamefont
  {Fl{\"u}hmann}}, \bibinfo {author} {\bibfnamefont {V.}~\bibnamefont
  {Negnevitsky}}, \bibinfo {author} {\bibfnamefont {M.}~\bibnamefont
  {Marinelli}}, \ and\ \bibinfo {author} {\bibfnamefont {J.~P.}\ \bibnamefont
  {Home}},\ }\href@noop {} {\bibfield  {journal} {\bibinfo  {journal} {Phys.
  Rev. X}\ }\textbf {\bibinfo {volume} {8}},\ \bibinfo {pages} {021001}
  (\bibinfo {year} {2018})}\BibitemShut {NoStop}%
\bibitem [{\citenamefont {Fl{\"u}hmann}\ \emph {et~al.}(2019)\citenamefont
  {Fl{\"u}hmann}, \citenamefont {Nguyen}, \citenamefont {Marinelli},
  \citenamefont {Negnevitsky}, \citenamefont {Mehta},\ and\ \citenamefont
  {Home}}]{fluhmann2018encoding}%
  \BibitemOpen
  \bibfield  {author} {\bibinfo {author} {\bibfnamefont {C.}~\bibnamefont
  {Fl{\"u}hmann}}, \bibinfo {author} {\bibfnamefont {T.~L.}\ \bibnamefont
  {Nguyen}}, \bibinfo {author} {\bibfnamefont {M.}~\bibnamefont {Marinelli}},
  \bibinfo {author} {\bibfnamefont {V.}~\bibnamefont {Negnevitsky}}, \bibinfo
  {author} {\bibfnamefont {K.}~\bibnamefont {Mehta}}, \ and\ \bibinfo {author}
  {\bibfnamefont {J.}~\bibnamefont {Home}},\ }\href@noop {} {\bibfield
  {journal} {\bibinfo  {journal} {Nature}\ }\textbf {\bibinfo {volume} {566}},\
  \bibinfo {pages} {513} (\bibinfo {year} {2019})}\BibitemShut {NoStop}%
\bibitem [{\citenamefont {Waldherr}\ \emph {et~al.}(2014)\citenamefont
  {Waldherr}, \citenamefont {Wang}, \citenamefont {Zaiser}, \citenamefont
  {Jamali}, \citenamefont {Schulte-Herbr{\"{u}}ggen}, \citenamefont {Abe},
  \citenamefont {Ohshima}, \citenamefont {Isoya}, \citenamefont {Du},
  \citenamefont {Neumann},\ and\ \citenamefont {Wrachtrup}}]{Waldherr2014}%
  \BibitemOpen
  \bibfield  {author} {\bibinfo {author} {\bibfnamefont {G.}~\bibnamefont
  {Waldherr}}, \bibinfo {author} {\bibfnamefont {Y.}~\bibnamefont {Wang}},
  \bibinfo {author} {\bibfnamefont {S.}~\bibnamefont {Zaiser}}, \bibinfo
  {author} {\bibfnamefont {M.}~\bibnamefont {Jamali}}, \bibinfo {author}
  {\bibfnamefont {T.}~\bibnamefont {Schulte-Herbr{\"{u}}ggen}}, \bibinfo
  {author} {\bibfnamefont {H.}~\bibnamefont {Abe}}, \bibinfo {author}
  {\bibfnamefont {T.}~\bibnamefont {Ohshima}}, \bibinfo {author} {\bibfnamefont
  {J.}~\bibnamefont {Isoya}}, \bibinfo {author} {\bibfnamefont {J.~F.}\
  \bibnamefont {Du}}, \bibinfo {author} {\bibfnamefont {P.}~\bibnamefont
  {Neumann}}, \ and\ \bibinfo {author} {\bibfnamefont {J.}~\bibnamefont
  {Wrachtrup}},\ }\href {\doibase 10.1038/nature12919} {\bibfield  {journal}
  {\bibinfo  {journal} {Nature}\ }\textbf {\bibinfo {volume} {506}},\ \bibinfo
  {pages} {204} (\bibinfo {year} {2014})}\BibitemShut {NoStop}%
\bibitem [{\citenamefont {Kelly}\ \emph {et~al.}(2015)\citenamefont {Kelly},
  \citenamefont {Barends}, \citenamefont {Fowler}, \citenamefont {Megrant},
  \citenamefont {Jeffrey}, \citenamefont {White}, \citenamefont {Sank},
  \citenamefont {Mutus}, \citenamefont {Campbell}, \citenamefont {Chen},
  \citenamefont {Chen}, \citenamefont {Chiaro}, \citenamefont {Dunsworth},
  \citenamefont {Hoi}, \citenamefont {Neill}, \citenamefont {O'Malley},
  \citenamefont {Quintana}, \citenamefont {Roushan}, \citenamefont
  {Vainsencher}, \citenamefont {Wenner}, \citenamefont {Cleland},\ and\
  \citenamefont {Martinis}}]{Kelly2015}%
  \BibitemOpen
  \bibfield  {author} {\bibinfo {author} {\bibfnamefont {J.}~\bibnamefont
  {Kelly}}, \bibinfo {author} {\bibfnamefont {R.}~\bibnamefont {Barends}},
  \bibinfo {author} {\bibfnamefont {A.~G.}\ \bibnamefont {Fowler}}, \bibinfo
  {author} {\bibfnamefont {A.}~\bibnamefont {Megrant}}, \bibinfo {author}
  {\bibfnamefont {E.}~\bibnamefont {Jeffrey}}, \bibinfo {author} {\bibfnamefont
  {T.~C.}\ \bibnamefont {White}}, \bibinfo {author} {\bibfnamefont
  {D.}~\bibnamefont {Sank}}, \bibinfo {author} {\bibfnamefont {J.~Y.}\
  \bibnamefont {Mutus}}, \bibinfo {author} {\bibfnamefont {B.}~\bibnamefont
  {Campbell}}, \bibinfo {author} {\bibfnamefont {Y.}~\bibnamefont {Chen}},
  \bibinfo {author} {\bibfnamefont {Z.}~\bibnamefont {Chen}}, \bibinfo {author}
  {\bibfnamefont {B.}~\bibnamefont {Chiaro}}, \bibinfo {author} {\bibfnamefont
  {A.}~\bibnamefont {Dunsworth}}, \bibinfo {author} {\bibfnamefont {I.-C.}\
  \bibnamefont {Hoi}}, \bibinfo {author} {\bibfnamefont {C.}~\bibnamefont
  {Neill}}, \bibinfo {author} {\bibfnamefont {P.~J.~J.}\ \bibnamefont
  {O'Malley}}, \bibinfo {author} {\bibfnamefont {C.}~\bibnamefont {Quintana}},
  \bibinfo {author} {\bibfnamefont {P.}~\bibnamefont {Roushan}}, \bibinfo
  {author} {\bibfnamefont {A.}~\bibnamefont {Vainsencher}}, \bibinfo {author}
  {\bibfnamefont {J.}~\bibnamefont {Wenner}}, \bibinfo {author} {\bibfnamefont
  {A.~N.}\ \bibnamefont {Cleland}}, \ and\ \bibinfo {author} {\bibfnamefont
  {J.~M.}\ \bibnamefont {Martinis}},\ }\href {\doibase 10.1038/nature14270}
  {\bibfield  {journal} {\bibinfo  {journal} {Nature}\ }\textbf {\bibinfo
  {volume} {519}},\ \bibinfo {pages} {66} (\bibinfo {year} {2015})}\BibitemShut
  {NoStop}%
\bibitem [{\citenamefont {Cramer}\ \emph {et~al.}(2016)\citenamefont {Cramer},
  \citenamefont {Kalb}, \citenamefont {Rol}, \citenamefont {Hensen},
  \citenamefont {Blok}, \citenamefont {Markham}, \citenamefont {Twitchen},
  \citenamefont {Hanson},\ and\ \citenamefont {Taminiau}}]{Cramer2016}%
  \BibitemOpen
  \bibfield  {author} {\bibinfo {author} {\bibfnamefont {J.}~\bibnamefont
  {Cramer}}, \bibinfo {author} {\bibfnamefont {N.}~\bibnamefont {Kalb}},
  \bibinfo {author} {\bibfnamefont {M.~A.}\ \bibnamefont {Rol}}, \bibinfo
  {author} {\bibfnamefont {B.}~\bibnamefont {Hensen}}, \bibinfo {author}
  {\bibfnamefont {M.~S.}\ \bibnamefont {Blok}}, \bibinfo {author}
  {\bibfnamefont {M.}~\bibnamefont {Markham}}, \bibinfo {author} {\bibfnamefont
  {D.~J.}\ \bibnamefont {Twitchen}}, \bibinfo {author} {\bibfnamefont
  {R.}~\bibnamefont {Hanson}}, \ and\ \bibinfo {author} {\bibfnamefont {T.~H.}\
  \bibnamefont {Taminiau}},\ }\href {\doibase 10.1038/ncomms11526} {\bibfield
  {journal} {\bibinfo  {journal} {Nat. Commun.}\ }\textbf {\bibinfo {volume}
  {7}},\ \bibinfo {pages} {11526} (\bibinfo {year} {2016})}\BibitemShut
  {NoStop}%
\bibitem [{\citenamefont {Ofek}\ \emph {et~al.}(2016)\citenamefont {Ofek},
  \citenamefont {Petrenko}, \citenamefont {Heeres}, \citenamefont {Reinhold},
  \citenamefont {Leghtas}, \citenamefont {Vlastakis}, \citenamefont {Liu},
  \citenamefont {Frunzio}, \citenamefont {Girvin}, \citenamefont {Jiang} \emph
  {et~al.}}]{ofek2016extending}%
  \BibitemOpen
  \bibfield  {author} {\bibinfo {author} {\bibfnamefont {N.}~\bibnamefont
  {Ofek}}, \bibinfo {author} {\bibfnamefont {A.}~\bibnamefont {Petrenko}},
  \bibinfo {author} {\bibfnamefont {R.}~\bibnamefont {Heeres}}, \bibinfo
  {author} {\bibfnamefont {P.}~\bibnamefont {Reinhold}}, \bibinfo {author}
  {\bibfnamefont {Z.}~\bibnamefont {Leghtas}}, \bibinfo {author} {\bibfnamefont
  {B.}~\bibnamefont {Vlastakis}}, \bibinfo {author} {\bibfnamefont
  {Y.}~\bibnamefont {Liu}}, \bibinfo {author} {\bibfnamefont {L.}~\bibnamefont
  {Frunzio}}, \bibinfo {author} {\bibfnamefont {S.}~\bibnamefont {Girvin}},
  \bibinfo {author} {\bibfnamefont {L.}~\bibnamefont {Jiang}},  \emph
  {et~al.},\ }\href@noop {} {\bibfield  {journal} {\bibinfo  {journal}
  {Nature}\ }\textbf {\bibinfo {volume} {536}},\ \bibinfo {pages} {441}
  (\bibinfo {year} {2016})}\BibitemShut {NoStop}%
\bibitem [{\citenamefont {Hu}\ \emph {et~al.}(2019)\citenamefont {Hu},
  \citenamefont {Ma}, \citenamefont {Cai}, \citenamefont {Mu}, \citenamefont
  {Xu}, \citenamefont {Wang}, \citenamefont {Wu}, \citenamefont {Wang},
  \citenamefont {Song}, \citenamefont {Zou} \emph {et~al.}}]{hu2019quantum}%
  \BibitemOpen
  \bibfield  {author} {\bibinfo {author} {\bibfnamefont {L.}~\bibnamefont
  {Hu}}, \bibinfo {author} {\bibfnamefont {Y.}~\bibnamefont {Ma}}, \bibinfo
  {author} {\bibfnamefont {W.}~\bibnamefont {Cai}}, \bibinfo {author}
  {\bibfnamefont {X.}~\bibnamefont {Mu}}, \bibinfo {author} {\bibfnamefont
  {Y.}~\bibnamefont {Xu}}, \bibinfo {author} {\bibfnamefont {W.}~\bibnamefont
  {Wang}}, \bibinfo {author} {\bibfnamefont {Y.}~\bibnamefont {Wu}}, \bibinfo
  {author} {\bibfnamefont {H.}~\bibnamefont {Wang}}, \bibinfo {author}
  {\bibfnamefont {Y.}~\bibnamefont {Song}}, \bibinfo {author} {\bibfnamefont
  {C.-L.}\ \bibnamefont {Zou}},  \emph {et~al.},\ }\href@noop {} {\bibfield
  {journal} {\bibinfo  {journal} {Nat. Phys.}\ }\textbf {\bibinfo {volume}
  {15}},\ \bibinfo {pages} {503} (\bibinfo {year} {2019})}\BibitemShut
  {NoStop}%
\bibitem [{\citenamefont {Albert}\ \emph {et~al.}(2018)\citenamefont {Albert},
  \citenamefont {Noh}, \citenamefont {Duivenvoorden}, \citenamefont {Young},
  \citenamefont {Brierley}, \citenamefont {Reinhold}, \citenamefont {Vuillot},
  \citenamefont {Li}, \citenamefont {Shen}, \citenamefont {Girvin},
  \citenamefont {Terhal},\ and\ \citenamefont {Jiang}}]{albert2017performance}%
  \BibitemOpen
  \bibfield  {author} {\bibinfo {author} {\bibfnamefont {V.~V.}\ \bibnamefont
  {Albert}}, \bibinfo {author} {\bibfnamefont {K.}~\bibnamefont {Noh}},
  \bibinfo {author} {\bibfnamefont {K.}~\bibnamefont {Duivenvoorden}}, \bibinfo
  {author} {\bibfnamefont {D.~J.}\ \bibnamefont {Young}}, \bibinfo {author}
  {\bibfnamefont {R.~T.}\ \bibnamefont {Brierley}}, \bibinfo {author}
  {\bibfnamefont {P.}~\bibnamefont {Reinhold}}, \bibinfo {author}
  {\bibfnamefont {C.}~\bibnamefont {Vuillot}}, \bibinfo {author} {\bibfnamefont
  {L.}~\bibnamefont {Li}}, \bibinfo {author} {\bibfnamefont {C.}~\bibnamefont
  {Shen}}, \bibinfo {author} {\bibfnamefont {S.~M.}\ \bibnamefont {Girvin}},
  \bibinfo {author} {\bibfnamefont {B.~M.}\ \bibnamefont {Terhal}}, \ and\
  \bibinfo {author} {\bibfnamefont {L.}~\bibnamefont {Jiang}},\ }\href
  {\doibase 10.1103/PhysRevA.97.032346} {\bibfield  {journal} {\bibinfo
  {journal} {Phy. Rev. A}\ }\textbf {\bibinfo {volume} {97}},\ \bibinfo {pages}
  {032346} (\bibinfo {year} {2018})}\BibitemShut {NoStop}%
\bibitem [{\citenamefont {{Noh}}\ \emph {et~al.}(2019)\citenamefont {{Noh}},
  \citenamefont {{Albert}},\ and\ \citenamefont {{Jiang}}}]{noh2019gkp}%
  \BibitemOpen
  \bibfield  {author} {\bibinfo {author} {\bibfnamefont {K.}~\bibnamefont
  {{Noh}}}, \bibinfo {author} {\bibfnamefont {V.~V.}\ \bibnamefont {{Albert}}},
  \ and\ \bibinfo {author} {\bibfnamefont {L.}~\bibnamefont {{Jiang}}},\ }\href
  {\doibase 10.1109/TIT.2018.2873764} {\bibfield  {journal} {\bibinfo
  {journal} {IEEE Transactions on Information Theory}\ }\textbf {\bibinfo
  {volume} {65}},\ \bibinfo {pages} {2563} (\bibinfo {year}
  {2019})}\BibitemShut {NoStop}%
\bibitem [{\citenamefont {Cahill}\ and\ \citenamefont
  {Glauber}(1969)}]{Cahill1969}%
  \BibitemOpen
  \bibfield  {author} {\bibinfo {author} {\bibfnamefont {K.~E.}\ \bibnamefont
  {Cahill}}\ and\ \bibinfo {author} {\bibfnamefont {R.~J.}\ \bibnamefont
  {Glauber}},\ }\href {\doibase 10.1103/PhysRev.177.1857} {\bibfield  {journal}
  {\bibinfo  {journal} {Phys. Rev.}\ }\textbf {\bibinfo {volume} {177}},\
  \bibinfo {pages} {1857} (\bibinfo {year} {1969})}\BibitemShut {NoStop}%
\bibitem [{\citenamefont {Aharonov}\ \emph {et~al.}(1969)\citenamefont
  {Aharonov}, \citenamefont {Pendleton},\ and\ \citenamefont
  {Petersen}}]{Aharonov1969}%
  \BibitemOpen
  \bibfield  {author} {\bibinfo {author} {\bibfnamefont {Y.}~\bibnamefont
  {Aharonov}}, \bibinfo {author} {\bibfnamefont {H.}~\bibnamefont {Pendleton}},
  \ and\ \bibinfo {author} {\bibfnamefont {A.}~\bibnamefont {Petersen}},\
  }\href {\doibase 10.1007/BF00670008} {\bibfield  {journal} {\bibinfo
  {journal} {Int. J. Theor. Phys.}\ }\textbf {\bibinfo {volume} {2}},\ \bibinfo
  {pages} {213} (\bibinfo {year} {1969})}\BibitemShut {NoStop}%
\bibitem [{\citenamefont {Popescu}(2010)}]{Popescu2010}%
  \BibitemOpen
  \bibfield  {author} {\bibinfo {author} {\bibfnamefont {S.}~\bibnamefont
  {Popescu}},\ }\href {\doibase 10.1038/nphys1619} {\bibfield  {journal}
  {\bibinfo  {journal} {Nat. Phys.}\ }\textbf {\bibinfo {volume} {6}},\
  \bibinfo {pages} {151} (\bibinfo {year} {2010})}\BibitemShut {NoStop}%
\bibitem [{\citenamefont {Reagor}\ \emph {et~al.}(2016)\citenamefont {Reagor},
  \citenamefont {Pfaff}, \citenamefont {Axline}, \citenamefont {Heeres},
  \citenamefont {Ofek}, \citenamefont {Sliwa}, \citenamefont {Holland},
  \citenamefont {Wang}, \citenamefont {Blumoff}, \citenamefont {Chou},
  \citenamefont {Hatridge}, \citenamefont {Frunzio}, \citenamefont {Devoret},
  \citenamefont {Jiang},\ and\ \citenamefont {Schoelkopf}}]{reagor2015quantum}%
  \BibitemOpen
  \bibfield  {author} {\bibinfo {author} {\bibfnamefont {M.}~\bibnamefont
  {Reagor}}, \bibinfo {author} {\bibfnamefont {W.}~\bibnamefont {Pfaff}},
  \bibinfo {author} {\bibfnamefont {C.}~\bibnamefont {Axline}}, \bibinfo
  {author} {\bibfnamefont {R.~W.}\ \bibnamefont {Heeres}}, \bibinfo {author}
  {\bibfnamefont {N.}~\bibnamefont {Ofek}}, \bibinfo {author} {\bibfnamefont
  {K.}~\bibnamefont {Sliwa}}, \bibinfo {author} {\bibfnamefont
  {E.}~\bibnamefont {Holland}}, \bibinfo {author} {\bibfnamefont
  {C.}~\bibnamefont {Wang}}, \bibinfo {author} {\bibfnamefont {J.}~\bibnamefont
  {Blumoff}}, \bibinfo {author} {\bibfnamefont {K.}~\bibnamefont {Chou}},
  \bibinfo {author} {\bibfnamefont {M.~J.}\ \bibnamefont {Hatridge}}, \bibinfo
  {author} {\bibfnamefont {L.}~\bibnamefont {Frunzio}}, \bibinfo {author}
  {\bibfnamefont {M.~H.}\ \bibnamefont {Devoret}}, \bibinfo {author}
  {\bibfnamefont {L.}~\bibnamefont {Jiang}}, \ and\ \bibinfo {author}
  {\bibfnamefont {R.~J.}\ \bibnamefont {Schoelkopf}},\ }\href {\doibase
  10.1103/PhysRevB.94.014506} {\bibfield  {journal} {\bibinfo  {journal} {Phys.
  Rev. B}\ }\textbf {\bibinfo {volume} {94}},\ \bibinfo {pages} {014506}
  (\bibinfo {year} {2016})}\BibitemShut {NoStop}%
\bibitem [{\citenamefont {Wallraff}\ \emph {et~al.}(2004)\citenamefont
  {Wallraff}, \citenamefont {Schuster}, \citenamefont {Blais}, \citenamefont
  {Frunzio}, \citenamefont {Huang}, \citenamefont {Majer}, \citenamefont
  {Kumar}, \citenamefont {Girvin},\ and\ \citenamefont
  {Schoelkopf}}]{Wallraff2004}%
  \BibitemOpen
  \bibfield  {author} {\bibinfo {author} {\bibfnamefont {A.}~\bibnamefont
  {Wallraff}}, \bibinfo {author} {\bibfnamefont {D.~I.}\ \bibnamefont
  {Schuster}}, \bibinfo {author} {\bibfnamefont {A.}~\bibnamefont {Blais}},
  \bibinfo {author} {\bibfnamefont {L.}~\bibnamefont {Frunzio}}, \bibinfo
  {author} {\bibfnamefont {R.~S.}\ \bibnamefont {Huang}}, \bibinfo {author}
  {\bibfnamefont {J.}~\bibnamefont {Majer}}, \bibinfo {author} {\bibfnamefont
  {S.}~\bibnamefont {Kumar}}, \bibinfo {author} {\bibfnamefont {S.~M.}\
  \bibnamefont {Girvin}}, \ and\ \bibinfo {author} {\bibfnamefont {R.~J.}\
  \bibnamefont {Schoelkopf}},\ }\href@noop {} {\bibfield  {journal} {\bibinfo
  {journal} {Nature}\ }\textbf {\bibinfo {volume} {431}},\ \bibinfo {pages}
  {162} (\bibinfo {year} {2004})}\BibitemShut {NoStop}%
\bibitem [{\citenamefont {Kitaev}(1995)}]{kitaev1995quantum}%
  \BibitemOpen
  \bibfield  {author} {\bibinfo {author} {\bibfnamefont {A.~Y.}\ \bibnamefont
  {Kitaev}},\ }\href@noop {} {\bibfield  {journal} {\bibinfo  {journal} {arXiv
  preprint quant-ph/9511026}\ } (\bibinfo {year} {1995})}\BibitemShut {NoStop}%
\bibitem [{\citenamefont {Svore}\ \emph {et~al.}(2013)\citenamefont {Svore},
  \citenamefont {Hastings},\ and\ \citenamefont {Freedman}}]{svore2013faster}%
  \BibitemOpen
  \bibfield  {author} {\bibinfo {author} {\bibfnamefont {K.~M.}\ \bibnamefont
  {Svore}}, \bibinfo {author} {\bibfnamefont {M.~B.}\ \bibnamefont {Hastings}},
  \ and\ \bibinfo {author} {\bibfnamefont {M.}~\bibnamefont {Freedman}},\
  }\href@noop {} {\bibfield  {journal} {\bibinfo  {journal} {arXiv preprint
  arXiv:1304.0741}\ } (\bibinfo {year} {2013})}\BibitemShut {NoStop}%
\bibitem [{\citenamefont {Weigand}\ and\ \citenamefont
  {Terhal}(2018)}]{weigand2018generating}%
  \BibitemOpen
  \bibfield  {author} {\bibinfo {author} {\bibfnamefont {D.~J.}\ \bibnamefont
  {Weigand}}\ and\ \bibinfo {author} {\bibfnamefont {B.~M.}\ \bibnamefont
  {Terhal}},\ }\href@noop {} {\bibfield  {journal} {\bibinfo  {journal} {Phys.
  Rev. A}\ }\textbf {\bibinfo {volume} {97}},\ \bibinfo {pages} {022341}
  (\bibinfo {year} {2018})}\BibitemShut {NoStop}%
\bibitem [{\citenamefont {Haroche}\ and\ \citenamefont
  {Raimond}(2006)}]{Haroche2006}%
  \BibitemOpen
  \bibfield  {author} {\bibinfo {author} {\bibfnamefont {S.}~\bibnamefont
  {Haroche}}\ and\ \bibinfo {author} {\bibfnamefont {J.-M.}\ \bibnamefont
  {Raimond}},\ }\href {\doibase 10.1093/acprof:oso/9780198509141.001.0001}
  {\emph {\bibinfo {title} {{Exploring the Quantum}}}}\ (\bibinfo  {publisher}
  {Oxford University Press},\ \bibinfo {year} {2006})\BibitemShut {NoStop}%
\bibitem [{\citenamefont {Walls}\ and\ \citenamefont
  {Milburn}(2007)}]{walls2007quantum}%
  \BibitemOpen
  \bibfield  {author} {\bibinfo {author} {\bibfnamefont {D.~F.}\ \bibnamefont
  {Walls}}\ and\ \bibinfo {author} {\bibfnamefont {G.~J.}\ \bibnamefont
  {Milburn}},\ }\href@noop {} {\emph {\bibinfo {title} {Quantum optics}}}\
  (\bibinfo  {publisher} {Springer Science \& Business Media},\ \bibinfo {year}
  {2007})\BibitemShut {NoStop}%
\bibitem [{\citenamefont {Glancy}\ and\ \citenamefont
  {Knill}(2006)}]{glancy2006}%
  \BibitemOpen
  \bibfield  {author} {\bibinfo {author} {\bibfnamefont {S.}~\bibnamefont
  {Glancy}}\ and\ \bibinfo {author} {\bibfnamefont {E.}~\bibnamefont {Knill}},\
  }\href {\doibase 10.1103/PhysRevA.73.012325} {\bibfield  {journal} {\bibinfo
  {journal} {Phys. Rev. A}\ }\textbf {\bibinfo {volume} {73}},\ \bibinfo
  {pages} {012325} (\bibinfo {year} {2006})}\BibitemShut {NoStop}%
\bibitem [{\citenamefont {Fukui}\ \emph {et~al.}(2018)\citenamefont {Fukui},
  \citenamefont {Tomita}, \citenamefont {Okamoto},\ and\ \citenamefont
  {Fujii}}]{fukui2018high}%
  \BibitemOpen
  \bibfield  {author} {\bibinfo {author} {\bibfnamefont {K.}~\bibnamefont
  {Fukui}}, \bibinfo {author} {\bibfnamefont {A.}~\bibnamefont {Tomita}},
  \bibinfo {author} {\bibfnamefont {A.}~\bibnamefont {Okamoto}}, \ and\
  \bibinfo {author} {\bibfnamefont {K.}~\bibnamefont {Fujii}},\ }\href@noop {}
  {\bibfield  {journal} {\bibinfo  {journal} {Phys. Rev. X}\ }\textbf {\bibinfo
  {volume} {8}},\ \bibinfo {pages} {021054} (\bibinfo {year}
  {2018})}\BibitemShut {NoStop}%
\bibitem [{\citenamefont {Vuillot}\ \emph {et~al.}(2019)\citenamefont
  {Vuillot}, \citenamefont {Asasi}, \citenamefont {Wang}, \citenamefont
  {Pryadko},\ and\ \citenamefont {Terhal}}]{vuillot2019quantum}%
  \BibitemOpen
  \bibfield  {author} {\bibinfo {author} {\bibfnamefont {C.}~\bibnamefont
  {Vuillot}}, \bibinfo {author} {\bibfnamefont {H.}~\bibnamefont {Asasi}},
  \bibinfo {author} {\bibfnamefont {Y.}~\bibnamefont {Wang}}, \bibinfo {author}
  {\bibfnamefont {L.~P.}\ \bibnamefont {Pryadko}}, \ and\ \bibinfo {author}
  {\bibfnamefont {B.~M.}\ \bibnamefont {Terhal}},\ }\href@noop {} {\bibfield
  {journal} {\bibinfo  {journal} {Phys. Rev. A}\ }\textbf {\bibinfo {volume}
  {99}},\ \bibinfo {pages} {032344} (\bibinfo {year} {2019})}\BibitemShut
  {NoStop}%
\bibitem [{\citenamefont {Touzard}\ \emph {et~al.}(2019)\citenamefont
  {Touzard}, \citenamefont {Kou}, \citenamefont {Frattini}, \citenamefont
  {Sivak}, \citenamefont {Puri}, \citenamefont {Grimm}, \citenamefont
  {Frunzio}, \citenamefont {Shankar},\ and\ \citenamefont
  {Devoret}}]{Touzard2019}%
  \BibitemOpen
  \bibfield  {author} {\bibinfo {author} {\bibfnamefont {S.}~\bibnamefont
  {Touzard}}, \bibinfo {author} {\bibfnamefont {A.}~\bibnamefont {Kou}},
  \bibinfo {author} {\bibfnamefont {N.~E.}\ \bibnamefont {Frattini}}, \bibinfo
  {author} {\bibfnamefont {V.~V.}\ \bibnamefont {Sivak}}, \bibinfo {author}
  {\bibfnamefont {S.}~\bibnamefont {Puri}}, \bibinfo {author} {\bibfnamefont
  {A.}~\bibnamefont {Grimm}}, \bibinfo {author} {\bibfnamefont
  {L.}~\bibnamefont {Frunzio}}, \bibinfo {author} {\bibfnamefont
  {S.}~\bibnamefont {Shankar}}, \ and\ \bibinfo {author} {\bibfnamefont
  {M.~H.}\ \bibnamefont {Devoret}},\ }\href {\doibase
  10.1103/PhysRevLett.122.080502} {\bibfield  {journal} {\bibinfo  {journal}
  {Phys. Rev. Lett.}\ }\textbf {\bibinfo {volume} {122}},\ \bibinfo {pages}
  {080502} (\bibinfo {year} {2019})}\BibitemShut {NoStop}%
\bibitem [{\citenamefont {Puri}\ \emph {et~al.}(2018)\citenamefont {Puri},
  \citenamefont {Grimm}, \citenamefont {Campagne-Ibarcq}, \citenamefont
  {Eickbusch}, \citenamefont {Noh}, \citenamefont {Roberts}, \citenamefont
  {Jiang}, \citenamefont {Mirrahimi}, \citenamefont {Devoret},\ and\
  \citenamefont {Girvin}}]{puri2018stabilized}%
  \BibitemOpen
  \bibfield  {author} {\bibinfo {author} {\bibfnamefont {S.}~\bibnamefont
  {Puri}}, \bibinfo {author} {\bibfnamefont {A.}~\bibnamefont {Grimm}},
  \bibinfo {author} {\bibfnamefont {P.}~\bibnamefont {Campagne-Ibarcq}},
  \bibinfo {author} {\bibfnamefont {A.}~\bibnamefont {Eickbusch}}, \bibinfo
  {author} {\bibfnamefont {K.}~\bibnamefont {Noh}}, \bibinfo {author}
  {\bibfnamefont {G.}~\bibnamefont {Roberts}}, \bibinfo {author} {\bibfnamefont
  {L.}~\bibnamefont {Jiang}}, \bibinfo {author} {\bibfnamefont
  {M.}~\bibnamefont {Mirrahimi}}, \bibinfo {author} {\bibfnamefont {M.~H.}\
  \bibnamefont {Devoret}}, \ and\ \bibinfo {author} {\bibfnamefont {S.~M.}\
  \bibnamefont {Girvin}},\ }\href@noop {} {\bibfield  {journal} {\bibinfo
  {journal} {arXiv preprint arXiv:1807.09334}\ } (\bibinfo {year}
  {2018})}\BibitemShut {NoStop}%
\bibitem [{\citenamefont {Grimm}\ \emph {et~al.}(2019)\citenamefont {Grimm},
  \citenamefont {Frattini}, \citenamefont {Puri}, \citenamefont {Mundhada},
  \citenamefont {Touzard}, \citenamefont {Mirrahimi}, \citenamefont {Girvin},
  \citenamefont {Shankar},\ and\ \citenamefont {Devoret}}]{Grimm2019}%
  \BibitemOpen
  \bibfield  {author} {\bibinfo {author} {\bibfnamefont {A.}~\bibnamefont
  {Grimm}}, \bibinfo {author} {\bibfnamefont {N.~E.}\ \bibnamefont {Frattini}},
  \bibinfo {author} {\bibfnamefont {S.}~\bibnamefont {Puri}}, \bibinfo {author}
  {\bibfnamefont {S.~O.}\ \bibnamefont {Mundhada}}, \bibinfo {author}
  {\bibfnamefont {S.}~\bibnamefont {Touzard}}, \bibinfo {author} {\bibfnamefont
  {M.}~\bibnamefont {Mirrahimi}}, \bibinfo {author} {\bibfnamefont {S.~M.}\
  \bibnamefont {Girvin}}, \bibinfo {author} {\bibfnamefont {S.}~\bibnamefont
  {Shankar}}, \ and\ \bibinfo {author} {\bibfnamefont {M.~H.}\ \bibnamefont
  {Devoret}},\ }\href@noop {} {\bibfield  {journal} {\bibinfo  {journal} {arXiv
  preprint arXiv:1907.12131}\ } (\bibinfo {year} {2019})}\BibitemShut {NoStop}%
\bibitem [{\citenamefont {Shi}\ \emph {et~al.}(2019)\citenamefont {Shi},
  \citenamefont {Chamberland},\ and\ \citenamefont {Cross}}]{shi2019fault}%
  \BibitemOpen
  \bibfield  {author} {\bibinfo {author} {\bibfnamefont {Y.}~\bibnamefont
  {Shi}}, \bibinfo {author} {\bibfnamefont {C.}~\bibnamefont {Chamberland}}, \
  and\ \bibinfo {author} {\bibfnamefont {A.~W.}\ \bibnamefont {Cross}},\
  }\href@noop {} {\bibfield  {journal} {\bibinfo  {journal} {arXiv preprint
  arXiv:1905.00903}\ } (\bibinfo {year} {2019})}\BibitemShut {NoStop}%
\bibitem [{\citenamefont {Gao}\ \emph {et~al.}(2018)\citenamefont {Gao},
  \citenamefont {Lester}, \citenamefont {Zhang}, \citenamefont {Wang},
  \citenamefont {Rosenblum}, \citenamefont {Frunzio}, \citenamefont {Jiang},
  \citenamefont {Girvin},\ and\ \citenamefont {Schoelkopf}}]{Gao2018}%
  \BibitemOpen
  \bibfield  {author} {\bibinfo {author} {\bibfnamefont {Y.~Y.}\ \bibnamefont
  {Gao}}, \bibinfo {author} {\bibfnamefont {B.~J.}\ \bibnamefont {Lester}},
  \bibinfo {author} {\bibfnamefont {Y.}~\bibnamefont {Zhang}}, \bibinfo
  {author} {\bibfnamefont {C.}~\bibnamefont {Wang}}, \bibinfo {author}
  {\bibfnamefont {S.}~\bibnamefont {Rosenblum}}, \bibinfo {author}
  {\bibfnamefont {L.}~\bibnamefont {Frunzio}}, \bibinfo {author} {\bibfnamefont
  {L.}~\bibnamefont {Jiang}}, \bibinfo {author} {\bibfnamefont {S.~M.}\
  \bibnamefont {Girvin}}, \ and\ \bibinfo {author} {\bibfnamefont {R.~J.}\
  \bibnamefont {Schoelkopf}},\ }\href {\doibase 10.1103/PhysRevX.8.021073}
  {\bibfield  {journal} {\bibinfo  {journal} {Phys. Rev. X}\ }\textbf {\bibinfo
  {volume} {8}},\ \bibinfo {pages} {021073} (\bibinfo {year}
  {2018})}\BibitemShut {NoStop}%
\bibitem [{\citenamefont {Baragiola}\ \emph {et~al.}(2019)\citenamefont
  {Baragiola}, \citenamefont {Pantaleoni}, \citenamefont {Alexander},
  \citenamefont {Karanjai},\ and\ \citenamefont
  {Menicucci}}]{baragiola2019all}%
  \BibitemOpen
  \bibfield  {author} {\bibinfo {author} {\bibfnamefont {B.~Q.}\ \bibnamefont
  {Baragiola}}, \bibinfo {author} {\bibfnamefont {G.}~\bibnamefont
  {Pantaleoni}}, \bibinfo {author} {\bibfnamefont {R.~N.}\ \bibnamefont
  {Alexander}}, \bibinfo {author} {\bibfnamefont {A.}~\bibnamefont {Karanjai}},
  \ and\ \bibinfo {author} {\bibfnamefont {N.~C.}\ \bibnamefont {Menicucci}},\
  }\href@noop {} {\bibfield  {journal} {\bibinfo  {journal} {arXiv preprint
  arXiv:1903.00012}\ } (\bibinfo {year} {2019})}\BibitemShut {NoStop}%
\bibitem [{\citenamefont {Frattini}\ \emph {et~al.}(2018)\citenamefont
  {Frattini}, \citenamefont {Sivak}, \citenamefont {Lingenfelter},
  \citenamefont {Shankar},\ and\ \citenamefont {Devoret}}]{frattini2018}%
  \BibitemOpen
  \bibfield  {author} {\bibinfo {author} {\bibfnamefont {N.~E.}\ \bibnamefont
  {Frattini}}, \bibinfo {author} {\bibfnamefont {V.~V.}\ \bibnamefont {Sivak}},
  \bibinfo {author} {\bibfnamefont {A.}~\bibnamefont {Lingenfelter}}, \bibinfo
  {author} {\bibfnamefont {S.}~\bibnamefont {Shankar}}, \ and\ \bibinfo
  {author} {\bibfnamefont {M.~H.}\ \bibnamefont {Devoret}},\ }\href {\doibase
  10.1103/PhysRevApplied.10.054020} {\bibfield  {journal} {\bibinfo  {journal}
  {Phys. Rev. A}\ }\textbf {\bibinfo {volume} {10}},\ \bibinfo {pages} {054020}
  (\bibinfo {year} {2018})}\BibitemShut {NoStop}%
\bibitem [{\citenamefont {Chow}\ \emph {et~al.}(2010)\citenamefont {Chow},
  \citenamefont {DiCarlo}, \citenamefont {Gambetta}, \citenamefont {Motzoi},
  \citenamefont {Frunzio}, \citenamefont {Girvin},\ and\ \citenamefont
  {Schoelkopf}}]{chow2010optimized}%
  \BibitemOpen
  \bibfield  {author} {\bibinfo {author} {\bibfnamefont {J.~M.}\ \bibnamefont
  {Chow}}, \bibinfo {author} {\bibfnamefont {L.}~\bibnamefont {DiCarlo}},
  \bibinfo {author} {\bibfnamefont {J.~M.}\ \bibnamefont {Gambetta}}, \bibinfo
  {author} {\bibfnamefont {F.}~\bibnamefont {Motzoi}}, \bibinfo {author}
  {\bibfnamefont {L.}~\bibnamefont {Frunzio}}, \bibinfo {author} {\bibfnamefont
  {S.~M.}\ \bibnamefont {Girvin}}, \ and\ \bibinfo {author} {\bibfnamefont
  {R.~J.}\ \bibnamefont {Schoelkopf}},\ }\href@noop {} {\bibfield  {journal}
  {\bibinfo  {journal} {Phys. Rev. A}\ }\textbf {\bibinfo {volume} {82}},\
  \bibinfo {pages} {040305} (\bibinfo {year} {2010})}\BibitemShut {NoStop}%
\bibitem [{\citenamefont {McClure}\ \emph {et~al.}(2016)\citenamefont
  {McClure}, \citenamefont {Paik}, \citenamefont {Bishop}, \citenamefont
  {Steffen}, \citenamefont {Chow},\ and\ \citenamefont {Gambetta}}]{IBMreset}%
  \BibitemOpen
  \bibfield  {author} {\bibinfo {author} {\bibfnamefont {D.~T.}\ \bibnamefont
  {McClure}}, \bibinfo {author} {\bibfnamefont {H.}~\bibnamefont {Paik}},
  \bibinfo {author} {\bibfnamefont {L.~S.}\ \bibnamefont {Bishop}}, \bibinfo
  {author} {\bibfnamefont {M.}~\bibnamefont {Steffen}}, \bibinfo {author}
  {\bibfnamefont {J.~M.}\ \bibnamefont {Chow}}, \ and\ \bibinfo {author}
  {\bibfnamefont {J.~M.}\ \bibnamefont {Gambetta}},\ }\href {\doibase
  10.1103/PhysRevApplied.5.011001} {\bibfield  {journal} {\bibinfo  {journal}
  {Phys. Rev. A}\ }\textbf {\bibinfo {volume} {5}},\ \bibinfo {pages} {011001}
  (\bibinfo {year} {2016})}\BibitemShut {NoStop}%
\bibitem [{\citenamefont {Nigg}\ \emph {et~al.}(2012)\citenamefont {Nigg},
  \citenamefont {Paik}, \citenamefont {Vlastakis}, \citenamefont {Kirchmair},
  \citenamefont {Shankar}, \citenamefont {Frunzio}, \citenamefont {Devoret},
  \citenamefont {Schoelkopf},\ and\ \citenamefont {Girvin}}]{nigg2012black}%
  \BibitemOpen
  \bibfield  {author} {\bibinfo {author} {\bibfnamefont {S.~E.}\ \bibnamefont
  {Nigg}}, \bibinfo {author} {\bibfnamefont {H.}~\bibnamefont {Paik}}, \bibinfo
  {author} {\bibfnamefont {B.}~\bibnamefont {Vlastakis}}, \bibinfo {author}
  {\bibfnamefont {G.}~\bibnamefont {Kirchmair}}, \bibinfo {author}
  {\bibfnamefont {S.}~\bibnamefont {Shankar}}, \bibinfo {author} {\bibfnamefont
  {L.}~\bibnamefont {Frunzio}}, \bibinfo {author} {\bibfnamefont
  {M.}~\bibnamefont {Devoret}}, \bibinfo {author} {\bibfnamefont
  {R.}~\bibnamefont {Schoelkopf}}, \ and\ \bibinfo {author} {\bibfnamefont
  {S.}~\bibnamefont {Girvin}},\ }\href@noop {} {\bibfield  {journal} {\bibinfo
  {journal} {Phys. Rev. Lett.}\ }\textbf {\bibinfo {volume} {108}},\ \bibinfo
  {pages} {240502} (\bibinfo {year} {2012})}\BibitemShut {NoStop}%
\bibitem [{\citenamefont {Johansson}\ \emph {et~al.}(2013)\citenamefont
  {Johansson}, \citenamefont {Nation},\ and\ \citenamefont
  {Nori}}]{johansson2013qutip}%
  \BibitemOpen
  \bibfield  {author} {\bibinfo {author} {\bibfnamefont {J.~R.}\ \bibnamefont
  {Johansson}}, \bibinfo {author} {\bibfnamefont {P.~D.}\ \bibnamefont
  {Nation}}, \ and\ \bibinfo {author} {\bibfnamefont {F.}~\bibnamefont
  {Nori}},\ }\href@noop {} {\bibfield  {journal} {\bibinfo  {journal} {Comp.
  Phys. Com.}\ }\textbf {\bibinfo {volume} {184}},\ \bibinfo {pages} {1234}
  (\bibinfo {year} {2013})}\BibitemShut {NoStop}%
\bibitem [{\citenamefont {Bravyi}\ and\ \citenamefont
  {Kitaev}(2005)}]{Bravyi2005MagicState}%
  \BibitemOpen
  \bibfield  {author} {\bibinfo {author} {\bibfnamefont {S.}~\bibnamefont
  {Bravyi}}\ and\ \bibinfo {author} {\bibfnamefont {A.}~\bibnamefont
  {Kitaev}},\ }\href {\doibase 10.1103/PhysRevA.71.022316} {\bibfield
  {journal} {\bibinfo  {journal} {Phys. Rev. A}\ }\textbf {\bibinfo {volume}
  {71}},\ \bibinfo {pages} {022316} (\bibinfo {year} {2005})}\BibitemShut
  {NoStop}%
\end{thebibliography}%

\bibliographystyle{apsrev4-1}

\clearpage
\onecolumngrid
\section*{\LARGE{Supplementary information}}

\renewcommand{\thefigure}{S\arabic{figure}}
\renewcommand{\thetable}{S\arabic{table}}

\setcounter{figure}{0}
\setcounter{table}{0}

\section*{Experimental design}
\begin{figure}
\begin{center}\includegraphics[scale=0.6]{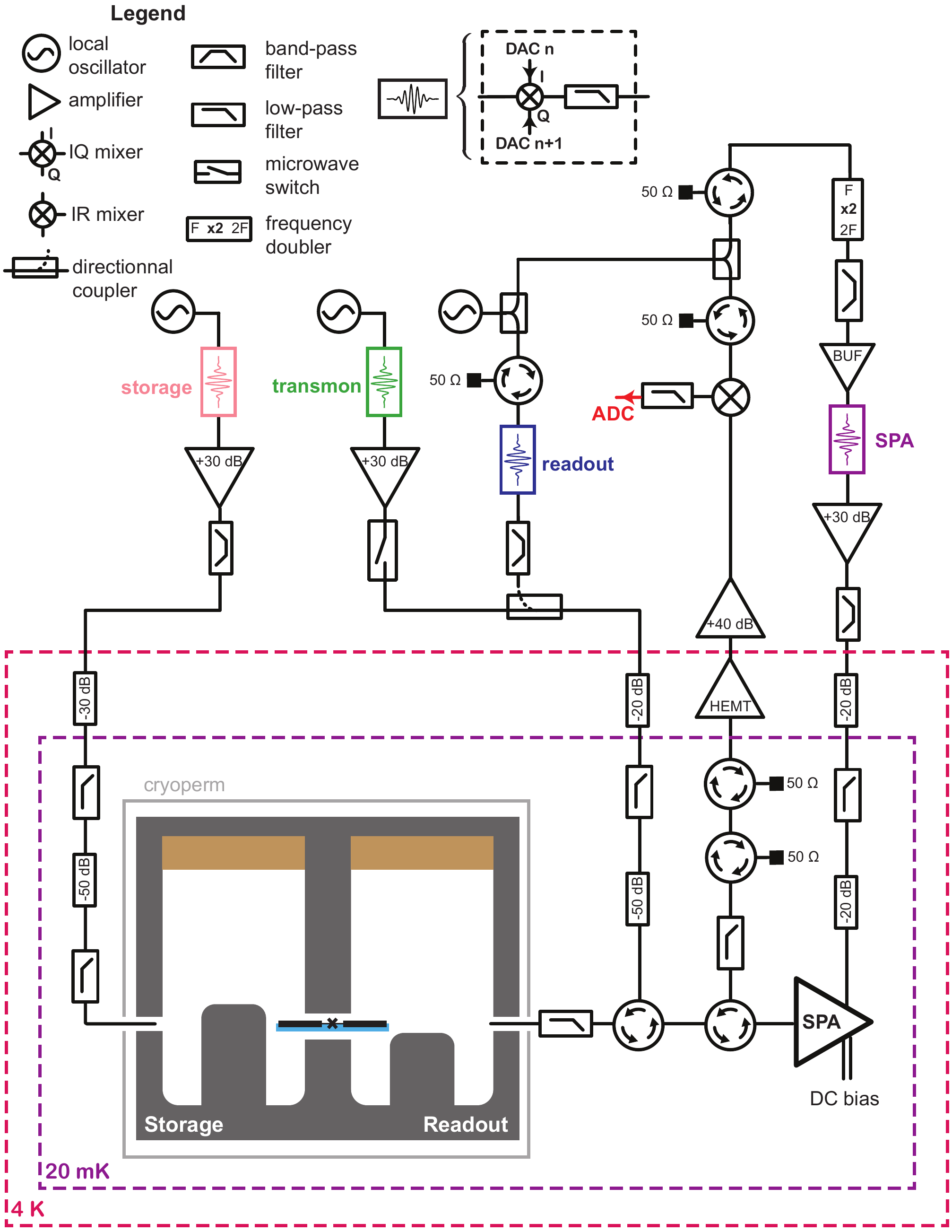}
    \caption{\label{fig:lines} {\bf Wiring diagram.} Two coaxial microwave resonators (gray) bridged by a single transmon superconducting qubit (black) fabricated on a sapphire chip (blue, see text for details) are anchored on a dilution refrigerator base plate. The cavities are machined out of a single aluminum block, and closed by a lid coated with copper powder mixed in \emph{Apiezon Wax W} (brown). Microwave pulses used to probe or control the system are generated by IQ mixing of a local oscillator (LO) with a low frequency signal with arbitrary envelope delivered by the DAC of an integrated FPGA system. The microwaves propagate down heavily attenuated lines to the system. Circulators are used to route the reflected readout signal to a near quantum limited SNAIL-parametric amplifier (SPA)~\cite{frattini2018} used in phase-sensitive mode (pulsed pump microwave at twice the readout signal frequency). The signal is  further amplified and down-converted at room-temperature before digitization by the FPGA board. 
    }
\end{center}
\end{figure} 

The experimental system is schematically represented in Fig.~\ref{fig:lines}, and its parameters are summarized on Table~1. The storage and readout oscillators are the fundamental modes of two rectangular coaxial microwave cavities machined out of a single 6061 aluminum alloy block. A section of rectangular waveguide with the same dimensions extends each cavity and is closed off by a top lid coated with copper powder mixed in \emph{Apiezon Wax W} (brown layer in Fig.~\ref{fig:lines}). We embed in the wax a copper braid attached to the base-plate of the refrigerator in order to thermally anchor this non-magnetic, highly dissipative material, which damps and thermalizes the higher frequency modes of the structure. The cavities are protected from external magnetic fields using an {\it Amumetal A4K} shield.\\

The cavities are bridged by a single transmon superconducting circuit, made of a double-angle-evaporated $\mathrm{Al/AlO_x/Al}$ Josephson junction (inductance 7.3~nH) bridging two 0.7 mm-by-0.4 mm rectangular aluminum pads. It is fabricated using the bridge-free fabrication technique on a double-side-polished 5 mm-by-37.5 mm chip of c-plane sapphire with a 0.43 mm thickness. The chip is clamped from two sides (out of the representation plane in Fig.~\ref{fig:lines}) by compressing it between thermally anchored copper blocks covered with a 200~$\mu$m thick indium foil.\\

The full wiring diagram of the experiment is depicted in Fig.~\ref{fig:lines}. The microwave lines are filtered using both homemade \emph{eccosorb}-based dissipative filters and commercial reflective \emph{K}\&\emph{L} filters. The pulses used to probe and control the system are generated at room temperature by IQ mixing of a local oscillator (LO) provided by a microwave source at $\omega+\omega_h$ with a low frequency signal at $\omega_h$ ($ 50~\mathrm{MHz}< \omega_h/2\pi < 100~\mathrm{MHz}$) with arbitrary envelope delivered by the Digital to Analog Converter (DAC) of an integrated FPGA system.  For reading out the transmon, the signal reflected off from the cavity is pre-amplified by reflection on a near quantum limited SNAIL-parametric amplifier (SPA)~\cite{frattini2018} used in phase-sensitive mode: a pulsed, high-power, pump at twice the readout signal frequency is generated by frequency-doubling and IQ-mixing the signal from the same LO. The readout signal is then further amplified, down-converted at room-temperature and digitized by the FPGA system.

\section*{The conditional displacement gate}
\label{sec:condis}

\begin{figure}
\begin{center}
    \includegraphics{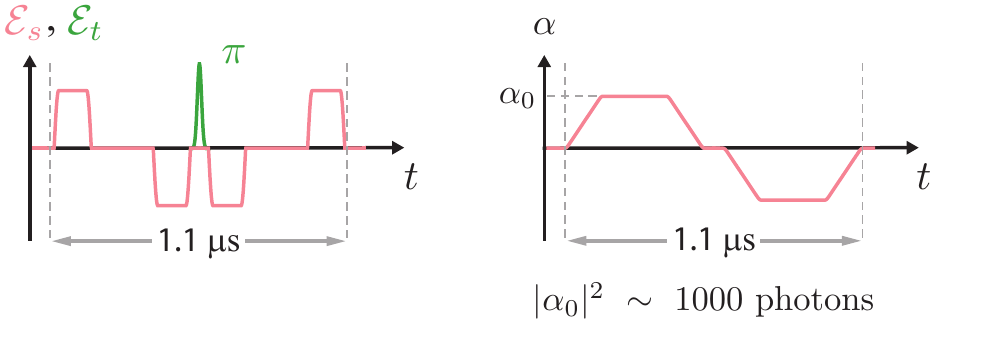}
    \caption{\label{fig:pulse_sequence} {\bf Conditional displacement sequence} Pulse sequence (left panel) and response of the storage mode (right panel) leading to a conditional displacement. The amplitude of the storage and transmon pulses are given by $\mathcal{E}_s$ and $\mathcal{E}_t$ respectively. The first two resonant pulses (pink) drive a large excursion of the storage state in phase space, parametrised by the amplitude $\alpha(t)$. The state of the transmon is then flipped with a rotation of angle $\pi$ (green), and two more pulses drive the storage mode along a symmetric trajectory. Overall, the storage mode returns to its initial state, but a transmon-dependent conditional displacement $\CD(\beta)$ emerges, where $\beta$ is proportional to the amplitude of the storage pulses (see text). The total duration of the sequence is \SI{1.1}{\micro \second}}
\end{center}
\end{figure} 

In this section, we detail the sequence that leads to the conditional displacement gate of the main paper. In the dispersive regime of cQED, the joint storage-transmon Hamiltonian is, in the doubly rotating frame,

 \begin{equation}
 \frac{{\bf H}(t)}{\hbar}=-\frac{\chi}{2} \adag \aop \sigmaz +i \mathcal{E}_s(t) \adag - i\mathcal{E}_s^{\ast}(t) \aop,
 \label{eq:hamil}
 \end{equation} 
 where $\aop=(\qop + i \pop)/\sqrt{2}$ is the annihilation operator of the storage mode, $\sigmaz$ is a Pauli operator of the transmon, $\chi$ is the dispersive shift, and the displacement rate $\mathcal{E}_s$ results from a resonant microwave drive applied to the storage mode through a weakly coupled port.\\
 The conditional displacement unitary results from a sequence of fast transmon rotations (30~ns long drive pulses shaped to minimize leakage to the second excited state~\cite{chow2010optimized}, not included in \eqref{eq:hamil}) and displacements of the storage oscillator induced by the resonant drive in \eqref{eq:hamil}. The effect of this drive can be accounted for by moving to a displaced frame via the transformation $\aop \rightarrow \aop+ \alpha(t)$, where $\alpha(t)$ is the response of the storage oscillator, which changes the Hamiltonian into

 \begin{equation}
 \label{eq:H_tilde}
 \frac{{\bf \tilde{H}}(t)}{\hbar}=-\frac{\chi}{2} \big(\alpha(t) \adag + \alpha^{\ast}(t) \aop \big)\sigmaz
 -\frac{\chi}{2}  \adag \aop \sigmaz  -\frac{\chi}{2} |\alpha(t)|^2\sigmaz.
 \end{equation}
 The first term of this expression is the generator of conditional displacements while the other two are spurious terms that must be suppressed. To this end, we perform a double echo decoupling sequence, where we synchronously change the sign of $\alpha(t)$ and flip the ancilla transmon, such that the desired coupling persists while the spurious terms are eliminated. To do so, the storage mode is first displaced by a large value $\alpha$ ($|\alpha|^2\simeq320$~photons when measuring the stabilizers, and up to $|\alpha|^2\simeq2500$~photons when measuring points of the characteristic function lying at corners of the colormaps presented in Figs.~2-4 of the main text) and brought back to the origin of phase-space after an arbitrary time. Next, the transmon state is flipped and the oscillator state is displaced in the opposite direction for the same duration, after which it is again displaced back to the origin. This sequence is represented in Fig.~\ref{fig:pulse_sequence}.

We now compute explicitly the evolution induced by the time-dependent Hamiltonian ${\bf \tilde{H}}$. We note $T_{\mathrm{int}}$ the duration of interaction, $\varepsilon =\frac{\chi T_{\mathrm{int}}}{2 N}$ and $t_k=k\varepsilon$. In the limit $N \rightarrow \infty$, we can write the evolution operator as
\begin{equation}
\U=\prod_{k=1}^N e^{i\varepsilon (\Ha(t_k) +\Hb(t_k)+\Hc(t_k))},
\end{equation}
where
\begin{equation}
\begin{split}
\Ha(t)=&\adag \aop ~s(t)\sigmaz \\
\Hb(t)=&(\alpha(t) \adag+\alpha(t)^{\ast}\aop)~s(t)\sigmaz \\
\Hc(t)=&|\alpha(t)|^2~s(t)\sigmaz \\
s(t) =& 1 - 2\Theta(t-T_{\mathrm{int}}/2)
\end{split}
\end{equation}
In these expressions, $s(t)=\pm 1$ accounts for the transmon echo at $T_{\mathrm{int}}/2$ (see Fig.~2b), and we use the ordering $\prod_{k=1}^N \U_k = \U_N \U_{N-1}... \U_1$ for the product so that the evolution operators are applied in chronological order.\
The terms $\Hc(t_k)$ commute with $\Ha(t_l)$, $\Hb(t_{l})$ and $\Hc(t_{l})$ at all times $t_l$, and can thus be factored out of the exponential and grouped in a separated product.   Moreover, $\prod_{k=1}^N e^{i\varepsilon \Hc(t_k)}=\I$ so that the corresponding evolution can be dropped out. We now use the Baker-Campbell-Hausdorff formula to expand the exponential terms in $\Ha+\Hb$ at each time $t_k$
\begin{equation}
\label{U2}
\U=\prod_{k=1}^N e^{i\varepsilon \Ha(t_k)} e^{i\varepsilon \Hb(t_k)}  \U_{nc}(t_k) ,
\end{equation}
where $\U_{nc}$ captures an infinite product of nested commutators exponentials. We verify that $\mathrm{log} (\U_{nc}(t_k))=O(\varepsilon^2) $ so that these evolutions can be neglected when integrating over $[0,T_{\mathrm{int}}]$.\\
We use again the Baker-Campbell-Hausdorff formula to rearrange the remaining terms in Eq.~(\ref{U2}):
\begin{equation}
\U=  \big(\prod_{k=1}^N \prod_{l=1}^{k-1} e^{-\varepsilon^2 \G(t_l)} \big)\big(\prod_{k=1}^N e^{i\varepsilon \Hb(t_k)} \big)\big(\prod_{k=1}^N e^{i\varepsilon \Ha(t_k)} \big).
\end{equation}
Here, we have defined $\G(t)=[\Ha(t),\Hb(t)]=\alpha(t)\adag-\alpha^{\ast}(t)\aop$, and commuted the terms $e^{-\varepsilon^2 \G(t_l)}$ through the others by neglecting $N^2/2$ evolution operators $\V_{nc}(t_k,t_l)$ with $\mathrm{log} (\V_{nc}(t_k))=O(\varepsilon^3)$.\\
Now using that $[\Ha(t_k),\Ha(t_l)]=[\Hb(t_k),\Hb(t_l)]=[\G(t_k),\G(t_l)]=0$ for all $k, l$ ($\alpha$ keeps the same phase modulo $\pi$ over the whole evolution) and going to the continuous limit we get
\begin{equation}
\U= e^{- \frac{\chi^2}{4} \int_{t=0}^{T_{\mathrm{int}}} \int_{t'=0}^t \G(t') \mathrm{d}t'\mathrm{d}t} e^{i \frac{\chi}{2} \int_{t=0}^{T_{\mathrm{int}}} \Hb(t) \mathrm{d}t} e^{i \frac{\chi}{2} \int_{t=0}^{T_{\mathrm{int}}} \Ha(t) \mathrm{d}t}.
\end{equation}
The last term in this product cancels due to the transmon echo at $T_{\mathrm{int}}/2$. The second term gives the conditional displacement $\CD(\beta)$ with $\beta=i\sqrt{2}\chi \Big( \int_t^{T_{\mathrm{int}}/2} \alpha(t)\mathrm{d}t -\int_{T_{\mathrm{int}}/2}^{T_{\mathrm{int}}} \alpha(t)\mathrm{d}t \Big) $ that we consider in the main paper. The first term gives a short unconditional displacement of the oscillator orthogonal to the desired conditional displacement. We finally get
\begin{equation}
\U=\D(\gamma) \CD(\beta) ,
\end{equation}
with $\gamma=-\frac{\chi^2}{2\sqrt{2}}\int_{t=0}^{T_{\mathrm{int}}} \int_{t'=0}^t \alpha(t')\mathrm{d}t'\mathrm{d}t$. Note that in the limit of a fast conditional displacement for which $\int |\alpha| \mathrm{d}t  \rightarrow \infty $ and $T_{\mathrm{int}} \rightarrow 0$, we have $|\gamma/\beta|\rightarrow 0$.\\

For the values of $\alpha$ and the interaction time we use for the square grid stabilization, we find $|\gamma|=0.04\sim |\beta|/60$. This small spurious shift of the oscillator state is compensated for when applying the feedback displacement following the interaction, which is also in the orthogonal direction to the conditional displacement. The feedback shifts after peak-sharpening steps are thus asymmetric and read $\delta_q = +0.04~\pm~0.2$ when sharpening the peaks along $q$, and $\delta_p = -0.04~\mp~0.2$ when sharpening the peaks along $p$. In practice, we vary the length of these corrective shifts around the predicted value and find $\delta_q = +0.06 \pm 0.2$ and $\delta_p = -0.06 \mp 0.2$ to empirically maximize the logical qubit coherence time (see Sec.~\ref{sec:kickbacks}). This small discrepancy can be attributed to experimental imperfections shifting the grid states during the other error-correction rounds: for comparison, a relative variation by $\sim 10^{-3}$ of the voltage of the DAC used to generate the storage mode displacements pulses (see Fig.~\ref{fig:lines}) between the first and second half of the conditional displacements is sufficient to create such systematic shifts. 

\section*{Experimental parameters characterization}

\begin{center}
\begin{table}
\begin{center}
\begin{tabular} {|c||c|c|}
\hline
\rowcolor{lightgray}
Storage oscillator single-photon lifetime & $T_s$ & $245~\mu\mathrm{s}$ \\
\hline
Storage oscillator frequency & $\omega_s/2\pi$ & 5.26~GHz \\ 
\hline
\rowcolor{lightgray}
Storage oscillator Kerr anharmonicity & $K_s/2\pi$ & 1~Hz \\ 
\hline
\rowcolor{lightgray}
Transmon energy lifetime & $T_1$ & $50~\mu\mathrm{s}$ \\ 
\hline
\rowcolor{lightgray}
Transmon coherence lifetime ~~~~~~(echo) & $T_{2e}$ & $60~\mu\mathrm{s}$ \\ 
\hline
Transmon coherence lifetime ~(Ramsey) & $T_{2R}$ & $8~\mu\mathrm{s}$ \\ 
\hline
Transmon resonance frequency & $\omega_t/2\pi$ & 6.71~GHz \\ 
\hline
Transmon anharmonicity & $K_t/2\pi$ & 193~MHz \\ \hline
Readout oscillator single-photon lifetime & $T_r$ & $65~\mathrm{ns}$ \\
\hline
Readout oscillator frequency & $\omega_r/2\pi$ & 8.2~GHz \\ 
\hline
Storage-transmon dispersive shift & $\chi/2\pi$ & 28~kHz\\ 
\hline
Readout-transmon dispersive shift & $\chi_r/2\pi$ & 1~MHz \\ 
\hline
\rowcolor{lightgray}
Jump rate to higher transmon levels during error-correction & $\Gamma_{\rightarrow f}$ & $(3~\mathrm{ms})^{-1}$ \\ 
\hline
\end{tabular}
\end{center}
\caption{ {\bf Measured experimental parameters.} Parameters entering the master-equation simulations described on Sec.~\ref{sec:MES} are highlighted in gray.
}
\end{table}
\end{center}

\subsection*{Transmon readout}
\begin{figure}
\begin{center}\includegraphics[scale=1]{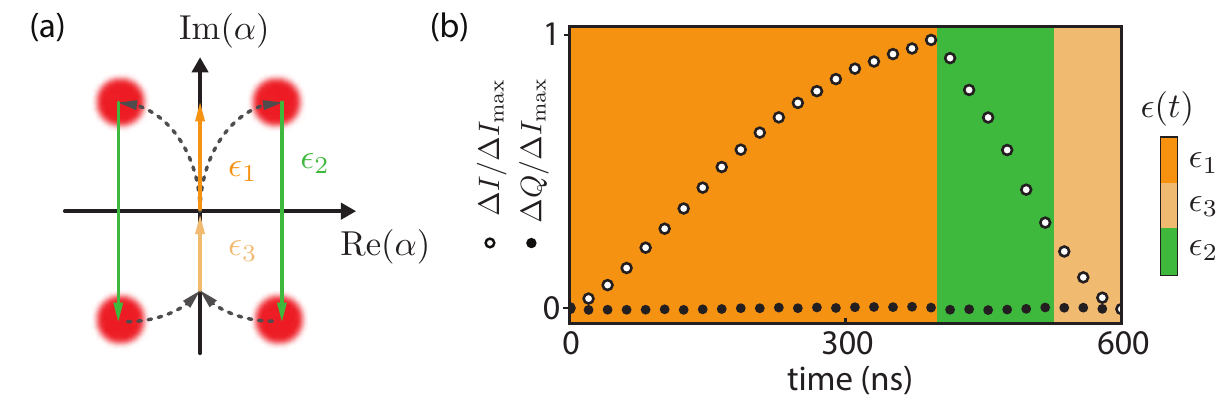}
    \caption{\label{fig:readout} {\bf Fast transmon readout} {\bf a)} Trajectories of the readout oscillator state in phase-space conditioned on the transmon state. By alternating positive and negative amplitude drives of the oscillator, the two trajectories separate then recombine (see text). {\bf b)} Measured differential signal when the transmon is prepared in $|\pm z\rangle$. The reflected signal from the readout resonator is amplified (near quantum limited amplifying chain including a phase-sensitive parametric amplifier~\cite{frattini2018}). The down-converted, digitized and window-integrated signal quadratures  $I$ and $Q$ reflect the intra-cavity field coordinates $\mathrm{Re}(\alpha)$ and $\mathrm{Im}(\alpha)$, respectively amplified and de-amplified. 
    }
\end{center}
\end{figure} 

The transmon is readout using a dedicated resonator (see Fig.~\ref{fig:lines}) overcoupled to an output line (photon exit rate $ \kappa_r = 2\pi \times 2.5~\mathrm{MHz}$) forming a near-quantum limited amplification chain. The transmon and the readout oscillator are dispersively coupled with a dispersive shift $\chi_r= 2\pi \times 1~\mathrm{MHz}$. We measure the transmon with the largest possible photon number without degrading its $T_1$. Note that the dispersive coupling between the readout mode and the storage mode $\chi_{sr}=\frac{\chi \chi_r}{2K_t}\sim 2\pi\times70~\mathrm{Hz}$ (see Table.~1) is small enough that the storage mode is not dephased by this measurement. \\

In order to shorten the ringdown time of the resonator, which is desired to limit the dead time before manipulating the transmon again, we use a fast unloading protocol~\cite{IBMreset}. This modified dispersive readout scheme is schematized in Fig.~\ref{fig:readout}a. It starts by driving the resonator with a positive amplitude $\epsilon_1$. The readout oscillator state gets displaced and separates along an angle given by the state of the transmon. We then revert the amplitude of the drive to a large negative value $\epsilon_2$ (green arrows) to displace the two possible coherent states to the opposite direction in phase-space. The two states then recombine and are brought back to the origin using a third drive with a positive amplitude $\epsilon_3$. One can understand this sequence by comparing it to the conditional displacement sequence described in Sec.~\ref{sec:condis}. However, here, we do not flip the transmon state in the middle of the sequence. Thus, the overall conditional displacement resulting from the action of the second term in {$\bf \tilde{H}$} is canceled: depending on the transmon state, the oscillator trajectories in phase-space first separate, then recombine. Integrating the transient signal yields the transmon state, with the oscillator left in the ground state at the end of the sequence.\\


A noticeable difference with the case of the storage mode conditional displacement is that the photon lifetime in the oscillator is, here, of the same order as the sequence duration. We optimize the exact amplitudes and phases of $\epsilon_{1, 2, 3}$ using an algorithm that minimizes the number of photons in the oscillator after 600~ns. As a result, the transmon can be rotated 100~ns later (see Fig.~2b) with no detectable spurious dephasing. The measured differential signal for the transmon prepared in $|\pm z\rangle$ is plotted on  Fig.~\ref{fig:readout}b. From its maximum value, the signal drops by more than $99~\%$ in about 200~ns, which is shorter than the time $\sim10T_r=650~$ns required for the oscillator field amplitude to decay by a comparable amount with passive photon damping. \\
Integrating the differential signal, we obtain a readout fidelity above $99~\%$ (this lower bound is inferred from the $99.5~\%$ contrast of a Rabi oscillation recorded to calibrate the experimental setup). Note that, in our QEC protocol, readout errors have a similar effect as transmon phase-flips (see Sec. S.~7.1), which do not impact significantly the error-correction performances (see Table 2). Readout errors are not included in master equation simulations, in which the readout sequence is modeled as a perfect instantaneous projection of the transmon state after 350~ns in master-equation simulations (see Sec.~\ref{sec:MES}).

\subsection*{Displacement length characterization }
\label{sec:discalib}
The transmission of the input lines cannot be precisely known, so the storage oscillator displacement rate for a given room-temperature pulse power is {\it a priori} unknown. We thus need to calibrate the displacement $\D(\delta)$ associated with a given driving pulse amplitude  (the pulses used for feedback displacements have a Gaussian temporal shape with a fixed standard deviation of 5~ns). Similarly, for the driving sequence represented in Fig.~\ref{fig:pulse_sequence}, we need to estimate the conditional displacement $\CD(\beta)$ corresponding to a given drive amplitude. Note that we could in principle deduce one of these scaling factors from the other (conditional displacements result from large unconditional ones), but such an estimate would be imprecise due to the vastly different timescales involved. \\

\begin{figure}
\begin{center}\includegraphics[scale=1.2]{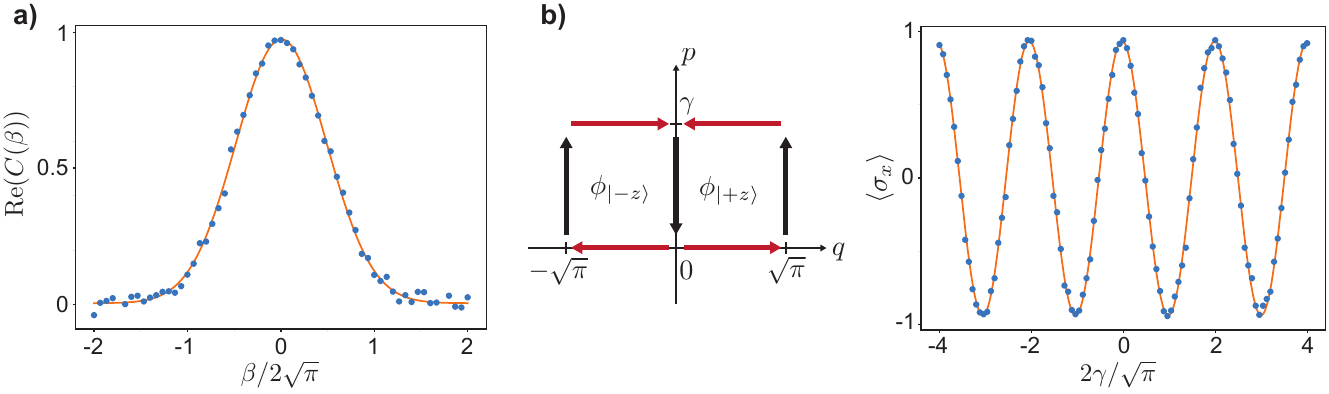}
    \caption{\label{fig:disp} {\bf Displacement length calibration.} {\bf a)} We adjust the scaling of the conditional displacements so that the measured characteristic function of the vacuum state (blue circles) is a Gaussian with standard deviation $\sqrt{2}$ (orange line). {\bf b)} Unconditional displacements are calibrated by applying a sequence of conditional displacements $\CD( 2\sqrt{\pi})$ (red arrows) and unconditional ones $\D(\pm \gamma)$ (black arrows) on the transmon prepared in $|+x\rangle$. At the end of the sequence, the transmon state has picked up a geometrical phase $\phi=-\phi_{|+z\rangle}+\phi_{|-z\rangle}$ where $\phi_{|\pm z\rangle}$ is the oriented area circled in the oscillator phase-space when the transmon is in $|\pm z \rangle$. The displacement scaling is adjusted so that varying $\gamma$ results in oscillations of $\langle \sigmax \rangle$ (blue circles) with a period $\sqrt{\pi}$ (orange line).
    }
\end{center}
\end{figure}

We start by estimating the conditional displacements scale. This is done by measuring the characteristic function of the vacuum state, which reads $C_{\mathrm{vac}}(\beta)=e^{-\frac{|\beta|^2}{4}}$ with our convention. We assume the oscillator to be in the vacuum state at equilibrium and, as described in Sec.~1,  we measure $C(\beta)$ (plotted in Fig.~\ref{fig:disp}a) by averaging a transmon measurement following a conditional displacement $\CD(\beta)$. Neglecting the transmon bit-flip errors during the conditional displacement and spurious thermal excitations of the oscillator, the scale of the conditional displacements is then set by adjusting $C(\beta)$ to a Gaussian with standard deviation $\sqrt{2}$. As for the other parameters relevant to the performance of our QEC protocol, this original calibration of the conditional displacements is finely tuned (variation of about $1~\%$)  by empirically maximizing the coherence of the logical qubit under QEC.\\

We now turn to calibrating the unconditional displacements length. This is done by performing the sequence of gates $\D(-i \gamma)\CD(-2\sqrt{\pi})\D(i \gamma)\CD(2\sqrt{\pi})$ on the oscillator initially in vacuum and the transmon in $|+x\rangle$. $\gamma$ is the {\it a priori} unknown displacement length we want to calibrate. As represented in Fig.~\ref{fig:disp}b, the storage oscillator state follows a transmon-dependent trajectory in phase space. At the end of the sequence, the two possible trajectories recombine, so that the oscillator and the transmon are disentangled. However, the two trajectories circle around two opposed oriented areas in phase-space, so that the transmon picks up a geometrical phase: at the end of the sequence, its state reads $e^{i\sqrt{\pi}\gamma}|+z\rangle+e^{-i\sqrt{\pi}\gamma}|-z\rangle$ up to normalization. When varying $\gamma$, a final $\sigma_x$ measurement of the transmon yields oscillations with a period $\sqrt{\pi}$. The recorded oscillations plotted in Fig.~\ref{fig:disp}c are used to calibrate the displacement length.

\subsection*{Single-photon lifetime in the storage mode}
\label{sec:storagelifetime}

We estimate the photon lifetime in the storage mode by recording the decay of a coherent state amplitude and neglecting any pure dephasing (see Sec.~\ref{sec:dephasing} for comments on this hypothesis). We start by displacing the oscillator state by $\delta_0$ (coherent state $|\delta_0/\sqrt{2}\rangle$ with our notation), the transmon being in its ground state. The free decay of the coherent field  is expected to be
\begin{equation}
    \delta(t)=\delta_0 e^{-i\Delta_s t-\frac{t}{2T_s}}.
\end{equation}
where $\Delta_s$ is the detuning between the microwave radiation used to displace the oscillator (loading and characteristic function measurement) and the resonance frequency. The characteristic function of the oscillator in this coherent state reads
\begin{equation}
\label{eq:characcoh}
C(\beta,t)=e^{-\frac{1}{4}|\beta|^2}e^{-i\mathrm{Im}(\delta(t) \beta^{\ast})}.
\end{equation} 

We measure $C(\beta,t)$ for $\beta \in \mathbb{R}$ and $\beta \in i\mathbb{R}$ and fit this signal to obtain $\delta(t)$. The fitted value as a function of time is represented in Fig.~\ref{fig:freedecay}a. The coherent state rotates in phase-space and its amplitude decays exponentially with a characteristic time $2T_{s}=490~\mu\mathrm{s}$. 

\subsection*{Storage mode resonance frequency and dispersive coupling to the transmon}
\label{sec:dispersive}
\begin{figure}
\begin{center}\includegraphics[scale=1.2]{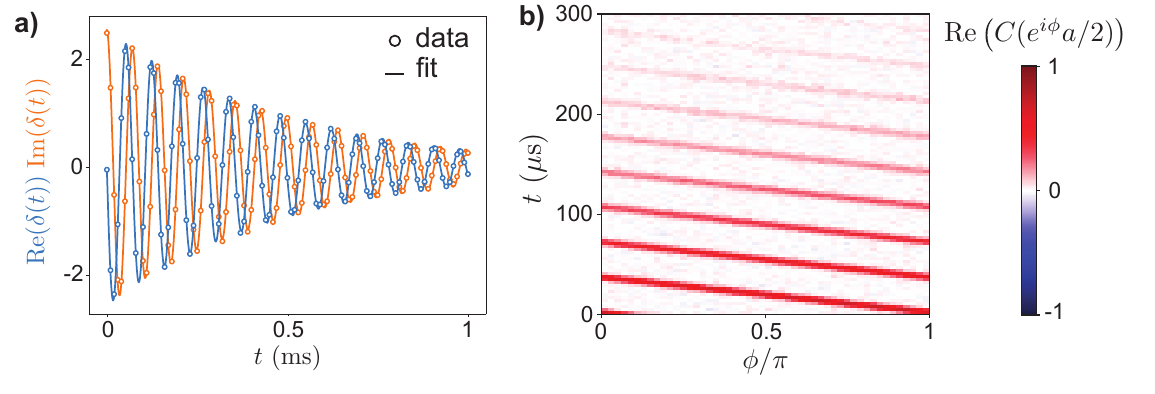}
    \caption{\label{fig:freedecay} {\bf Storage oscillator characterization}  {\bf a)} With the transmon in its ground state, the free decay of a coherent state $|\delta(t)/\sqrt{2}  \rangle$ in the storage oscillator is monitored by measuring the characteristic function at a varying time $t$ and fitting it using Eq.~(\ref{eq:characcoh}). The state amplitude decays exponentially with characteristic time $2T_s=490~\mu$s and it rotates at $-\chi/2$, which is the detuning between the rotating frame frequency and the storage oscillator frequency dressed by the transmon in $|g\rangle$. {\bf b)} After preparing the logical qubit in $|+X_L\rangle$, the free decay of the $\X$ component of the logical Bloch vector (QEC turned off) is monitored in time. To this end, we measure the real part of the characteristic function for points lying on a circle with radius $a/2$. The value of $C(e^{i\phi}a/2)$ for $0<\phi<\pi$ is represented in color at various times. At $t=0$, the  function takes its maximum value at $\phi=0$ where it coincides with $\langle \mathrm{Re}(\X) \rangle$. For $t>0$, the transmon being in its ground state, the storage oscillator frequency is detuned from the rotating frame frequency and the function takes its maximum value at an angle $\phi_{\mathrm{max}}=-\chi t/2$. }
\end{center}
\end{figure}

From the same field amplitude decay measurement described in the previous section, we can extract the storage mode resonance frequency $\omega_s - \frac{\chi}{2} $ when the transmon is in its ground state (see Eq.~1). We repeat this measurement after preparing the transmon in its excited state, in which case the mode resonates at $\omega_s + \frac{\chi}{2} $. We thus get a first estimate of the oscillator  frequency $\omega_s$ (the rotating frame frequency of the Hamiltonian (2)), and of the dispersive shift $\chi$.\\ 

However, transmon relaxation events during the field decay and during the characteristic function measurement sequence limit the precision of this calibration. A more precise estimate of $\omega_s$  within $\pm 100$~Hz is obtained by varying the frequency of all the oscillator control pulses during stabilization and empirically maximizing the logical qubit coherence time. $\chi$ is also more finely estimated by considering the decay of the logical Pauli operators expectation value when the stabilization is OFF. For a varying time $t$ after preparing the logical qubit in $|+X_L\rangle$ and resetting the transmon in its ground state, we measure the real part of the characteristic function for points lying on a circle with radius $a/2$: in Fig.~\ref{fig:freedecay}b, we represent in color $\text{Re}\left(C(\frac{a}{2}e^{i\phi})\right)$ for $\phi \in [0,\pi]$ at various times. At $t=0$, the function takes its maximal value, corresponding to $\langle \mathrm{Re}(\X) \rangle$, at $\phi=0$. For $t>0$, the position of the maximum rotates at $-\frac{\chi}{2}$, which is the detuning between the working frequency and the oscillator frequency for the transmon in its ground state. We thus get $\chi= 2\pi \times 28~$kHz, which is the value used to estimate the maximum photon number of $|\alpha^{\mathrm{max}}|^2=320$ reached when performing the conditional displacements for the measurement of the stabilizers. We can also estimate the storage oscillator Kerr anharmonicity inherited from its hybridization to the transmon mode~\cite{nigg2012black} to be $K_s=\frac{\chi^2}{4 K_t}=2\pi \times1~$Hz, where $K_t=2\pi\times 193~$MHz is the transmon anharmonicity.\\

Extracting the maximum value of the circular characteristic function cuts at all times after preparing $|+X_L\rangle$, $|+Y_L\rangle$ and $|+Z_L\rangle$, we reconstruct the decay of the three components of the logical Bloch vector $\langle \mathrm{Re}(\X) \rangle_{\mathrm{OFF}}$, $\langle \mathrm{Re}(\Y) \rangle_{\mathrm{OFF}}$ and $\langle \mathrm{Re}(\Z) \rangle_{\mathrm{OFF}}$, used as a baseline to evaluate the quantum error-correction performances of our protocol (color crosses in Fig.~3c and Fig.~4c).

\subsection*{Excitation to higher levels of the transmon}
\label{sec:fpop}

Excitations of the transmon to its second excited state $|f\rangle$ lead to a depolarization of the logical qubit. Indeed, while our protocol includes resets of the transmon every correction round by non-demolition readout and feedback, the control pulses on the $|g\rangle \leftrightarrow |e\rangle$ transition are ineffective when the transmon is in $|f\rangle$. Thus, when the transmon excites to $|f\rangle$, the oscillator state rotates at $\Delta_s=-\frac{3}{2}\chi$, which is the detuning between the working frequency and the dispersively shifted oscillator frequency, until the transmon spontaneously relaxes back to $|e\rangle$ at a random time. Such relaxation happens after a typical time $T_1/2$, which is much longer than $1/\Delta_s$, so that by the time the transmon eventually relaxes to the $\{|g\rangle,|e\rangle\}$ manifold, the oscillator state has equal chance to be decoded as, say, $|+Z_L\rangle$ as $|-Z_L\rangle$. Thus, the leakage rate $\Gamma_{\rightarrow f}$ of the transmon to the $|f\rangle$ state directly translates into a depolarization rate for the logical qubit.\\

Such excitations of the transmon can be either thermally activated (the equilibrium occupation of the first excited state of the transmon is found to be $\sim~1~\%$) or triggered by the fast control pulses applied on the $|g\rangle \leftrightarrow |e\rangle$ transition. Indeed, the total duration of these pulses is 30~ns, which corresponds to a spectral width of the same order as the transmon anharmonicity. We limit these undesired excitations by using derivative removal via adiabatic gate (DRAG) pulse shaping~\cite{chow2010optimized}. \\
\begin{figure}
\begin{center}\includegraphics[scale=1.2]{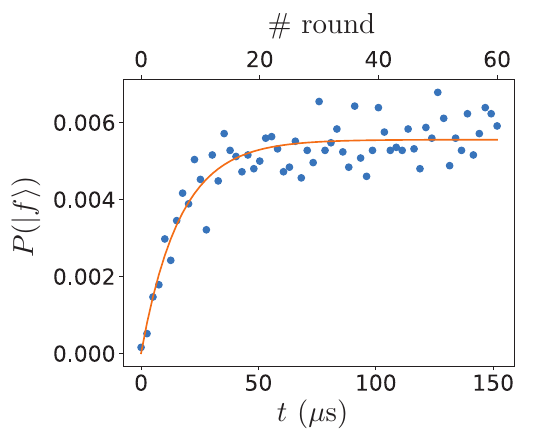}
    \caption{\label{fig:fpop} {\bf Excitation to $|f\rangle$.} From the thermal equilibrium state of the transmon (negligible occupation of the $|f\rangle$ level), the same pulse sequence used in the QEC protocol is applied to the system. The occupation of $|f\rangle$ is measured as a function of round number using a modified dispersive readout. The finite readout fidelity $F\approx0.9$ is calibrated and corrected for. The orange line is an exponential fit with characteristic time $T=16.5~\mu\mathrm{s}\lesssim T_1/2$.
    }
\end{center}
\end{figure} 

We estimate the rate $\Gamma_{\rightarrow f}$ of transmon excitation to $|f\rangle$ by measuring the $|f\rangle$ level occupation as a function of round number during correction (see Fig.~\ref{fig:fpop}), and using a hidden Markov model. Before the protocol is turned on, the probability of occupation of $|f\rangle$ is negligible (thermal equilibrium). When the protocol is on, the average occupation of the first excited state instantaneously becomes $P(|e\rangle)= 0.5$, and the $|f\rangle$ level starts to fill. $P(|f\rangle)$ reaches a new equilibrium $P_{eq}(|f\rangle)=P(|e\rangle)\frac{\Gamma_{ef}}{\Gamma_{fe}}$ with a characteristic time $1/(\Gamma_{ef}/2+\Gamma_{fe})$, where $\Gamma_{ef}$ (respectively $\Gamma_{fe}$) is the excitation rate to $|f\rangle$ (respectively de-excitation rate to $|e\rangle$) when the transmon is in $|e\rangle$ (respectively $|f\rangle$). Fitting $P(|f\rangle)$ with an exponential, we extract $\Gamma_{ef}$ and then get $\Gamma_{\rightarrow f}=P(|e\rangle)\Gamma_{ef}=(3~\mathrm{ms})^{-1}$.

\subsection*{Master equation simulations}
\label{sec:MES}
In this section, we briefly describe the  master-equation simulations  reproducing all experimental data presented in the main text. The parameters entering the simulations are summarized in Tab.~I and are independently calibrated.\\

The joint storage-transmon state is represented by a $300\times 300$ density matrix ($150\times2$-dimension Hilbert space). Its dynamics is found by solving the Lindblad master-equation
\begin{equation}
    \frac{\mathrm{d}\rhoo}{\mathrm{d}t}=-\frac{i}{\hbar}[{\bf \tilde{H}},\rhoo]+\frac{1}{T_s}\mathcal{D}[\aop]\rhoo +\frac{1}{T_1}\mathcal{D}[\sigmam]\rhoo+\frac{1}{2T_{\phi}}\mathcal{D}[\sigmaz]\rhoo,
\end{equation}
where $\mathcal{D}[\LL]\rhoo=\LL \rhoo \LL^{\dag}-\frac{1}{2}(\LL^{\dag}\LL\rhoo+\rhoo \LL^{\dag}\LL)$ and $T_{\phi}=1/(1/T_2-1/2T_1 )$ is the transmon pure dephasing time. The Hamiltonian ${\bf \tilde{H}}$ is taken in the displaced frame during the conditional displacements and following Eq.~\ref{eq:H_tilde}, it reads
\begin{equation}
\label{eq:Hdisp}
\frac{{\bf \tilde{H}}}{\hbar}=-\frac{\chi}{2}  (\adag+\alpha^{\ast})(\aop+\alpha)  \sigmaz - \frac{K_s}{2} \Big((\adag+\alpha^{\ast}) (\aop+\alpha)\Big)^2,
\end{equation}
where we have included for completeness the Kerr anharmonicity inherited by the oscillator.\\

We numerically solve this equation using {\it Qutip}~\cite{johansson2013qutip} during each round of QEC. The transmon control pulses and the feedback displacements of the storage oscillator are modeled as instantaneous unitary evolutions (rotation and displacement operators applied on the density matrix). The transmon readout is modeled as a perfect, instantaneous projection taking place at the middle of the actual readout pulse (time $t_{1}$). We compute the two possible un-normalized density matrices resulting from this measurement $\rhoo_{\pm}(t_1)=\M_{\pm}\rhoo(t_1) \M_{\pm}^{\dag}$ corresponding to the two possible measurement outcomes, with $\M_{\pm}=|\pm z\rangle \langle \pm z|$. We simulate separately the evolution of these two matrices. After a time $t_{2}$ including the second half of the readout pulse and the delay required for the fast-electronics board to process the measurement signal, we apply the feedback operations $\U_{\pm}$ (oscillator displacement and rotation of the transmon) corresponding to each measurement outcome and sum the two matrices $\rhoo(t_1+t_2)=\U_+\rhoo_+(t_1+t_2)\U_+^{\dag}+\U_-\rhoo_-(t_1+t_2)\U_-^{\dag}$.\\

These simulations allow us to reproduce quantitatively the expectation values of the stabilizers measured experimentally (see Figs.~3a-5a and Fig.~5a) as well as the preparation fidelity of all logical states. Occupation of the $|f\rangle$ level of the transmon (below 1~\%) is then neglected. On the other hand, when considering the lifetime of the components of the logical qubit Bloch vector, the depolarization induced by transmon excitations to $|f\rangle$ with rate $\Gamma_{\rightarrow f}$ (see Sec.~\ref{sec:fpop}) cannot be neglected: the decaying coherence signal returned by simulations is fitted with an exponential law, and the supplementary dephasing rate $\Gamma_{\rightarrow f}$ is added to the simulated decay rate to get the predicted decay plotted in Figs.~4b-5c.\\

Overall, the simulations reproduce experimental data within a $\sim2~\%$ margin, except for the expectation values of the hexagonal code stabilizers, for which the mismatch increases to $5~\%$. This discrepancy is probably explained by the fact that this particular data set was recorded several days later than the rest of the experimental results presented in the paper, without re-tuning the experimental setup parameters (in particular compensating for drifts in the power of various pulses). \\

We also use master equation simulations to estimate the impact of the various error channels on the error-correction performances. The results of these simulations are summarized on Table~2. We do not give a quantitative error budget as some errors appear to slightly compound each other and their effect cannot be quantitatively estimated independently. Qualitatively, propagation of transmon bit-flips errors (see Sec.~\ref{sec:bitflips}) and errors linked to the intrinsic storage photon dissipation (see Sec.~\ref{sec:optenv}) dominate the error budget and each account for about half of the logical qubit errors. As expected, phase-flips of the transmon only marginally affect the QEC performances (see Sec.~\ref{sec:phaseflips}). 

\begin{center}
\begin{table}
\begin{center}
\begin{tabular} {|c||c|c||c|c||c|c|}
\hline
 & \multicolumn{2}{c|}{$\X$,$\Z$ square code} & \multicolumn{2}{c|}{$\Y$ square code}& \multicolumn{2}{c|}{$\X$,$\Y$,$\Z$ hex. code} \\
\hline
{\bf Error channel} & $T~(\mu\mathrm{s})$&$\Gamma/2\pi$~(kHz)& $T~(\mu\mathrm{s})$&$\Gamma/2\pi$~(kHz)& $T~(\mu\mathrm{s})$&$\Gamma/2\pi$~(kHz)\\
\hline

Dissip. &  890&0.18& 450& 0.36& 600& 0.27 \\
\hline
Bit-flips & 650& 0.24& 340& 0.47& 470& 0.34 \\
\hline
Phase-flips & 12000& 0.013& 6500& 0.025& 8000& 0.020 \\
\hline
Dissip. \& Bit-flips  & 340& 0.47& 180& 0.88& 250& 0.64 \\
\hline
Leakage to $|f\rangle$  & 3000& 0.053& 3000& 0.053& 3000& 0.053 \\
\hline
\hline

{\bf All}  & 295& 0.54& 165& 0.96& 220& 0.72 \\

\hline

\end{tabular}
\end{center}
\caption{ {\bf Impact of the various error channels on error-correction performances.} We use master equation simulations to estimate the lifetime of the logical qubit Bloch vector components in presence of each noise process separately (storage mode photon dissipation, bit-flips and phase-flips of the transmon and excitation of the transmon to the $|f\rangle$ level). The corresponding decay rate is given in each case. The storage oscillator photon dissipation and the transmon bit-flips dominate the error budget and slightly compound each other. 
}
\end{table}
\end{center}

\subsection*{Non-linearity and pure dephasing of the storage mode}
\label{sec:dephasing}

A crucial feature of our experimental system is the low value of the storage oscillator non-linearity inherited from its hybridization to the transmon mode. As mentioned is Sec.~\ref{sec:dispersive}, residual Kerr non-linearity is estimated to be $K_s\approx 2\pi \times 1~$Hz. A larger value could limit the coherence of the logical qubit in two ways. First, Kerr non-linearity distorts the grid states, which can be understood at first order as a position and momentum dependent rotation term in the oscillator phase-space. This effect gets stronger as the grid state envelope gets broader, which makes the envelope-trimming procedure described in Sec.~2 even more relevant. Second, spurious resonant terms appear in the displaced frame-Hamiltonian given in Eq.~(\ref{eq:Hdisp}). These terms of order $K_s|\alpha|^2$ lead to distortion of the grid state when performing too fast conditional displacements. In the experiment, we used a maximum value $|\alpha_{\mathrm{max}}|^2=$320~photons. For twice faster conditional displacements (more than 4 times larger maximum photon number), we observed a lower logical qubit coherence time, which is reproduced by simulations. Note that overall, designing the experiment with a lower transmon-storage dispersive coupling should decrease the oscillator Kerr non-linearity while retaining the ability to perform conditional displacements with the same speed: the conditional displacement rate scales as $\chi |\alpha|$ while the spurious non-linear terms appear at a given value of $\chi^2 |\alpha|^2$.\\

One can transpose this reasoning to the case of higher order non linearities and decoherence processes of the resonator. In particular, we can place an upper bound on the storage mode pure dephasing rate $\kappa_{\phi} \lesssim 2\pi\times 1$~Hz. Indeed, adding the decoherence operator $\kappa_{\phi}\mathcal{D}[\adag \aop]\rho$ in the Lindblad master equation also results  in spurious terms of order $|\alpha|^2$ when moving to the displaced frame, and simulations indicate that a larger pure dephasing rate would result in a shorter logical qubit coherence time than reported in the experiment. This upper bound justifies our approach for calibrating the storage oscillator energy relaxation time as described in Sec.~\ref{sec:storagelifetime}.

\section*{Ideal GKP codes}


Infinitely extended GKP codes are defined by at least two stabilizers $\SX = \D(a)$ and $\SZ = \D(b)$, where $a$ and $b$ are complex numbers. One can geometrically verify that these displacement operators commute using $\D(-\beta)\D(-\alpha)\D(\beta)\D(\alpha)=e^{i\mathcal{A}}$, where $\mathcal{A}=\mathrm{Im}(\alpha^{\ast}\beta)$ is the oriented surface circled in phase-space when applying sequentially the four displacement operators. Each choice of $a$ and $b$ defines a new code and, therefore, there are an infinite number of possible GKP codes. In the rest of this section, we consider two of these GKP codes.

\subsection*{Square code}
\begin{figure}
\begin{center}\includegraphics[scale=0.8]{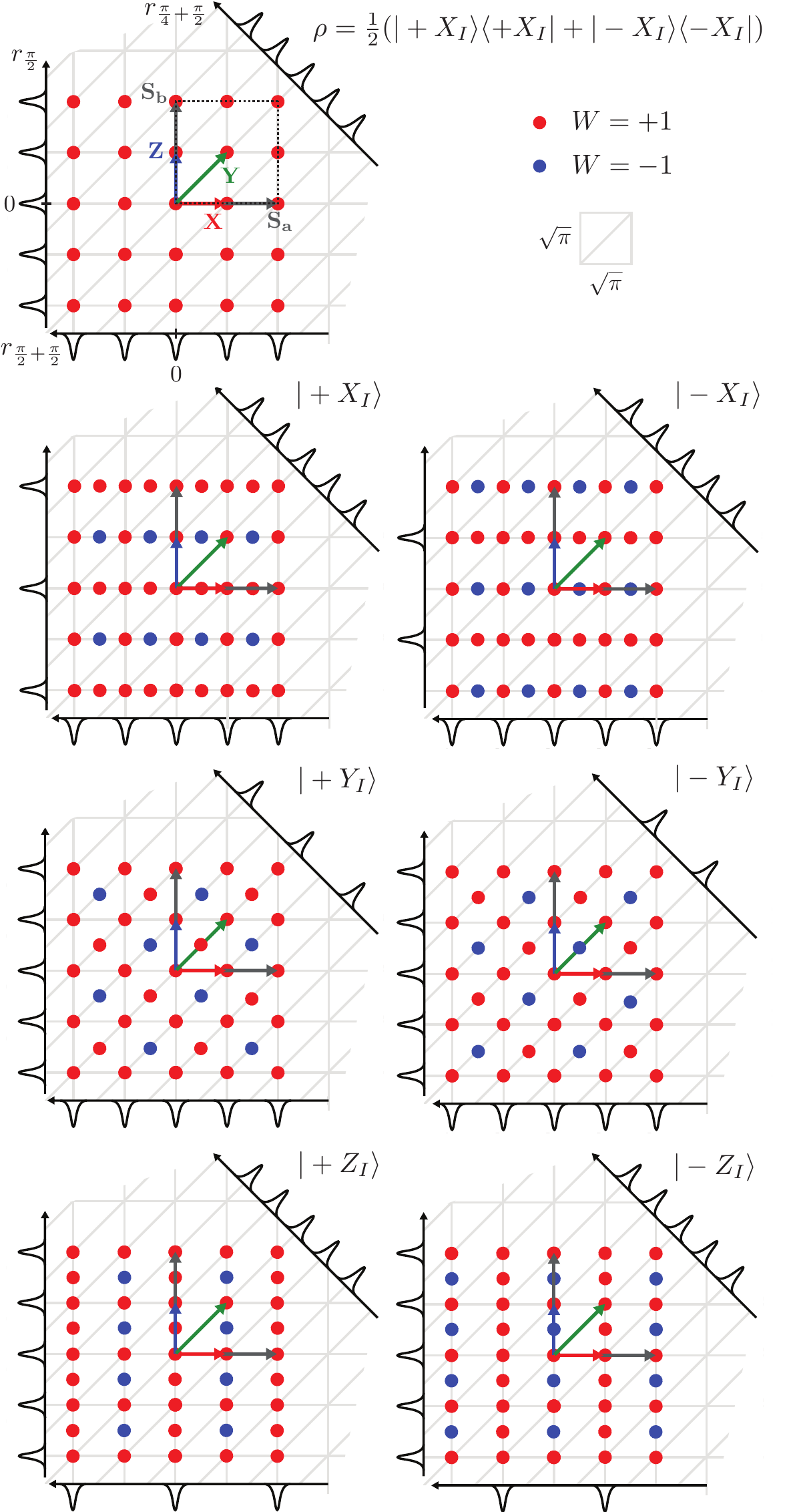}
    \caption{\label{fig:squarewigner} {\bf Logical states in the ideal square code}. Wigner quasi-probability distributions and their marginals along axes orthogonal to $a$, $b$, and $a+b$ are schematically represented for the maximally mixed logical state--revealing the square lattice structure, highlighted with gray lines---and the eigenstates of each Pauli operator. We give quasi-probability peaks (red and blue dots) as well as peaks of the marginals (black lines) a finite extension for better visualization. The distribution extends infinitely in phase-space and is obtained by tessellating the pattern of the unit cell delimited by a  black dashed line (area $4\pi$). 
    }
\end{center}
\end{figure} 

The square GKP code is stabilized by the two operators $\SX=\D(a=2\sqrt{\pi})$ and $\SZ=\D(b=ia)$~\cite{gottesman2001encoding}. In Fig.~\ref{fig:squarewigner}, we represent the Wigner quasi-probability distributions of the eigenstates of each Pauli operator as well as the maximally mixed logical state. The code words are +1 eigenstates of the stabilizers, which can be viewed as displacement operators by $a$ along $q=r_0$ and $p=r_{\pi/2}$, so that all the Wigner functions are $a$-periodic in $r_0$ and $r_{\pi/2}$ (we use the notation  $r_{\theta}$ for the generalized phase-space coordinate along the axis forming an angle $\theta$ with the $q$-axis). Alternatively, one can view the stabilizer $\SX=e^{-i a  {\bf r_{\pi/2}} }$ (respectively $\SZ=e^{-i a  {\bf r_{\pi}} }$) as a function of ${\bf r_{\pi/2}}$ (respectively ${\bf r_{\pi}}$) of modulus 1. Thus, for all the ideal logical states, the marginal probability distributions $P_{\theta}=\int_{r_{\theta+\pi/2}} W(r_{\theta},r_{\theta +\pi/2})\mathrm{d}r_{\theta+\pi/2}$  along the directions of the reciprocal lattice $\theta=\pi/2$ and $\theta=\pi$ (black lines in Fig.~\ref{fig:squarewigner}) have a support limited to the antecedents of +1 by this function. This function is $(2\pi)/a$-periodic and these antecedents verify $r_{\theta}=0~\mathrm{mod}~2\pi/a$.   \\

The logical Pauli operators are $\X=\D(a/2)$, $\Y=\D((a+b)/2)$ and $\Z=\D(b/2)$. They commute with the  stabilizers and verify the Pauli group composition rules inside the code. Following similar arguments as for the stabilizers, we show that the Wigner functions of the states $|\pm X_I \rangle $, $|\pm Y_I \rangle $ and $|\pm Z_I \rangle $ are $a/2$-periodic along $r_{\theta}$  with respectively $\theta=0, \pi/4, \pi/2$, and one peak out of two vanishes for their probability distribution $P_{\theta+\pi/2}$ along the reciprocal axis. In particular, we notice that the probability peaks of $|+ Y_I\rangle$ are $\sqrt{2}$ closer to those of $|- Y_I\rangle$ than for $|\pm X_I\rangle$ or $|\pm Z_I\rangle$, which explains the shorter lifetime of the $\Y$ component of the logical Bloch vector. \\

Finally, let us note that the non-zero values of the Wigner functions are points on a square lattice with cell side twice smaller than that of the lattice defining the actual probability peaks of $P_{\pi/2}$ and $P_{\pi}$ (highlighted with gray lines in Fig.~\ref{fig:squarewigner}).  Thus, by analogy with Schr\"{o}dinger cat states~\cite{walls2007quantum}, we can interpret the non-zero values of the Wigner function on the larger lattice as ``probability blobs", when those shifted by half the lattice period are ``interference fringes".

\subsection*{Hexagonal code}

\begin{figure}
\begin{center}\includegraphics[scale=0.8]{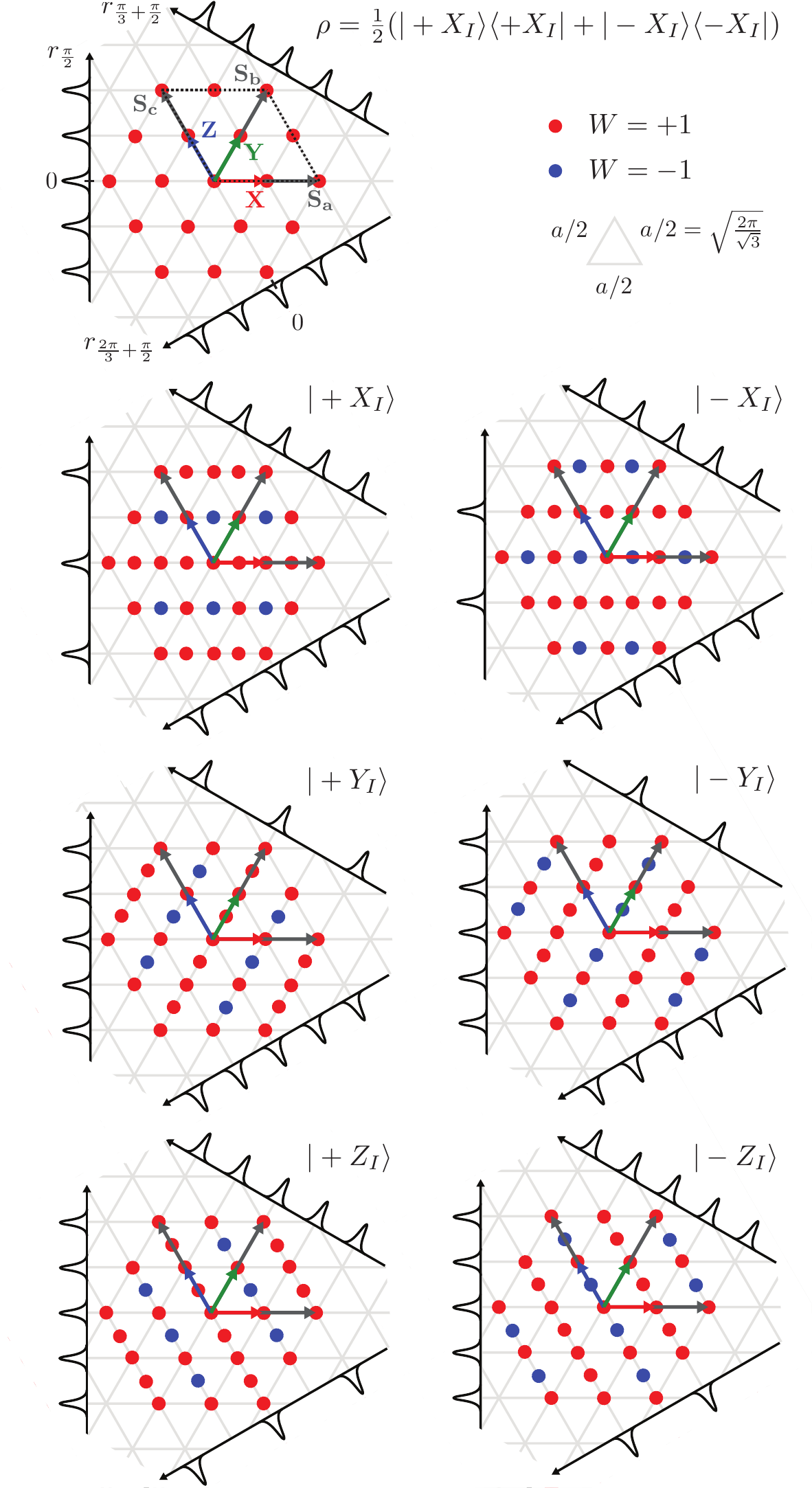}
    \caption{\label{fig:hexwigner} {\bf Logical states in the ideal hexagonal code}. Wigner quasi-probability distributions and their marginals along axes orthogonal to $a$, $b$, and $c$ are schematically represented for the maximally mixed logical state--revealing the hexagonal lattice structure, highlighted with gray lines---and the eigenstates of each Pauli operator. We give quasi-probability peaks (red and blue dots) as well as peaks of the marginals (black lines) a finite extension for better visualization. The distribution extends infinitely in phase-space and is obtained by tessellating the pattern of the unit cell delimited by a  black dashed line (area $4\pi$). 
    }
\end{center}
\end{figure} 

The hexagonal GKP code is stabilized by the two commuting displacement operators $\SXH=\D(a=\sqrt{\frac{8\pi}{\sqrt{3}}})$ and $\SYH=\D(b=ae^{i\frac{\pi}{3}})$~\cite{gottesman2001encoding}. We also consider a third displacement operator, which commutes with the two previous ones, $\SZH=\D(c=ae^{i\frac{2\pi}{3}})$. Any two of these operators can define unambiguously the hexagonal code, which is the intersection of their manifold with eigenvalue +1. In practice, our QEC protocol employs symmetrically measurements of the three stabilizers in order for the peaks of the corrected grid states to verify rotational symmetry. This is needed in order for the decay of the logical qubit Bloch vector to be isotropic (see Sec.~\ref{sec:hexastab}). \\

The logical Pauli operators are $\X=\D(a/2)$, $\Y=\D(b/2)$ and $\Z=\D(c/2)$. The Wigner function of the eigenstates of each Pauli operator are represented in Fig.~\ref{fig:hexwigner}. From similar arguments given for the square code, we find that these functions are all $a$-periodic along $r_0$, $r_{\pi/3}$ and $r_{2\pi/3}$, with an extra $a/2$-periodicity along one of these axes for each Pauli eigenstate. Moreover, on the reciprocal lattice, the marginals $P_{\theta}$ with $\theta=\pi/2$, $\theta=\pi/3+\pi/2$ and $\theta=2\pi/3+\pi/2$ of all states have non-zero values only at $r_{\theta}=2\pi/a$. For the states $|\pm X_I\rangle$, $|\pm Y_I\rangle$ and $|\pm Z_I\rangle$, every other peak vanishes respectively for $\theta=\pi/2, \pi/3+\pi/2, 2\pi/3+\pi/2$.

\section*{Finitely squeezed GKP code}

In this section, we study the properties of the QEC code in Wigner space. In order to avoid aberrations linked to maximum-likelihood reconstruction of the grid states Wigner functions from the measured characteristic functions, we extrapolate the code properties from master-equation simulations reproducing the data presented in the main paper with independently measured parameters. Thus, the figures of merit given in this section should be considered with caution: they characterize realistic grid states  with the same envelope and peak width as measured in our experiment, but may not reflect the unknown experimental imperfections that are not captured by simulations.

\subsection*{Optimal envelope size}
\label{sec:optenv}

In this section, we estimate the optimal envelope size for the GKP code in presence of storage oscillator photon loss only. Following Ref.~\cite{walls2007quantum}, the dynamics of a harmonic oscillator entailed by photon damping at rate $\kappa_s$ is modeled by a Fokker-Planck equation on its Wigner quasi-probability distribution. With the conventions used in the main text,
\begin{equation}
    \label{eq:fokker}
    \frac{\partial W}{\partial t}=\frac{\kappa_s}{2} \Big( \frac{\partial (q W)}{\partial q} + \frac{\partial (p W)}{\partial p} + \frac{1}{2}\frac{\partial^2 W}{\partial q^2} + \frac{1}{2}\frac{\partial^2 W}{\partial p^2} \Big).
\end{equation}
 One can integrate this equation over, say, $p$ to get an equation on the probability distribution $P(q)$
\begin{equation}
\label{eq:FPP}
    \frac{\partial P}{\partial t}=\frac{\kappa_s}{2} \Big( \frac{\partial (q P)}{\partial q} + \frac{1}{2}\frac{\partial^2 P}{\partial q^2}  \Big).
\end{equation}
We then have 
\begin{equation}
\begin{split}
\frac{\mathrm{d} \langle q \rangle }{\mathrm{d}t}&=\int_q q\frac{\partial P}{\partial t} ~~\mathrm{d}q\\
&=\frac{\kappa_s}{2} \int_q q\frac{\partial (q P)}{\partial q} + \frac{1}{2} q\frac{\partial^2  P}{\partial q^2} ~~\mathrm{d}q\\
&=\frac{\kappa_s}{2} (-\int_q q P(q)~\mathrm{d}q + 0)\\
&=-\frac{\kappa_s}{2} \langle q \rangle,
\end{split}
\end{equation}
and 
\begin{equation}
\begin{split}
\label{eq:diff}
\frac{\mathrm{d} \langle q^2 \rangle }{\mathrm{d}t}&=\int_q q^2\frac{\partial P}{\partial t} ~~\mathrm{d}q\\
&=\frac{\kappa_s}{2} \int_q q^2\frac{\partial (q P)}{\partial q} + \frac{1}{2} q^2\frac{\partial^2  P}{\partial q^2} ~~\mathrm{d}q\\
&=\frac{\kappa_s}{2} \int_q -2q^2 P(q) +P(q)  ~~\mathrm{d}q \\
&=\kappa_s ( - \langle q^2 \rangle + \frac{1}{2}).
\end{split}
\end{equation}
Thus, the first two terms in Eq.~(\ref{eq:fokker}) are drifts at speed $-\kappa_s q/2$ (deterministic contraction  of the probability distribution with rate $\kappa_s$), while the last two correspond to diffusion with a constant $\kappa_s/2$ of the probability distribution. In steady-state, they compensate each other and the oscillator is in the vacuum state characterized by $\langle q \rangle = \langle p \rangle=0$ and $\langle q^2 \rangle = \langle p^2 \rangle=\frac{1}{2}$.\\

Both drift and diffusion terms contribute to distorting the grid states and can lead to logical flips after some finite time. Indeed, the evolution of the quasi-probability distribution is continuous and the probability of a logical flip is the fraction of the distribution that has traveled by more than $a/4$ in phase-space. We estimate the optimal envelope width for the GKP code by requiring that the drift and the diffusion of the probability distribution result in the same average traveled distance during the typical time $T$ of QEC (defined more rigorously in the next paragraph), in the limit of short time $\kappa_s T \ll 1$. We consider a square grid state whose envelope width is $\Delta$, and, for clarity, we consider the evolution of the $q$-probability distribution only.  Over the time $T$, an infinitely  squeezed state at position $q$ drifts by a length $d_{\mathrm{drift}}(q)=|q| (1-e^{-\kappa_s T/2 })$ so that the average traveled distance for the whole distribution is
\begin{equation}
\begin{split}
  D_{\mathrm{drift}}=&~\int_q d_{\mathrm{drift}}(q) P(q) \mathrm{d}q\\
  =&~\Delta (1-e^{-\kappa_s T/2 })\\
  \approx&~\Delta \frac{\kappa_s T}{2}.
  \end{split}
\end{equation}
Following Eq.~(\ref{eq:diff}), photon dissipation also leads to a uniform diffusion of probability distribution in phase-space with a constant $\kappa_s/2$. Thus, an infinitely  squeezed state at position $q$ spreads over the time $T$ to a peak with width $d_{\mathrm{diff}}(q)=\sqrt{\frac{\kappa_s T}{2}}$. The average traveled distance for the whole distribution is
\begin{equation}
\begin{split}
  D_{\mathrm{diff}}=&~\int_q d_{\mathrm{diff}}(q) P(q) \mathrm{d}q\\
  =&~\sqrt{\frac{\kappa_s T}{2}}.
  \end{split}
\end{equation}
For the optimal envelope size, we get 
\begin{equation}
    \begin{split}
        D_{\mathrm{drift}}=D_{\mathrm{diff}}&= \sqrt{\frac{\kappa_s T}{2}}\\
        \Delta &= \sqrt{\frac{2}{\kappa_s T}}.
    \end{split}
\end{equation}
These results can be understood with a simple model. We assume that after time-intervals of duration $T$, we perform instantaneous phase-estimation~\cite{kitaev1995quantum, svore2013faster, terhal2016encoding} of the stabilizers to project the oscillator state in a grid state with envelope $\Delta$ and peak width $\sigma=\frac{1}{2\Delta}=\frac{1}{2}\sqrt{\frac{\kappa_s T}{2}}$ (note that in general, the grid state probability distribution can have a peak width $\tilde{\sigma}  >\sigma$ if the oscillator state has been randomly shifted out of the code manifold, whereas  $\Delta$ does not vary). During the next time interval, diffusion and drift of the quasi-probability inflate the peaks by $\sim\sqrt{\frac{\kappa_s T}{2}} =2 \sigma$ so that their width remains of order $\sigma$. Thus, the envelope size is optimal: if one were to change phase-estimation parameters to target grid states with thinner peaks (larger envelope), $D_{\mathrm{drift}}$ would increase and the peaks would re-inflate to a width larger than $\sigma$. On the other hand, if the target states have broader peaks (smaller envelope), orthogonal logical states overlap more than necessary, resulting in an increased logical flip rate. The shorter the QEC time $T$, the thinner the peaks of the target states can be.\\

In the discrete-time Markovian protocol used in the experiment, we can replace $T$ by the characteristic convergence time $\sim 25~\mu$s to reach steady-state when turning on the QEC (see Fig.~2a). We then estimate an optimal envelope size $\Delta \sim 4$. This estimation only provides an order of magnitude as a rigorous treatment should include details of the QEC scheme, higher moments of the probability distribution induced by drifts and other decoherence mechanisms beyond photon loss.  We find quantitatively the optimal envelope width by maximizing the coherence time of the logical qubit in simulations, yielding $\Delta=3.2$ (see Fig.~\ref{fig:assignement}a).

\subsection*{Characteristics of the error-corrected code}
\label{sec:codecharac}
\begin{figure}
\begin{center}\includegraphics[scale=1]{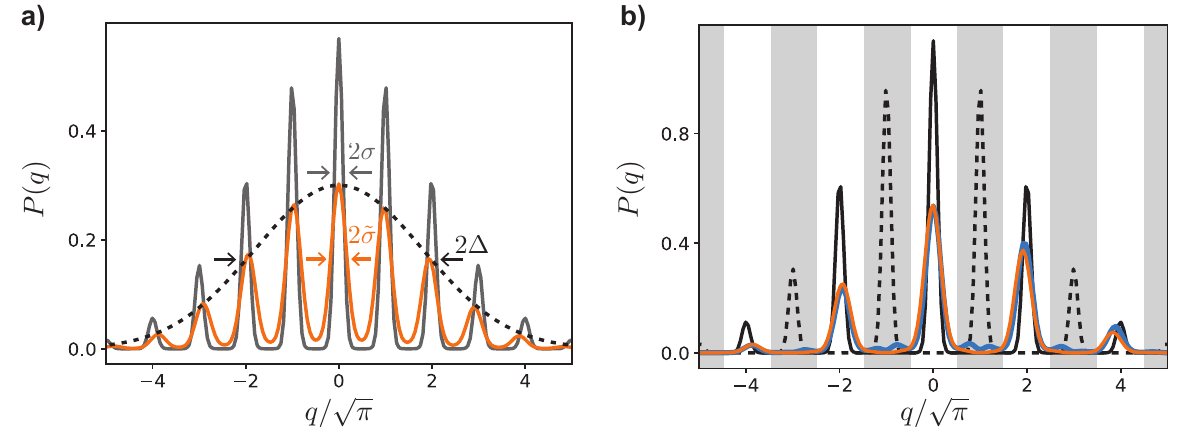}
    \caption{\label{fig:assignement}  {\bf Simulation of the experimentally error-corrected code} {\bf a)} The probability distribution of the steady-state of the QEC protocol (orange line) has an envelope width $\Delta=3.2$, chosen to maximize the logical qubit coherence time. The peaks are broadened to a width $\tilde{\sigma}=0.29 >\sigma=1/2\Delta$ by dissipation compared to the maximally mixed logical state strictly inside the code (gray line). {\bf b)} Probability distribution of the state prepared when aiming for $|+X_L\rangle$. The blue line corresponds to a single $\mathrm{Re}(\X)$ measurement conditioning a feedback $\Z$-gate, and the orange line to the same protocol followed by a second $\mathrm{Re}(\X)$ measurement to herald a higher fidelity state. By integrating the distribution on $[-\sqrt{\pi}/2,\sqrt{\pi}/2]~\mathrm{mod}~2\sqrt{\pi}$ (gray stripes), we estimate that a perfect homodyne measurement along the $q$ quadrature would assign the two prepared states to the target state $|+X_L\rangle$ with respective probability $F_2=0.9, 0.97$. However, the relative distance of these states to $|+X_L\rangle$ (plain black line) with respect to the orthogonal state $|-X_L\rangle$ (dashed black line) defines another preparation fidelity with higher value $F_3=0.976, 0.996$.
    }
\end{center}
\end{figure} 
In this section, we summarize the properties of the experimentally error-corrected GKP code, extracted from master-equation simulations (see Sec.~\ref{sec:MES}). For simplicity sake, we consider the square code only during the remainder of this section. As noted in the main text, the properties of the error-corrected hexagonal code are similar. In particular, the error-corrected grid states are characterized by the same squeezing of the peaks for both encodings  in steady state. \\

In Fig.~\ref{fig:assignement}a, we represent the $q$-probability distribution in steady-state of the QEC protocol (mixed logical state whose characteristic function is represented in Fig.~3b), right before a $q$-peak sharpening round. This is the time when the peaks of the $q$-probability distribution are the widest, and we would find the same distribution on $p$ before a $p$-peak sharpening round. The grid state envelope has a width $\Delta=3.2$, which is chosen to maximize the logical qubit coherence time. As detailed in Sec.~\ref{sec:backaction}, this width is set by the length of the conditional displacements used in envelope trimming rounds. The peaks of the grid state have a standard deviation $\tilde{\sigma}=0.3 >1/2\Delta$, which corresponds to a squeezing of $20~\mathrm{Log}_{10}(\frac{\delta q_{\mathrm{vac}}}{\tilde{\sigma}})=7.4~$dB ($\delta q_{\mathrm{vac}}=1/\sqrt{2}$ is the standard deviation of the vacuum state probability distribution). Due to dissipation acting on the oscillator and other decoherence mechanisms, the peaks are thus broader than those of the actual code states, for which $\sigma=1/2\Delta=0.15$ (13.4~dB squeezing). This is consistent with the oscillator state purity which is smaller than that of the fully mixed logical qubit state 
$\mathrm{Tr}(\rho_{ss}^2)=0.17<0.5=\mathrm{Tr}(\rho_{\mathrm{mix}}^2)$. As mentioned in the main text, the peaks of the distribution are thinner after a $q$-peak sharpening round, with a standard deviation corresponding to a squeezing of 9.5~dB.\\

We can now rigorously define the code states by 
\begin{equation}
    \begin{split}
        \langle q |+ X_L\rangle \propto & \sum_{n\in 2\mathbb{Z}} e^{-\frac{(q-n\sqrt{\pi})^2}{4\sigma^2}-\frac{n^2\pi}{4\Delta^2}}\\
        \langle q |- X_L\rangle \propto & \sum_{n\in 2\mathbb{Z}+1} e^{-\frac{(q-n\sqrt{\pi})^2}{4\sigma^2}-\frac{n^2\pi}{4\Delta^2}}\\
        \langle r |+ Y_L\rangle \propto & \sum_{n\in 2\mathbb{Z}} e^{-\frac{(r-n\sqrt{\pi/2})^2}{4\sigma^2}-\frac{ n^2\pi/2}{4\Delta^2}}\\
        \langle r |- Y_L\rangle \propto & \sum_{n\in 2\mathbb{Z}+1} e^{-\frac{(r-n\sqrt{\pi/2})^2}{4\sigma^2}-\frac{ n^2\pi/2}{4\Delta^2}}\\
        \langle p |+ Z_L\rangle \propto & \sum_{n\in 2\mathbb{Z}} e^{-\frac{(p-n\sqrt{\pi})^2}{4\sigma^2}-\frac{n^2\pi}{4\Delta^2}}\\
        \langle p |- Z_L\rangle \propto & \sum_{n\in 2\mathbb{Z}+1} e^{-\frac{(p-n\sqrt{\pi})^2}{4\sigma^2}-\frac{n^2\pi}{4\Delta^2}}
    \end{split}
\end{equation}
where $|q\rangle$, $|p\rangle$ and $|r\rangle$ respectively represent infinitely squeezed states along $q$, $p$ and the axis $r$ rotated from $q$ at $45^{\circ}$. Note that in these expressions, the width of the wavefunction peaks is $\sqrt{2}\sigma$, larger than that of the probability distribution peaks. We estimate the overlap between two states representing orthogonal logical states to be $|\langle +X_L |-X_L\rangle |^2=|\langle +Z_L |-Z_L\rangle |^2 \ll |\langle +Y_L |-Y_L\rangle |^2 =4.10^{-7}$. This negligible overlap allows us to rigorously define a logical qubit in the finitely squeezed code.

\subsection*{State preparation fidelity}

We now use simulations described in Sec.~\ref{sec:MES} to estimate the preparation fidelity for the eigenstates of each logical Pauli operator (see Sec.~3). In Fig.~\ref{fig:assignement}b, we represent the simulated $q$-probability distribution of the  state generated when aiming for $|+X_L\rangle$. The blue line represents the  state $\rho_1$ obtained with a single-round measurement of $\mathrm{Re}(\X)$ followed by a feedback $\Z$-gate if the outcome is $-1$. Again, note that  $\X$ corresponds to the Pauli operator of the ideal, infinitely squeezed GKP code, so that $\mathrm{Re}(\X)$  is strictly bi-valued only in that case. We reiterate that in the case of the finitely squeezed code, the distribution of the operator $\X$, whose complex eigenvalues lie on the unit circle,  is still sharply peaked near +1 and -1. Thus, extracting one bit of information is sufficient to prepare {\it approximately} $|+X_L\rangle$ (plain black line in Fig.~\ref{fig:assignement}b). As described before, this measurement is performed with a $\sigmax$ readout of the transmon, following its preparation in $|+x\rangle$ and a conditional displacement $\CD(a/2)$. Note that this conditional displacement is followed by a shift to recenter the grid at $q=0~\mathrm{mod}~2\pi/a$ (see Sec.~\ref{sec:paulimeas}). This shift, as well as the feedback $\Z$-gate, displaces the grid state envelope, which translates into an asymmetry of the $q$-probability distribution, or equivalently a non-zero value of the characteristic function imaginary part (not shown). Subsequent envelope-trimming steps corral the envelope back to the center of phase-space, restoring the symmetry of the grid.\\

We can give a first definition of the preparation fidelity as the expectation value of the real part of the ideal code Pauli operator, re-scaled to belong to [0,1]: $F_1=(\langle \mathrm{Re}(\X) \rangle +1)/2=0.86$. This definition is relevant as it corresponds to the result of a subsequent logical qubit readout along $\X$ as performed in the experiment (neglecting dissipation during the conditional displacement and transmon errors). However, it does not reflect fairly the preparation fidelity to the finitely squeezed target state $|+X_L\rangle$. Indeed, this fidelity would be smaller than one even when perfectly preparing the target state since the logical states that we consider here are not eigenstates of $\X$ and thus  $\langle+X_L |\mathrm{Re}(\X)|+X_L\rangle<1$. Similar arguments show that a protocol based on a single-round measurement of $\mathrm{Re}(\X)$ and feedback cannot perfectly prepare $|+X_L\rangle$. \\

A second definition of the preparation fidelity is given by integrating the $q$-probability distribution of the generated state over regions $[-\sqrt{\pi}/2,+\sqrt{\pi}/2]+2n\sqrt{\pi}$ (gray stripes in Fig.~\ref{fig:assignement}b). We thus find the fidelity $F_2=0.90$. It corresponds to the probability that an ideal homodyne detection along the $q$ quadrature, which collapses the grid state to an infinitely squeezed state, assigns the generated state to the target state $|+X_L\rangle$.\\

We can go one step further and consider the best possible guess in assigning the generated state to $|+X_L\rangle$ or $|-X_L\rangle$ from the full information of the simulated density matrix. This third definition yields a higher fidelity to the target state $F_3=\frac{\mathrm{Tr}(\rho_1~|+X_L\rangle \langle +X_L|)}{\mathrm{Tr}(\rho_1~|+X_L\rangle \langle +X_L|)+\mathrm{Tr}(\rho_1~|-X_L\rangle \langle -X_L|)}=0.976$. It is the value that one would get by performing a perfect projection of the state on the code manifold before measuring the bi-valued operator $(|+X_L\rangle \langle +X_L|+1)/2$. Note that such an operation is for the moment hypothetical as it would require to measure the actual stabilizer of the finite width code, keeping in post-selection only the states belonging to the code manifold, 
before performing a single-shot measurement of the finite code Pauli operator $|+X_L\rangle\langle +X_L|-|-X_L\rangle\langle -X_L|$. This operator is rigorously defined since $|+X_L\rangle$ and $|-X_L\rangle$ are orthogonal up to a very good approximation (see Sec.~\ref{sec:codecharac}), and it should also be possible to construct the stabilizers of the finite code, but no measurement protocol for these operators exists as of yet. \\

In the experiment, we boost the preparation fidelity by performing a second $\mathrm{Re}(\X)$ measurement and post-selecting the cases when the outcome is +1. We reproduce the generated state $\rho_2$ in simulations and, using the definitions given above, we find the preparation fidelities $F_1=0.90$, $F_2=0.97$ and $F_3=0.998$.  The reason for $F_2<F_3$ is that the peaks of the generated state when targetting $|+X_L\rangle$ (orange line in Fig.~\ref{fig:assignement}b) are broadened enough by dissipation for them to non-negligibly overlap with the regions colored in gray. Thus, a single-shot detection of $q$ can assign the state to $|-X_L\rangle$ with non-negligible probability $1-F_2$. However, the peak are not broad enough to significantly overlap with the actual pure logical state $|-X_L\rangle$ of the code (dashed black line in Fig.~\ref{fig:assignement}b), so that the thought-experiment described above, which includes post-selection, would assign the state to $|+X_L\rangle$ with higher probability $F_3$. Interestingly, one could similarly obtain a higher preparation fidelity $F_{2'}\simeq F_3$ using an homodyne detection of $q$ at the expense of post-selecting out the most ambiguous results when $q~\mathrm{mod}~\sqrt{\pi}\simeq \sqrt{\pi}/2 $ (obviously, this strategy cannot improve the {\it readout fidelity} of the logical qubit for this $q$-detection based scheme). Although $1-F_3 = 0.2~\%$, this metric is irrelevant for assessing quantum computational resources.  However, it still gives us the indication that the state preparation fidelity is limited by the broadening of the grid states peaks only. The fact that $F_3$ is close to unity indicates that preparation cannot further improved by repeating again the $\mathrm{Re}(\X)$ measurement. \\

Finally, it is important to note that the decay of the components of the logical qubit Bloch vector, as presented in Figs.~3-4, is independent of the used definition.

\section*{Details of the QEC protocol}

\subsection*{Discrete-time Markovian feedback QEC in the square code}
\label{sec:backaction}
We now write explicitly the Kraus operators acting on the oscillator state when measuring the transmon on a round of QEC. Each round starts with a preparation of the transmon in $|+x\rangle$ (see Fig.~2b). We re-write the following conditional displacement as $\CD(\beta)=e^{i\betap \frac{\sigmaz}{2}}$, where we have defined $\betap = -\mathrm{Re}(\beta) \pop + \mathrm{Im}(\beta) \qop$. Finally, the transmon is measured along $\sigmay$. If we neglect decoherence during the sequence and the short unconditional displacement described in the previous section, the Kraus operators acting on the storage mode associated with a transmon detection in $|\pm y\rangle$ is
\begin{equation}
\begin{split}
\label{eq:kraus}
\M_{\pm}=&\langle \pm y | e^{i\betap \frac{\sigmaz}{2}} |+x\rangle  \\
=& \langle +x | e^{i(\betap \pm \frac{\pi}{2} ) \frac{\sigmaz}{2}}  |+x\rangle\\
=& \cos \big( (\betap \pm \frac{\pi}{2})/2 \big).
\end{split}
\end{equation}

\begin{figure}
\begin{center}\includegraphics[scale=0.8]{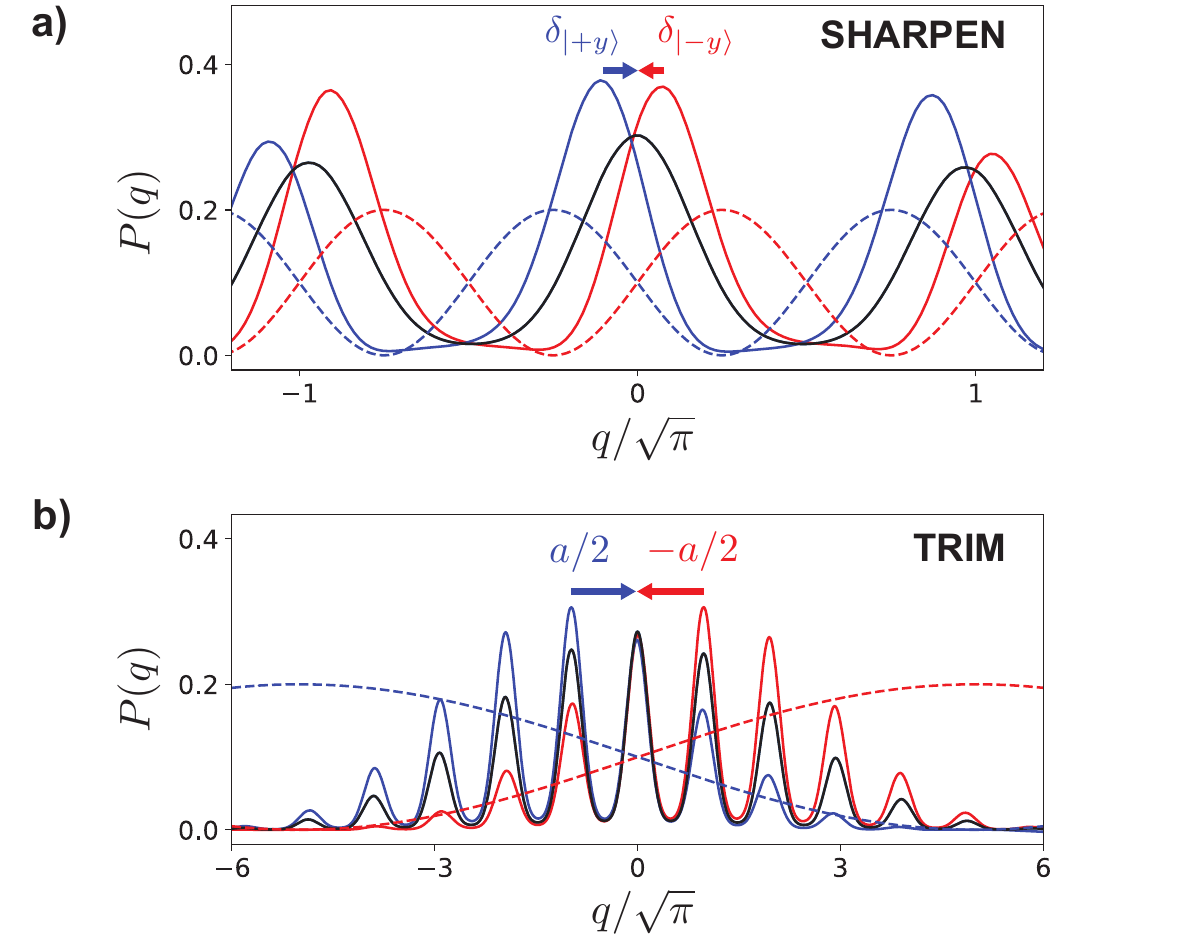}
    \caption{\label{fig:kraus} {\bf Simulated backaction of the transmon readout on the oscillator probability distribution.} {\bf a)} During a $q$-peak sharpening round occurring in steady-state (logical qubit in a fully mixed state), the backaction of the measurement of the transmon in $|\pm y\rangle$  multiplies the initial $q$-probability distribution (black line) by $\cos^2 \big((q a\pm \frac{\pi}{2})/2 \big)$ (blue and red dashed lines, scaled down for convenient representation). The  peaks thus get sharper and offset (blue and red plain lines). When measuring the transmon in $|+ y\rangle$ or $|- y\rangle$, this offset is  asymmetric with respect to the origin due to a displacement term appearing in the oscillator state evolution during the sharpening round (see Sec.~\ref{sec:condis}). Thus, the feedback shifts (arrows) re-centering the grid back towards $q=0~\mathrm{mod}~2\pi/a$ are asymmetric and read $\delta_{|\pm y\rangle}= \pm 0.2+0.04$. {\bf b)} During a $q$-envelope trimming step, the initial distribution is multiplied by $\cos^2 \big(q (\epsilon \pm \frac{\pi}{2})/2 \big)$ with $\epsilon\approx a/20$ (same color code). The probability distribution is partly collapsed towards positive or negative values and a feedback displacement by $a/2$ shifts it back towards the origin without changing the code stabilizers value.
    }
\end{center}
\end{figure}

In Fig.~\ref{fig:kraus}, we represent the backaction of the transmon readout, including the effect of decoherence during the transmon-oscillator interaction, found using the master-equation simulations described in Sec.~S2. After a large number of QEC rounds (steady-state of Fig.~3), we consider a  $q$-peak sharpening round of the square code for which $\betap=a\qop$ (top panel). The black line represents the $q$-probability distribution $P_0(q)=\int_p W_0(q,p)\mathrm{d}p$ for the oscillator state $\rho_0$ right before performing the sharpening round. The plain red and blue curves represent the probability distribution $P_{\pm}(q)$ of the state $\rho_{\pm}$ after performing the conditional displacement and measuring the transmon in $|\pm y\rangle$ (neglecting decoherence during the interaction and the small unconditional displacement described in the previous section, $\rho_{\pm}= \frac{M_{\pm} \rho_0 M^{\dag}_{\pm}}{\mathrm{Tr}(M_{\pm} \rho_0 M^{\dag}_{\pm})}$). The dashed lines represent $\cos^2 \big((q a\pm \frac{\pi}{2})/2 \big)$ (scaled down for convenient representation). Qualitatively, we verify that $P_{\pm}(q)\propto P_{0}(q) \cos^2 \big((q a\pm \frac{\pi}{2})/2 \big)$. The measurement backaction thus partly collapses the peaks of the $q$-probability distribution, which get sharpened. Their center of mass are offset, which is corrected for by a small feedback displacement (red and blue arrows). The peaks  also become skewed, but quickly retrieve a Gaussian shape under the effect of dissipation.\\

One also needs to consider the measurement backaction on the $p$-probability distribution (not represented). The Kraus operators of Eq.~(\ref{eq:kraus}) can be re-written as
\begin{equation}
\label{eq:kraus2}
\M_{\pm}=\frac{1}{2} \Big( e^{\pm i\frac{\pi}{4}} \D(\beta/2) + e^{\mp i\frac{\pi}{4}} \D(-\beta/2)  \Big),
\end{equation}
with $\beta=b=ia$ for a $q$-peak sharpening round. Along $p$, the probability distribution expands with the generation of a new outward peak. It also gets shifted by $\frac{a}{2}$~modulo~$a$, which deterministically flips the logical qubit ($\Z$-gate for the $q$-peak sharpening rounds, $\X$-gate for the $p$-peak sharpening rounds). Finally, interference between the two displaced versions of the state with a relative $\frac{\pi}{2}$-phase distorts the $p$-probability distribution envelope, which can be seen as a consequence of the skewness acquired by the $q$-peaks by virtue of the Fourier transform properties.\\

To mitigate this undesired expansion of the envelope, we alternate rounds dedicated to sharpening the peaks with rounds dedicated to trimming the envelope. In this case, we use a shorter conditional displacement $\CD(\epsilon)$, with $|\epsilon| \approx a/20$. In Fig.~\ref{fig:kraus}b, we represent the backaction of the transmon $\sigmay$ readout at the end of a $q$-envelope trimming step for which $\betap=\epsilon \qop$. Similarly to the peak sharpening step, this backaction can be understood as a multiplication of $P(q)$ by $\cos^2 \big( (q\epsilon \pm \frac{\pi}{2})/2 \big)$, which partly collapses the distribution towards positive or negative $q$-values. A following feedback displacement (blue and red arrows) shifts the whole distribution back towards the origin. These displacements by $a/2$ are large enough to compensate for the envelope expansion induced by peak-sharpening rounds. They commute with the code stabilizers and thus do not modify their values. The logical qubit is deterministically flipped (the feedback displacement performs a $\X$-gate for the $q$-envelope trimming rounds, and a $\Z$-gate for the $p$-envelope trimming rounds).\\

The value of $\epsilon$ sets the ``strength" of the measurement. When $\epsilon \rightarrow 0$, the multiplying cosine has a period much larger than the represented $q$-probability distribution. The backaction would then only marginally modify the state, and the envelope would keep expanding during peak-sharpening rounds until it reaches a steady-state with a much larger envelope. In the opposite limit, a larger $\epsilon$ value would result in a stronger collapse of the probability distribution and a smaller envelope in steady-state.\\ 

\subsection*{Optimization of the feedback displacements}
\label{sec:kickbacks}
\begin{figure}
\begin{center}\includegraphics[scale=1.2]{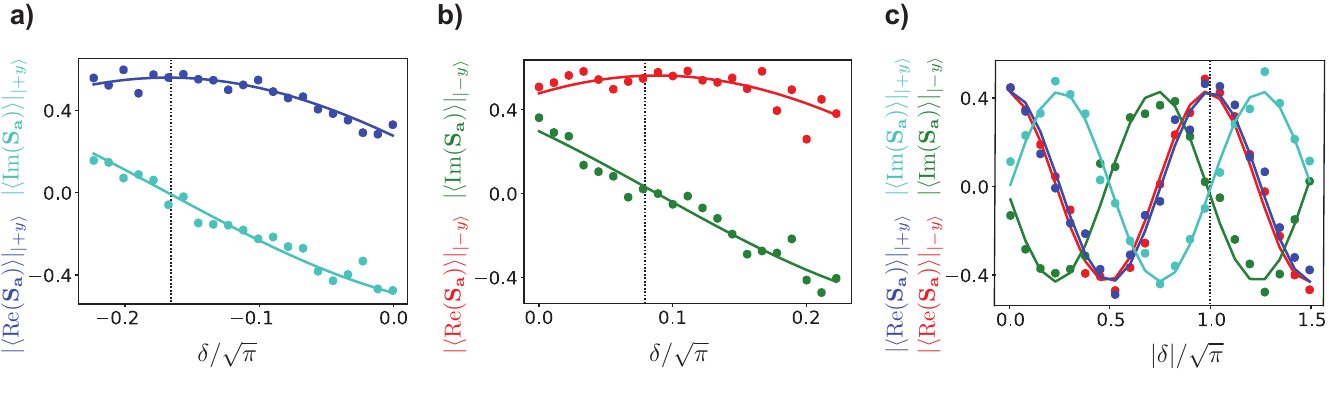}
    \caption{\label{fig:kickcalib} {\bf Tuning of the feedback displacements.} {\bf a)} After reaching the steady-state of the QEC protocol, we perform a last $p$-peak sharpening round for which we vary the length $\delta$ of the feedback shift when the transmon is measured in $|+y\rangle$, and record the stabilizer expectation value $\langle \SX \rangle_{|+y\rangle} $ conditioned on this same outcome (dots). The phase of $\langle \SX \rangle_{|+y\rangle}$ oscillates with a period $2\pi/a$ (lines represent a {\it sine} and {\it cosine} function with this period), and the optimal feedback shift (vertical dashed line) cancels its imaginary part. The grid state is then centered in $p~\mathrm{mod}~2\pi/a=0$. {\bf b} Idem when measuring the transmon in $|-y\rangle$. {\bf c} Same measurement after a $p$-envelope trimming round. The optimal feedback shifts length is $a/2= 2\pi/a$.
    }
\end{center}
\end{figure}

The optimal length of the feedback displacements applied at the end of each round is first estimated in simulations, and then finely adjusted by maximizing the real part of the stabilizers expectation value in steady-state. \\

The general procedure is the following. Starting from an initial guess for all the feedback shifts, we apply our QEC protocol for a large number of rounds in order to reach steady-state. For the last round only, we vary independently the length of the shift applied when the transmon is detected in $|+y\rangle$ or $|-y\rangle$. Conditioned on this last measurement outcome, we then  measure the expectation value of the stabilizers (this measurement is performed by resetting the transmon and averaging the result of a subsequent measurement of $\SX$ or $\SZ$ as described in Sec.~1). The conditional expectation values  $\langle \SX \rangle _{|\pm y\rangle}$ when tuning the feedback shift at the end of a $p$-peak sharpening round are represented in Fig.~\ref{fig:kickcalib}a-b. Varying the length $\delta$ of a shift displaces the generated grid state along $p$, which rotates the phase of $\langle \SX \rangle _{|\pm y\rangle}$ by an angle $a \delta $. We fit the recorded real and imaginary parts (circles) with a {\it cosine} and a a {\it sine} function (lines)  with period $2\pi/a$. The phase-offset of these oscillations directly indicates the optimal feedback shift (vertical dashed line) following a transmon measurement in $|+y\rangle$ or $|-y\rangle$: after applying these shifts, $\langle \SX \rangle$ becomes real and the grid state is centered in $p~\mathrm{mod}~2\pi/a=0$. Note that the shifts corresponding to the $|\pm y\rangle$ outcomes are asymmetric, which is expected (see Sec.~\ref{sec:condis}). \\

We perform the same tuning for the feedback shifts following a $p$-envelope trimming round (see Fig.~\ref{fig:kickcalib}c). Here, as expected, the optimal shifts are symmetric and their length is close to $a/2$ (enforcing that they are strictly $a/2$ can be seen as a more precise calibration of the displacements length compared to the method described in Sec.~\ref{sec:discalib}). We also tune the shifts following $q$-peak sharpening and $q$-envelope trimming rounds  (nullifying $\langle \SZ \rangle _{|\pm y\rangle}$, not shown). Since the steady-state grid used as an input to this optimization uses an imprecise initial guess for the feedback shifts and can thus be offset, we then iterate the whole protocol. The shifts are adjusted one final time to maximize the Pauli operators lifetime, and we obtain the optimal shift value $\delta_q=+0.06 \pm 0.2$ after a $q$-peak sharpening round and $\delta_p=-0.06 \mp 0.2$ after a $p$-peak sharpening round, in good agreement with our model for conditional displacements (see Sec.~\ref{sec:condis}). Note that for simplicity, we do not mention the shifts asymmetry in the main text.

\subsection*{Measurement of the logical Pauli operators in the square code}
\label{sec:paulimeas}

Measurement of $\mathrm{Re}(\X)$ (respectively $\mathrm{Re}(\Y)$ and $\mathrm{Re}(\Z)$) is performed by replacing a $q$-peak sharpening step with a conditional displacement $\CD(\beta)$ with $\beta=a/2$  (respectively $(a+b)/2, b/2$) followed by a $\sigmax$ readout of the transmon. The Kraus operators corresponding to the two measurement outcomes are
\begin{equation}
\begin{split}
\label{eq:kraus3}
\N_{+}=& \cos  (\betap /2 )=\frac{1}{2} \Big(  \D(\beta/2) +  \D(-\beta/2)  \Big)\\
\N_{-}=& \sin  (\betap /2 )=\frac{1}{2i} \Big(  \D(\beta/2) -  \D(-\beta/2)  \Big)
\end{split}
\end{equation}
where $\betap=-\frac{a}{2}\pop$ ~(respectively $\betap=\frac{a}{2}(\qop-\pop)$,~ $\betap=\frac{a}{2}\qop$). The measurement backaction thus multiplies the $\beta_{\perp}$-probability distribution by $\frac{1 \pm \cos \beta_{\perp}}{2}$, which damps one peak of the distribution out of two. On the conjugate variable, it displaces the state by $\pm \beta/2$, which reverses the sign of one or both of the stabilizers (shifts the grid state by half the lattice period). We flip back the stabilizer sign by applying an unconditional displacement $\D(-\beta/2)$ at the end of sequence.\\

\subsection*{QEC and measurement of the Pauli operators in the hexagonal code}~
\label{sec:hexastab}
\begin{figure}
\begin{center}\includegraphics[scale=1.3]{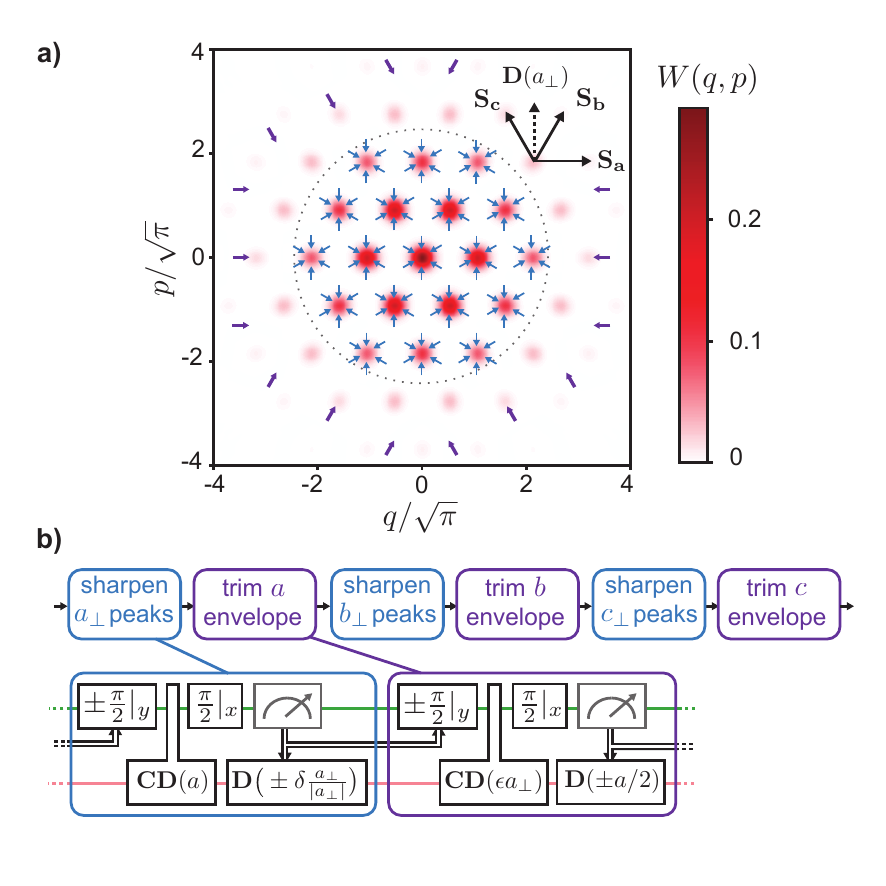}
    \caption{ \label{fig:hexa}{\bf Hexagonal code QEC.} {\bf a)} Simulated Wigner function of the logical mixed state for an hexagonal code with similar envelope width as in the experiment. The Markovian feedback stabilization generates position and momentum dependent shifts from 6 directions: displacements along $a_{\perp}$, $b_{\perp}$ and $c_{\perp}$ prevent the grid state peaks from spreading (blue arrows), while displacements along $a$, $b$ and $c$ prevent the envelope from expanding (purple arrows). {\bf b)} Repeated QEC sequence. Peak-sharpening steps consist of conditional displacements with same direction and length as the displacements stabilizing the code, followed by a transmon measurement triggering short feedback displacements on the orthogonal direction. An envelope-trimming step consists of a short conditional displacement orthogonal to the axes of the stabilizers, followed by a transmon readout triggering a long feedback displacement (half a stabilizer length).  }
\end{center}
\end{figure} 
We recall that the hexagonal code words are approximate eigenstates with eigenvalue 1 of the stabilizers $\SXH=\D(a=\sqrt{\frac{8\pi}{\sqrt{3}}})$, $\SYH=\D(b=ae^{i\frac{\pi}{3}})$ and $\SZH=\D(c=ae^{2i\frac{\pi}{3}})$. These stabilizers are redundant for the infinitely squeezed code as any two of them can define it unambiguously, but as explained below, our QEC protocol based on measurement of two stabilizers only would result in grid states with peaks without rotational symmetry. Then, the lifetime of the three Pauli operators $\X=\D(a/2)$, $\Y=\D(b/2)$ and $\Z=\D(c/2)$ would not be equal.\\

The discrete-time Markovian feedback protocol is directly adapted from the square code case, and is schematized in Fig.~\ref{fig:hexa}. The peaks of the grid state quasi-probability distribution are sharpened from the three directions $a_{\perp}=ae^{i\frac{\pi}{2}}$, $b_{\perp}e^{i\frac{\pi}{2}}$ and $c_{\perp}e^{i\frac{\pi}{2}}$ orthogonal to the stabilizers displacement. Indeed, as detailed in Sec.~\ref{sec:backaction}, the peak-sharpening step corresponding to the measurement of ${\bf{S_{\beta}}}$ ($\beta=a, b, c$) is based on the conditional displacement
$\CD(\beta)=e^{i\betap \frac{\sigmaz}{2}}$, with $\betap = -\mathrm{Re}(\beta) \pop + \mathrm{Im}(\beta) \qop$. The effect of this entangling gate is to map the oscillator state probability distribution $P(\beta_{\perp})=\int_{\beta} W(\beta, \beta_{\perp})\mathrm{d}\beta$ onto the statistics of the transmon Bloch vector longitudinal angle. The subsequent transmon $\sigmay$ measurement backaction multiplies $P(\beta_{\perp})$ by an oscillating function which sharpens its peaks by partly collapsing them. A feedback shift along $\beta_{\perp}$ with length $\pm \delta =\pm 0.2$ recenters the peaks at $\beta_{\perp}=0~\mathrm{mod}~ \frac{2\pi}{k}$. Overall, the repeated feedback shifts generate an effective dissipation preventing the peaks from spreading (blue arrows in Fig.~\ref{fig:hexa}a). Note that the conditional displacement $\CD(a), \CD(b), \CD(c)$ applied during each QEC round flips the logical qubit in a deterministic way ($\X$, $\Y$, $\Z$-gate).  \\

As represented in Fig.~\ref{fig:hexa}, the repetition of peak sharpening steps generates position and momentum dependent kicks in phase-space along $a_{\perp}$, $b_{\perp}$ and $c_{\perp}$. It is here clear that the same QEC protocol based on two stabilizers measurement only would result in ellipse-shaped peaks. For instance, not sharpening the peaks along $a_{\perp}=p$ would result in peaks more elongated along this direction than along the orthogonal direction $q$.\\

In practice, to trim the envelope of the grid state, we perform short conditional displacements along $a_{\perp}$, $b_{\perp}$, $c_{\perp}$ followed by transmon readout along $\sigmay$ and  feedback displacements respectively by $\pm a/2$, $\pm b/2$, $\pm c/2$. These trimming directions are chosen so that feedback displacements do not modify the stabilizers value, even if they flip the logical qubit in a deterministic way (applying a $\X$, $\Y$ and $\Z$ gate respectively). As in the square code case, the feedback displacements could be made longer by an integer factor with no added complexity, but simulations indicate that these choices would result in a larger envelope in steady-state and a shorter logical qubit coherence time.\\

Finally, let us note that similarly to the initialization of the square code logical qubit described in Sec.~\ref{sec:paulimeas}, measurement of $\mathrm{Re}(\X)$, $\mathrm{Re}(\Y)$ or $\mathrm{Re}(\Z)$ is done by modifying a peak-sharpening round of the corresponding stabilizer $\SXH$, $\SYH$ or $\SZH$: we set the conditional displacement to be twice shorter as $\CD(a/2)$, $\CD(b/2)$ and $\CD(c/2)$, and the transmon is readout along $\sigmax$. An additional unconditional displacement $\D(-a/2)$, $\D(-b/2)$, $\D(-c/2)$ is applied to recenter the grid state, which is otherwise shifted by half a stabilizer period by the conditional displacement.

\section*{Transmon errors and feedback performances}

In this section we estimate the convergence rate towards the code manifold under our QEC protocol and discuss the impact of transmon errors on our scheme. For simplicity, we reason in the framework of the square code, but the qualitative results which are given also apply for the hexagonal code, unless otherwise noted. 

\subsection*{Readout errors and phase-flips of the transmon}
\label{sec:phaseflips}
The interaction Hamiltonian used to engineer the conditional displacements acts on the transmon via the $\sigmaz$ operator only (see Eq.~\ref{eq:H_tilde}). Phase-flips of the transmon ($\sigmaz$-gates applied at random times) thus commute with the interaction Hamiltonian: they have the same effect on the system that they occur before, during, or after the interaction. In our Markovian feedback protocol, transmon phase-flips are thus equivalent to transmon readout errors and simply lead to a feedback displacement applied in the wrong direction. We prove in this section that the convergence rate toward the stabilized manifold decreases only linearly with the probability of error: if this rate remains much larger than the dissipation rate $\kappa_s$, the QEC performance is only marginally affected.\\

To prove this, let us suppose that there exists in the oscillator a highly squeezed GKP state (peak width $\sigma \ll a$). We define $x = \Big((q+\pi/a)~\mathrm{mod}~2\pi/a \Big)-\pi/a$. The distribution $P(x)$ is a centered Gaussian with width $\sigma$. We note that the $p$-peak sharpening rounds and the envelope trimming rounds do not modify $P(x)$ (in the limit of large envelope). Moreover, during the $q$-sharpening rounds, the measurement backaction (multiplication by $\cos ( \frac{a}{2}x \pm \frac{\pi}{4})$ according to Eq.~(\ref{eq:kraus})) does not modify $x$ in the limit of large squeezing. $x$ can thus be understood as encoding the position of a classical particle performing a 1D random walk. In absence of transmon phase-flips, its position is shifted each $q$-sharpening round by $\pm \delta$, with respective probability 
\begin{equation}
\begin{split}
    P^0_{\pm}(x)&=\cos^2(\frac{a}{2}x \pm \frac{\pi}{4})\\
    &\approx \frac{1}{2}(1 \mp 2ax)),
\end{split}
\end{equation}
where we have used that $x \ll a$. Now considering phase-flips of the transmon or readout errors taking place with probability $\epsilon$ each round, we get the new probabilities for the left and right jumps of the particle
\begin{equation}
    P_{\pm}(x)= \frac{1}{2}(1 \mp (1-2\epsilon a x).
\end{equation}
The continuous version of this random walk is a process corresponding to a Fokker-Planck equation on the distribution $P(x)$
\begin{equation}
    \label{eq:FP2}
    \frac{\partial P}{\partial t}=\Gamma  \frac{\partial (x P)}{\partial x} + D\frac{\partial^2 P}{\partial x^2}  
\end{equation}
with $\Gamma=\delta (1-2\epsilon)a/\tau$ and $D=\delta^2/\tau$ (for $x \ll a$). Here, $\tau=4T_{\mathrm{round}}$ is the time between two $q$-peak sharpening rounds. Similarly to the case of photon dissipation (see Eq.~(\ref{eq:FPP})), the distribution $P(x)$ converges with rate $\Gamma$ toward a Gaussian centered in 0 and with variance $\sigma^2_{\infty}=\frac{D}{\Gamma}=\frac{\delta}{(1-2\epsilon)a}$. Thus, in the limit $\delta \rightarrow 0$, one can achieve infinite squeezing of the GKP state despite an arbitrary large phase-flip rate or an arbitrary low transmon readout fidelity (as long as this readout does provide some information so that $\epsilon>0.5$). \\

Let us  comment on this result. 
\begin{itemize}
    \item This short feedback displacements strategy is qualitatively equivalent to performing phase-estimation~\cite{kitaev1995quantum,svore2013faster, terhal2016encoding} of the stabilizers based on a great number of redundant transmon measurements before applying a feedback displacement: the repeated transmon measurements mitigate any infidelity of the readout and allow one to achieve arbitrary precision on the phase-estimation.
    \item The convergence rate $\Gamma$ toward the code manifold is proportional to the transmon readout contrast $1-2\epsilon$. It is thus simply linear in the information extraction rate from the system.
    \item $\Gamma \propto \delta$ so that by decreasing $\delta$ to reduce $\sigma_{\infty}$, one sacrifices on the stabilization rate. When considering photon loss at rate $\kappa_s$ in Eq.~(\ref{eq:FP2}), one gets the steady-state peak width $\sigma \approx \sqrt{\frac{D+\kappa_s/2}{\Gamma}}$. Here, we have used an effective diffusion coefficient $\kappa_s/2$ for the photon dissipation, which assumes an optimal choice of envelope size (see Sec.~\ref{sec:optenv}). For a given stabilization time $T_{\mathrm{round}}$, the optimal value of $\delta$ is thus the result of a trade-off: long displacements are desired to counteract dissipation-induced diffusion, but need to be shortened when the transmon readout is inaccurate.
    \item Following Eq.~(\ref{eq:kraus2}), the envelope trimming rounds (respectively in $q$, $p$) can be seen as conditional displacements by $\epsilon$, and also contribute to broadening the peaks (respectively in $p$, $q$).  
\end{itemize}

In our experiment, the transmon pure dephasing rate is $\Gamma_{\phi}=(1/T_2-1/2T_1)/2=(140~\mu\mathrm{s})^{-1}$ and readout errors are negligible (readout fidelity above 99.5~\%). As can be seen from Table.~2, the impact of transmon phase-flips on the error-correction performances is negligible: for our experimental parameters, simulations indicate that they only  account for $\sim 1.5~\%$ of the logical errors.

\subsection*{Bit-flips of the transmon}

\label{sec:bitflips}
Contrary to phase-flips, bit-flips of the transmon (random $\sigmax$-gate) do not commute with the interaction Hamiltonian and can thus perturb the oscillator state. Let us for now set aside the details of our protocol for performing  conditional displacements, and assume that we have a controllable interaction Hamiltonian ${\bf H}=g(t) {\bf r}   \sigmaz$ (with ${\bf r}=\qop$ or $\pop$). We note $T_{\mathrm{int}}$ the duration of the transmon oscillator interaction ($T_{\mathrm{int}}=1.1~\mu\mathrm{s}$ in the experiment, see Fig.~2b).  For the peak sharpening rounds, we have $\int_{t=0}^{T_{\mathrm{int}}} s(t)g(t)\mathrm{d}t = a/2$, where $s(t)=\pm 1$ accounts for the transmon echo pulse at $T_{\mathrm{int}}/2$. During these rounds, the oscillator state is conditionally displaced by $\pm a/2$. As mentioned in Sec.~\ref{sec:backaction}, the logical qubit is thus deterministically flipped (the conditional displacement by $\pm a/2$ performs an unconditional $\X$-gate when sharpening the peaks along $p$, and the conditional displacement by $\pm b =\pm ia$ a  $\Z$-gate when sharpening along $q$). If a transmon bit-flip occurs during the interaction, $s(t)$ changes sign at a random time and part of the evolution is canceled. The oscillator state is then displaced out of the code by $\pm d$ with $d<a/2$. Subsequent error-correction rounds recenter the grid state in the code manifold, but  if $d<a/4$, a logical flip has occurred (the expected $\X$ or $\Z$-gate has not been applied). If on the other hand $d>a/4$ or if the transmon bit-flip happens after the interaction with the oscillator, no logical flip occurs (the following feedback displacement can still be in the wrong direction similarly to what happens after a transmon phase-flip).\\

Therefore, transmon bit-flips induce $\X$ (respectively $\Z$, $\Y$) logical errors only if they take place during a $p$-peak sharpening round  (respectively a $q$-peak sharpening round, a $p$-peak or $q$-peak sharpening round) at a time $t$ with
$\frac{T_{\mathrm{int}}}{4}\lesssim t\lesssim \frac{3T_{\mathrm{int}}}{4}$ (see Fig.~2b, the inequality would be exact if $|g(t)|$ were constant). 
Moreover, in the experiment, the transmon decays to a cold bath so that bit-flips originate from an amplitude damping channel at rate $1/T_1$. Regular echoes in the error-correction sequence (see Fig.~2b) tend to symmetrize this channel, but the average occupation of the excited level is only 0.5, so that bit-flips happen at rate $1/2T_1$. We can then estimate the rates of logical flips induced by transmon bit-flips as $\Gamma_{X}=\Gamma_{Y}/2=\Gamma_{Z}\approx\frac{16}{T_1}\frac{T_{\mathrm{int}}}{ T_{\mathrm{round}}}$, which limit the lifetime of the three components of the logical Bloch vector to $T_X, T_Z \lesssim 8T_1\frac{ T_{\mathrm{round}}}{T_{\mathrm{int}}}$ and $T_Y \lesssim 4T_1\frac{ T_{\mathrm{round}}}{T_{\mathrm{int}}}$. With a similar reasoning, we can show that $T_X, T_X, T_Z \lesssim 6T_1\frac{ T_{\mathrm{round}}}{T_{\mathrm{int}}}$ for the hexagonal code. \\

This picture is only slightly modified when considering the real interaction Hamiltonian (see Eq.~\ref{eq:H_tilde}). In that case, beyond the random displacements that we just described, transmon bit-flips also lead to spurious rotations of the oscillator state. Such rotations arise from the dispersive transmon-oscillator coupling, which are normally echoed out by regular $\pi$-rotations of the transmon applied every $T_e\sim T_{\mathrm{int}}/2$. A transmon bit-flip acts as a supplementary echo which breaks the regularity of the echo sequence and can result in a rotation by an angle $\phi \leq \chi T_e \approx 10^{\circ}$ in our experiment. These rotations increase slightly the logical error rate, but could be mitigated by performing more frequent echoes of the transmon during the conditional displacements (reversing accordingly the sign of the frame shift $\alpha$ in Eq.~2) and the readout. Such echoes become even more relevant as one considers stabilizing grid states with larger envelopes, for which the logical qubit is more sensitive to small rotations.  With our current error-correction parameters, master-equation simulations  reproduce qualitatively the error rate induced by transmon bit-flips as given in the previous paragraph (see Table.~2).

\section*{Teleported gates and arbitrary state preparation}

\begin{figure*}
\begin{center}
\includegraphics{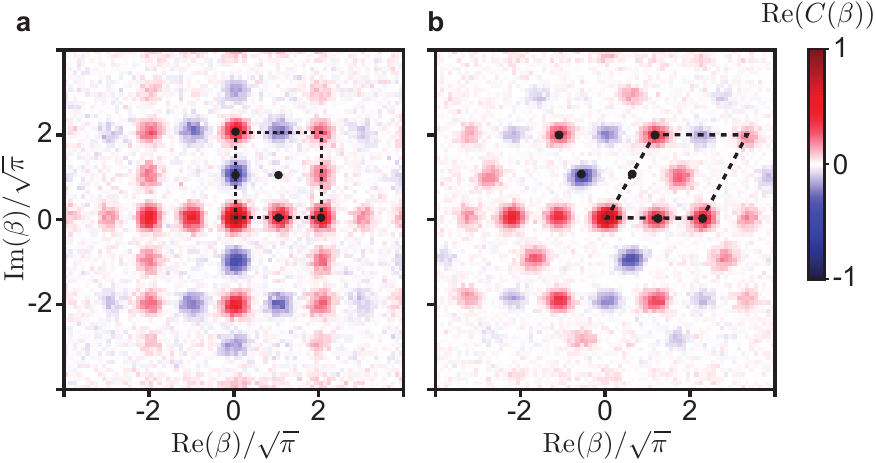}
    \caption{\label{fig5} {\bf Arbitrary state preparation} in the square {\bf a,} and hexagonal {\bf b,} codes. The sequence $\CD(a/2)$ applied on the storage-transmon $|+Z_L\rangle|+x\rangle$ state prepares the entangled Bell state $(|+X_L\rangle|+x\rangle+|-X_L\rangle|-x\rangle)/\sqrt{2}$, up to a re-centering displacement. Measuring the transmon in any state $|\psi\rangle$ then heralds the preparation of the state $|\Psi_L^{\ast}\rangle$ in the storage oscillator (logical Bloch vector with same polar angle and opposed azimuth to the transmon Bloch vector). Here $|\Psi_L^{\ast}\rangle=|M_L\rangle= \cos(\pi/8) | + X_L\rangle - \sin(\pi/8) | - X_L\rangle$.
    }
\end{center}
\end{figure*} 

In this section, we detail a protocol allowing one to perform an arbitrary manipulation of the logical qubit via a teleported gate~\cite{fluhmann2018encoding}, which allows for the preparation of arbitrary logical states. Here, for simplicity, we consider the infinitely squeezed code and identify the logical Pauli operators to the displacement operators of the oscillator.\\

In all generality, we consider the case of a target rotation around $\V$, where $\V=\X,~\Y$ or $\Z$. We can write any state $|\Psi_L\rangle$ of the logical qubit on the $|\pm V_L\rangle$ basis as $|\Psi_L\rangle=\alpha |+V_L\rangle +\beta |-V_L\rangle$.\\

The teleported gate starts with preparing the transmon in $|+x\rangle$. We then apply a conditional displacement $\CD(\gamma)$, with $\gamma$ chosen so that $\V=\D(\gamma)$ (same conditional displacement as if measuring $\V$). Finally, we apply an unconditional displacement  $\D(-\gamma/2)$ to recenter the code (see Sec.~\ref{sec:paulimeas}). At the end of the sequence, the joint transmon-oscillator state reads
\begin{equation}
\begin{split}
|\psi_{tot}\rangle &=\frac{ \alpha}{2} (\I+\V) |+V_L\rangle |+x\rangle +\frac{\alpha}{2} (\I-\V) |+V_L\rangle |-x\rangle \\
&~~~~+ \frac{\beta}{2} (\I+\V) |-V_L\rangle |+x\rangle +\frac{\beta}{2} (\I-\V) |-V_L\rangle |-x\rangle  \\
    &= \alpha |+V_L\rangle |+x\rangle +\beta |-V_L\rangle |-x\rangle.
\end{split}
\end{equation}
We then perform a transmon readout along an axis rotated by $(\theta, \phi)$ from $\sigmax$: the two pointer states of the measurement are $|\theta,\phi\rangle$ and $|\theta+\pi,\phi\rangle$ with the definition $|\theta,\phi\rangle=e^{i\phi/2}\cos \frac{\theta}{2}|+x\rangle +e^{-i\phi/2}\sin \frac{\theta}{2}|-x\rangle$. The Kraus operators acting on the oscillator state depending on the measurement outcome are $\M_{\theta,\phi}$ and $\M_{\theta+\pi,\phi}$ with 
\begin{equation}
    \M_{\theta,\phi}=e^{i\phi/2} \cos \frac{\theta}{2} |+V_L\rangle \langle +V_L| + e^{-i\phi/2}\sin \frac{\theta}{2} |-V_L\rangle \langle -V_L|.
\end{equation}
If we choose $\theta=\frac{\pi}{2}$, these operators are unitary up to a scaling factor. Applying a feedback $\V$-gate if the transmon is found in $(\theta+\pi,\phi)$, we find that 
\begin{equation}
   \frac{\M_{\theta,\phi} |\Psi_L\rangle}{\mathrm{Tr}(\M_{\theta,\phi} |\Psi_L\rangle)}=-\V \frac{\M_{\theta+\pi,\phi} |\Psi_L\rangle}{\mathrm{Tr}(\M_{\theta+\pi,\phi} |\Psi_L\rangle)} = \R_{V}(-\phi) |\Psi_L\rangle
\end{equation}
so that the whole sequence performs an unconditional rotation $\R_{V}(-\phi)$ of the logical qubit around $\V$ by an angle $-\phi$ with
\begin{equation}
   \R_V(-\phi)= e^{i\phi/2}  |+V_L\rangle \langle +V_L| + e^{-i\phi/2} |-V_L\rangle \langle -V_L|.
\end{equation}


We use this protocol to create the logical state $|M_L\rangle= \cos(\pi/8) | + X_L\rangle - \sin(\pi/8) | - X_L\rangle$ in both the square and hexagonal codes. To this end, we choose $\V=\X$ and $(\theta,\phi)=(-\frac{\pi}{4},0)$. When this sequence is applied  to the state $|+Z_L\rangle$ and when the transmon is measured in $|-\frac{\pi}{4},0\rangle$ (respectively in $|\frac{3\pi}{4},0\rangle$), it rotates the logical state by $3\pi/4$ around $\Y$ (respectively $-\pi/4$ around $\Y$). In the experiment, we herald the preparation of $|M_L\rangle = \cos \frac{\pi}{8}|+X_L\rangle -\sin \frac{\pi}{8} |-X_L\rangle$ when the transmon is measured in $|-\frac{\pi}{4},0\rangle$, but the preparation can be made deterministic by applying a $\Y$-gate when the transmon is measured in the orthogonal state. The characteristic function of the prepared states are represented in Fig.~\ref{fig5}. Note that preparing this state can be seen as the first step towards so-called magic state distillation \cite{Bravyi2005MagicState}.

\makeatother

\end{document}